\documentclass[10pt,leqno]{amsart}
\usepackage{amsmath,amsthm,amsfonts}
\usepackage {latexsym}
\usepackage{amssymb}
\usepackage[dvips]{epsfig}
\renewcommand{\thesection}{\arabic{section}}
\newcounter{num}

\newtheorem{theorem}{Theorem}[section]
\newtheorem{lemma}[theorem]{Lemma}
\newtheorem{prop}[theorem]{Proposition}

\newtheorem{corollary}[theorem]{Corollary}
\theoremstyle{definition}
\theoremstyle{remark}
\newtheorem{remark}[theorem]{Remark}
\newcommand{\definition}{\noindent{\it Definition.\quad}}

\renewcommand{\theequation}{\thesection .\arabic{equation}}
\let\subs\subsection
\renewcommand\subsection{\setcounter{equation}{0}
\gdef\theequation{\thesubsection \arabic{equation}}\subs}
\let\sect\section
\renewcommand\section{\setcounter{equation}{0}
\gdef\theequation{\thesection .\arabic{equation}}\sect}

\newcommand{\cD}{{\mathcal{D}}}
\newcommand{\cE}{{\mathcal{E}}}

\newcommand{\cJ}{{\mathcal{J}}}

\newcommand{\cM}{{\mathcal{M}}}

\newcommand{\cS}{{\mathcal{S}}}
\newcommand{\cF}{{\mathcal{F}}}

\newcommand{\cL}{{\mathcal{L}}}

\newcommand{\IC}{{\mathbb{C}}}

\newcommand{\IN}{{\mathbb{N}}}
\newcommand{\IR}{{\mathbb{R}}}
\newcommand{\IQ}{{\mathbb{Q}}}
\newcommand{\TT}{{\mathbb{T}}}

\newcommand{\tor}{\TT}
\newcommand{\IZ}{{\mathbb{Z}}}

\def\w{\omega}
\def\ab{(a, b)}
\def\xw{(x,\w)}
\def\xwn{(x, \w)(n)}
\def\xwo{(x, \w)(0)}
\def\xwe{(x, \w, E)}

\def\hab{{H_{[a, b]}}}
\def\hac{{H_{[a, c]}}}
\def\mab{{M_{[a, b]}}}
\def\mac{{M_{[a, c]}}}

\def\mes{\mathop{\rm{mes}}}
\def\compl{\mathop{\rm{compl}}}
\def\dist{\mathop{\rm{dist}}}
\def\diag{\mathop{\rm{diag}}}
\def\ran{\mathop{\rm{Ran}}}
\def\rsp{\mathop{\rm{sp}}}
\def\rank{\mathop{\rm{rank}}}
\def\ree{\mathop{\rm{Re}}}
\def\sgn{\mathop{\rm{sgn}}}
\def\la{\langle}
\def\ra{\rangle}
\def\ve{\varepsilon}
\def\vp{\varphi}
\def\vr{\varrho}
\def\tcup{\mathop{\textstyle{\bigcup}}\limits}
\def\tcap{\mathop{\textstyle{\bigcap}}\limits}
\def\tvp{\tilde{\vp}}
\def\vt{\vartheta}

\def\ka{\kappa}
\def\K{\kappa}

\def\tb{{\widetilde B}}
\def\te{{\widetilde E}}

\def\tf{{\tilde f}}
\def\th{{\widetilde H}}
\def\tn{{\widetilde N}}
\def\tm{{\widetilde M}}
\def\tw{{\tilde \w}}
\def\tx{{\tilde x}}

\def\tv{{\widetilde V}}
\def\hv{{\widehat V}}
\def\hh{{\widehat H}}
\def\he{{\widehat E}}
\def\O{\Omega}
\def\zero{{(0)}}

\def\one{{(1)}}
\def\oone{{\O^\one}}
\def\two{{(2)}}
\def\three{{(3)}}
\def\so{{(s+1)}}
\def\smo{{(s-1)}}

\def\spo{{{s+1}}}

\def\bw{{\bar \w}}
\def\uw{{\underline \w}}
\def\be{{\overline E}}
\def\ue{{\underline E}}

\def\ijk{ijk}

\def\pjk{pjk}

\def\hmap{{H_{[-N_1, N_1]}}}
\def\htmap{{H_{[-\tn, N]}}}

\def\fmap{{f_{[a,b]}}}
\def\facmap{{f_{[a, c]}}}
\def\ftmap{{f_{[-\tn, N]}}}
\def\emap{E^\one, \dots, E^{(s)}}
\def\vpmap{\vp^\one, \dots, \vp^{(s)}}
\def\xmap{{x_0 < x_1 < \dots < x_n}}
\def\lmap{{\lambda_1 \le \dots \le \lambda_n}}
\def\mmap{{\mu_1 \le \dots \le \mu_n}}

\def\pamap{\bigl(A'(x_0)\vp_0,\vp_0\bigr)}
\def\ppmap{\bigl(P'(x_0)\vp_0,\vp_0\bigr)}
\def\amap{\bigl(A(x_0)\vp_0, \vp_0\bigr)}
\def\pmap{\bigl(P(x_0)\vp_0, \vp_0\bigr)}

\title{}
\begin{document}

\title{Method of variations of potential of quasi-periodic Schr\"{o}dinger equation}
\author{Jackson Chan \\ University of Toronto}
\date{}
\maketitle
\begin{abstract}
We study the one-dimensional discrete quasi-periodic Schr\"{o}dinger equation
$$
-\vp(n+1)-\vp(n-1)+\lambda V(x+n\w)\vp(n)=E\vp(n), \qquad n \in \mathbb{Z}
$$
We introduce the notion of variations of potential and use it to define ``typical'' potential.  We show that for ``typical'' $C^3$ potential $V$, if the coupling constant $\lambda$ is large, then for most frequencies $\w$, the Lyapunov exponent is positive for all energies $E$.
\end{abstract}


\section{Introduction}
Given any function $V: \tor \rightarrow \IR$, we have a family of quasi-periodic discrete Schr\"{o}dinger equations
\begin{equation}
-\vp(n+1)-\vp(n-1)+\lambda V(x+n\w)\vp(n)=E\vp(n), \qquad n \in \mathbb{Z}
\end{equation}
where $(x,\w) \in \tor \times \tor$ are parameters.

Equation (1.1) can be rewritten as a first order difference equation:
$$
\left( \begin{array}{c} \vp(n+1) \\ \vp(n) \end{array} \right) = 
\left( \begin{array}{cc} \lambda V(x+n\w)-E & -1 \\ 1 & 0 \end{array} \right)
\left( \begin{array}{c} \vp(n) \\ \vp(n-1) \end{array} \right).
$$
The \emph{monodromy} matrix of this equation is
$$
M_{[a,b]}(x,\w,E) = 
\left( \begin{array}{cc} \lambda V(x+b\w)-E & -1 \\ 1 & 0 \end{array} \right)
\ldots
\left( \begin{array}{cc} \lambda V(x+a\w)-E & -1 \\ 1 & 0 \end{array} \right).
$$
To study the properties of the spectrum and eigenfunctions of equation (1.1) for generic $(x,\w)$, we make use of the \emph{Lyapunov exponent}.  It is defined by
$$
L(\w,E) = \lim_{n \rightarrow \infty} {1 \over n} 
\int_\tor \left\| M_{[1,n]}(x,\w,E) \right\| dx
$$
By Kingman's subadditive ergodic theorem, if the shift $x \mapsto x+\w$ is ergodic, then the limit exists and 
$$
L(\w,E) = \lim_{n \rightarrow \infty} {1 \over n} 
\left\| M_{[1,n]}(x,\w,E) \right\|
$$
for almost all $x \in \tor$.

For analytic potential $V$, Bourgain, Goldstein[BG] and Goldstein, Schlag[GS1] showed that the Lyapunov exponent is positive for large $| \lambda |$.  Furthermore, if $L(\w_0,E_0) >0$, then for most $\w$ close to $\w_0$, there exists $\delta>0$ such that
\begin{quote}
the spectrum of (1.1) in $(E_0-\delta, E_0+\delta)$ is pure point and the corresponding eigenfunctions decay exponentially.
\end{quote}
This property is called \emph{Anderson localization}.  (See the recent monograph by Bourgain[Bo] for more details.)  Generalization of the method of [BG] and [GS] for potential in a Gevrey-class was established by Klein[Kl].  

In this paper, we use Goldstein and Schlag's methods introduced in a recent work[GS2] to study the spectrum and eigenfunctions of equation (1.1).  We develop these methods for smooth potentials in the perturbative regime.

It is well known that the monodromy matrix is of the form
$$
M_{[a,b]}(x,\w,E) = \left( \begin{array}{cc}
f_{[a,b]}(x,\w,E) & -f_{[a+1,b]}(x,\w,E) \\
f_{[a,b-1]}(x,\w,E) & -f_{[a+1,b-1]}(x,\w,E) \end{array} \right)
$$
where
$$
f_{[p,q]}(x,\w,E) = \det \bigl[ H_{[p,q]}(x,\w) - E \bigr]
$$
\vspace{2 mm}
$$
H_{[p, q]}(x,\w) = \begin{pmatrix}
\lambda V(x+p\w) & -1 &&\\[6pt]
-1 & \lambda V\bigl(x+ (p+1)w\bigr) & -1 &\\[6pt]
& -1 &&\ddots\\[6pt]
&\ddots && -1\\[6pt]
&& -1 & \lambda V(x + q\w)\end{pmatrix}
$$
The spectrum of the Schr\"{o}dinger equation
$$
-\vp(n+1)-\vp(n-1)+\lambda V(x+n\w)\vp(n)=E\vp(n),
$$
with zero boundary conditions $\vp(a-1)=0$ and $\vp(b+1)=0$, consists of the 
eigenvalues of $H_{[a,b]}(x,\w)$.

If $V(x+m\w) \approx V(x+n\w)$ for some $m,n \in [a,b]$, $m \ne n$, 
one says that the equation exhibits ``resonance''. 
Difficulties in studying the spectrum of $H_{[a,b]}(x,\w)$ arise when 
resonances occur; two eigenvalues of $H_{[a,b]}(x,\w)$ can be very close and 
the corresponding eigenfunctions need not be ``localized''.

We develop methods to deal with resonances by defining 
some functions $E^{(s)}(x,\w)$ which take their values in 
$\rsp H_{[-N_s,N_s]}(x,\w)$, for appropriate $N_{s+1} \asymp e^{N_s^\tau}$, 
$s=1,2, \ldots$  
In Part I, for $C^1$ potential, we eliminate a small part of $\tor \times \tor$
from the domain of $E^{(s)}(x,\w)$ at each step.  This allows us to exclude the case when two eigenvalues of $H_{[a,b]}(x,\w)$, $b-a \asymp N_{s+1}$, become 
too close.  We are then able to define exponentially 
decaying eigenfunctions.  The result is the following theorem.

\begin{theorem} Given $V \in C^3(\tor)$, let $\cJ = V(\tor)$.  There exists $\lambda_0 = \lambda_0(V)$ 
such that, for any $| \lambda | > \lambda_0$, the following holds: \newline
There is $\O = \O (V,\lambda) \in [0, 1]$, 
$\mes \bigl([0,1]\setminus \O\bigr) \lesssim \lambda^{-1/2}$,
such that for any $\w \in \O$ there exists $\cE_\w \subset \lambda \cJ$, 
$\mes \bigl(\cJ \setminus \lambda^{-1}\cE_\w\bigr) \lesssim \lambda^{-1/2}$, 
with $L(\w,E) \gtrsim \log \lambda$ for all $E \in \cE_\w$. 
Also, for any $E \in \cE_\w$, there is $x \in \tor$ and $\{\vp(n)\}$, 
$|\vp(n)| \lesssim e^{-c|n|}$ such that 
$$ -\vp(n+1) - \vp(n-1) + V(x + n\w)\vp(n) = E\vp(n). $$
\end{theorem}

In a recent paper, Bjerkl\"{o}v[Bj] has obtained similar results.

The major drawback of the method in Part I is that part of the spectrum is 
also eliminated.  To include all spectral values, we use variations of potential in Part II.  Given a $C^3$ potential $V$, any $C^3$ function $\tilde V$ satisfying the conditions
\begin{align*}
\max_{x\in\tor} | V(x) - \tilde V(x) | & < \delta \\
\max_{x\in\tor} | V'(x) - \tilde V'(x) | & < \delta \\
\max_{x\in\tor} | V''(x) - \tilde V''(x) | & < \delta
\end{align*}
can be written, near $x=0$, in the form
$$ 
\tilde V(x) = V(x) + \eta + \xi x + {1\over 2} \theta x^2 + x^3 R(x)
$$
where $|\eta|, |\xi|, |\theta| < \delta$, $R \in 
C^3 (\tor \setminus \{ 0 \})$, 
$|\partial^\alpha R| \lesssim 1$  for any index $ |\alpha| \le 2$.  More generally, since $\tor$ is compact, we can find large integer $T$ so that 
\begin{equation}
\tilde V(x) = V(x) + \sum\limits_{m=1}^{T} \Bigl[ \eta_m + \xi_m \bigl(
x-{m\over T} \bigr) + {1\over 2} \theta_m \bigl( x-{m\over T} \bigr)^2 + 
\bigl( x- {m\over T} \bigr)^3 R_m \bigl( x-{m\over T} \bigr) \Bigr]
\end{equation}
for all $x \in \tor$, where $\eta=(\eta_1, \ldots , \eta_T),\ \xi=(\xi_1, \ldots , \xi_T),\ \theta=(\theta_1, \ldots , \theta_T) \in \prod\limits_1^T [-\delta,\delta]$, and $R_m \in C^3(\tor \setminus \{ 0 \})$, $|\partial^\alpha R_m| \lesssim 1$ for any index $ |\alpha| \le 2$.
This motivate the following definition.

\definition  Let $T$ be a large integer, $0< \delta \ll {1\over T^5}$.  Suppose $R_m(\eta_m,\xi_m,\theta_m;x)$ are $C^3$ functions, $m=1,2, \ldots, T$, $(\eta,\xi,\theta) \in \prod\limits_1^{3T} [-\delta,\delta]$, $x \in \tor$, satisfying the following conditions:
\begin{align}
|\partial_\alpha R_m(\eta_m,\xi_m,\theta_m;x)| \lesssim  {1\over T} & 
\qquad \text{for any index $|\alpha|\le 3$} \\
R_m(0,0,0;x) \equiv 0 & \\
R_m(\eta_m,\xi_m,\theta_m;x)=-x^{-3}\bigl( \eta_m + \xi_m x + {1\over 2}\theta_m x^2 \bigr)& 
\qquad \text{for $|x| \ge {1\over 2T}$}
\end{align} 
Define a $(T,\delta)$--variation of potential by
$$ 
W(\eta,\xi,\theta,\{ R_m \} ; x) = \sum\limits_{m=1}^{T} 
v_m\Bigl(\eta_m,\xi_m,\theta_m;x-{m\over T} \Bigr) 
$$
where
$$ v_m(\eta_m,\xi_m,\theta_m;x) = \eta_m + \xi_m x + {1\over 2} \theta_m x^2 + 
x^3 R_m (\eta_m,\xi_m,\theta_m;x) $$
 
By (1.4) and (1.5),
\begin{align*}
v_m(0, 0, 0; x) & \equiv 0 \\ 
\intertext{and}
v_m(\eta_m, \xi_m, \theta_m; x) & = 0\qquad \text{for $|x| \ge {1\over 2T}$}\ .
\end{align*}

Denote the collection of $(T,\delta)$--variations of potential by $\cS(T,\delta)$.  The set of parameters $(\eta, \xi, \theta)$ has measure $(2 \delta)^{3T}$.  We want to define a notion of ``typical'' potential by using the normalized measure on this set of parameters.  Hence, a set $S \subset \cS(T,\delta)$ is called $(1-\ve)$-typical if 
$$
|S| := \min\limits_{ \{ R_m \} } {1\over (2\delta)^{3T}}
\mes\bigl\{(\eta, \xi, \theta) \in [-\delta,\delta]^{3T}: 
W(\eta, \xi, \theta, \{ R_m \} ; . ) \in S \bigr\} \ge 1 - \ve
$$

The main result in this paper, summarized in the following theorem, is that for typical $C^3$ potential $\tilde V$, we have positive Lyapunov exponent for all energies $E$.

\begin{theorem} Given any $V \in C^3(\tor)$, $|V'(x)|+|V''(x)| \ge c > 0$, there is $\lambda_0=\lambda_0(V)$ such that for $|\lambda| > \lambda_0$, one has a collection of variations $\{ S_\ell=S_\ell(V,\lambda) \}_{\ell=1}^\infty$, $S_\ell \subset \cS(T^{(\ell)},\delta_\ell)$, $\log T^{(\ell+1)} \asymp \bigl( T^{(\ell)} \bigr)^\alpha$, $0< \alpha \ll 1$, $\sum\limits_{\ell=1}^{\infty} \bigl(1-|S_\ell|\bigr) \le \lambda^{-\beta}$, so that for any potential 
$$ \tv(x) = V(x) + \sum\limits_{\ell=1}^\infty 
 W^{(\ell)}\bigl(\eta^{(\ell)}, \xi^{(\ell)}, \theta^{(\ell)}, \{ R^{(\ell)}_m \} ; x \bigr) $$ 
where $W^{(\ell)} \in S_\ell$, there exists $\Omega=\Omega(\lambda,\tilde V)$, $\mes ( \tor \setminus \Omega ) \le \lambda^{-\beta}$, so that the Lyapunov exponent $L(\w,E) \ge {1 \over 4} \log \lambda$ for any $\w \in \Omega,\ E\in \IR$.
\end{theorem}

There are two central technical problems which one has to deal with in order to establish Theorem 1.2.  The first one consists of the splitting of eigenvalues of equation (1.1) on a finite interval $[-N,N]$.  The technology for this splitting was developed in the recent work by Goldstein and Schlag[GS2] in the case of analytic potential.  It is based on avalanche principle[GS1], elimination of resonances and localized eigenfunctions on a finite interval.  We modify this method for smooth potential and prove the following result:

\begin{prop}  Using the notation of Theorem 1.2, there exists integers $T'_s$, $\log T'_s \asymp \log T^{(s)}$, such that for any nested sequence of intervals $\cF_{s,k_s} = [ {k_s \over T_s}, {k_s + 1 \over T_s} )$, and $x\in\tor, \w\in\O$, there is a sequence integers 
$\{ N_s = N_s(x,\w) \}$, with $\log N_{s} \asymp \log T'_s$, so that 
\begin{equation} |E_1-E_2| > \exp (-N_s^{\tau}) \end{equation}
for distinct eigenvalues $E_1,E_2 \in \bigl( \rsp H_{[-N_s,N_s]} (x,\w) \bigr) \cap \cF_{s,k_s}$.
\end{prop}

The second problem is as follows.  The eigenvalues of the problem (1.1) on a finite interval [1,N] have a smooth parameterization as $E_1(x) < E_2(x) < \ldots 
< E_N(x),\ x \in \tor$.  This general result is due to the self adjointness of the the problem (1.1) and non-degeneracy of the the eigenvalues of (1.1) restricted on a finite interval.  The problem is how to eliminate multiple resonance, i.e. $|E_i(x)-E_j(x+m\w)|<\delta$ and $|E_i(x)-E_k(x+n\w)|<\delta$ for $m\ne n$.  To deal with this problem, we need to evaluate the quantity
\begin{equation}
|\partial_x E_j| + |\partial_{xx} E_j|
\end{equation}
from below.  

This problem was also studied in [GS2];  for analytic potential, the problem was solved using discriminant of polynomials and Sard-type arguments.  This method has no modification for smooth potentials.  To solve this problem, we need to introduce variations of the potential.  The most basic idea of our method is as follows.

``Typical'' smooth functions $F(x)$ are Morse functions, i.e. the quantity
\begin{equation}
|\partial_x F| + |\partial_{xx} F|
\end{equation}
has a ``good'' lower bound, gauged according to the ``size'' of $F$.  On the other hand, there is a basic relation between $\partial_x E_j$ and the potential $V(x)$:
\begin{equation}
\partial_x E_j = \sum\limits_{k=1}^{N} V'(x+k\w) |\vp_j(x)(k)|^2
\end{equation}
where $\vp_j(x)(.)$ is a normalized eigenfunction of (1.1) on the interval $[1,N]$ corresponding to $E_j(x)$.  Relation (1.9) enables one to express the ``genericity'' of the potential $V$ in terms of the lower bound for (1.7), provided $\vp_j(x)(.)$ is exponentially localized.  We use Sard-type arguments to show that the total mass of those $\w$ for which there is no ``response'' in (1.9) under the variations of $V$, is extremely small.

We will use a KAM-type approach to prove Theorem 1.2.  At each scale, we need to establish the following inductive hypothesis:
\begin{list}{h\arabic{num}.}{\usecounter{num}}
\item Localization in finite interval.  This means that at each scale, we can find $N_s$ such that the eigenfunctions of $H_{[-N_s,N_s]}$ decay exponentially for $|n|>\sqrt{N_s}$
\item Estimate the number of eigenvalues and separation of eigenvalues.
\item $|\partial_x E|+|\partial_{xx} E|> exp(-N_s^\sigma)$
\item Elimination of multiple resonances.
\end{list}

In section 6, we first use a Diophantine condition to obtain an upper bound of the number of entries that can lead to resonance.  Using this bound, we can find $N_1$ such that the entries where resonance occur is far away from the edge of $[-N_1,N_1]$.  This will give us exponentially decaying eigenfunctions, hence (h1) in the first scale.  We show how (h1) implies (h2) in the first scale in Section 7; this will be the base case for Proposition 1.3.

In section 8, we introduce variations of potential.  Using the results from Appendix F, we show that for typical potential, if the eigenvalues are separated, then (1.7) has a ``good'' lower bound.  Due to its technical nature, we defer the proof of (h4) in Appendix E.  There, we show that if (h3) is satisfied, then multiple resonances do not occur for most frequencies $\w$.  This forms a major part of the inductive hypothesis used in Section 9.  With no multiple resonances, we can define inductively $C^3$ functions $E^{(s)}_k(x,\w)$ which take values in $\rsp H_{[-N_s,N_s]}(x,\w)$, with exponentially decaying eigenfunctions.

In Section 10, we take $\tilde V$ as a limit of the varied potential.  We show that lower estimate for (1.7), which we have established for $E^{(s)}(x,\w)$, still holds for the eigenvalues of $H_{[-N_s,N_s]}(x,\w)$ with the limiting potential $\tilde V$.  Then, we will be able to  prove Theorem 1.2.

Author is thankful to M. Goldstein for his suggestion to study the case of smooth quasi-periodic potential using the variations of the potential and for advising author during the work on this project.  Author also benefited a great deal from learning the material of the work by Goldstein and Schlag[GS2], while it was in progress during the last few years.


\part{}
\setcounter{section}{1}
\section{Elimination of flat slope}
In Part I, we consider a $C^1$ potential $V(x)$.  It is well known that problems in studying the eigenvalues and eigenfunctions for quasi-periodic potential, to large extent, arises from the small denominator in the Green's function.  The KAM type approach to this problem developed by Goldstein, Schlag[GS1, GS2] consists of so-called multi-scale analysis which relate the eigenvalues of a relatively short interval with those of a longer interval.  

At initial scale, the small denominator is of the form $V(x+m\w)-V(x+n\w)$, $-N_1 \leq m,n \leq N_1$.  Clearly, when $|V'(x)|$ is small, this brings additional difficulties to the problem, since $|V(x)-V(x+n\w)|$ will be small for most frequencies $\w$.  In this section, we eliminate the part of the torus where the slope of $V$ is flat, i.e. $|V'(x)|$ is small.  We then eliminate a set of frequencies that lead to resonance.

Throughout Part I, let $V \in C^{1}(\mathbb{T})$, $\ \max\limits_{x \in \mathbb{T}}\, V(x) - \min\limits_{x\in \mathbb{T}}\, V(x) \le \max\limits_{x \in T} |V'(x)| \le C_0$; furthermore, assume $V$ has $m_0$ monotonicity intervals.  
Set $A = \{x \in \mathbb{T}: |V'(x)| < \ve\}$.  Next two lemmas give an 
upper bound of the measure of the spectral values that we have to exclude due 
to the elimination of flat slope.

\begin{lemma}\label{lem:2-1}
$\mes V(A) \le \ve$
\end{lemma}

\begin{proof} $\ve \ge \int_A |V'(x)| dx = \int_{\IR} 
\#\{x \in A: V(x) = E\}dE \ge \mes V(A)$.
\end{proof}

Take $\cJ = [\min\limits_{x \in \mathbb{T}} V(x), 
\max\limits_{x \in \mathbb{T}} V(x)]$.  Let 
$\cJ \setminus V(A) = \bigcup\limits_i[\alpha_i, \beta_i]$, 
$h_i = \beta_i - \alpha_i$.  Define
$h: [0, \infty) \to [0, \infty)$ by $h(y) = \sum_{i: h_i \le y}\, h_i$.  $h$ is increasing, continuous from the right, and $h(0) = 0$.  Hence, 
there is $\delta > 0$ such that $h(\delta) < \ve$.  Fix 
$0 < \delta < \ve^{10} \ll {1\over C_0}$ such that $h(\delta) < \ve$.  Set 
$R = \{i: h_i > \delta\}$.  Let 
$$ 
\cE^\zero = 
\bigcup\limits_{i \in R}[\alpha_i + 2 \delta^2,\ \beta_i - 2 \delta^2] 
$$

\begin{lemma}\label{lem:2-2}
$\#R \le C_0 \delta^{-1}$, $\mes(\cJ \setminus \cE^\zero) \le 2 \ve + 
4C_0\delta$.
\end{lemma}

\begin{proof} 
$$ (\#R)\delta \le \sum\limits_{i \in R}\, h_i \le C_0 $$
$$
\mes(\cJ \setminus \cE^\zero) = \mes V(A) + h(\delta) + (\#R)(4\delta^2) 
\le 2 \ve + 4C_0 \delta
$$
\end{proof}

Next, we eliminate some frequencies to avoid resonances.  Having 
eliminate flat slope, we can obtain an upper bound of the measure of 
the set of frequencies we eliminate.

For any $i \in R$, let $\alpha_i = y_{i0} < y_{i1} < \dots < 
y_{i,n_i} = \beta_i$, \, $\delta^2 \le y_{ik} - y_{i,k-1} < 2 \delta^2$ 
for $2 <k < n_i -1$ and 
$y_{i1} - y_{i0} = y_{i2} - y_{i1} = \delta^2 = y_{i,n_i -1} - y_{i,n_i -2} = 
y_{i,n_i} - y_{i,n_i -1}$.  For $1 \leq k \leq n_i$, consider 
$T_{ik} = V^{-1}\bigl([y_{i,k-1}, y_{ik}]\bigr)$.  Since $V$ has $m_0$ monotonicity intervals, $T_{ik}$ consists of at 
most $m_0$ intervals, each with length less than 
$(2\delta^2) \ve^{-1} < 2\delta$.

Say $T_{ik} = \bigcup\limits^{J_k}_{j=1} [a_{ikj}, b_{ikj}]$, $J_k \le m_0$.  
Set $\tilde B_{ikj} = \bigcup\limits_{|\ell -k| \le 2}\, 
\bigcup\limits_{1 \le m \le J_\ell}\bigl\{[a_{i \ell m}, b_{i \ell m}] - 
[a_{ikj}, b_{ikj}]\bigr\}$,
\begin{align*}
\tilde B_{ikj}(n) & = \left\{\w \in [0, 1]: \{ n\w\} \in \tb_{ikj}\right\},\\[6pt]
B_{ikj} & = \bigcup\limits_{0 <|n| \le N^2_1}\, \tb_{ikj}(n)\qquad 
\text{where}\ N_1 = \lfloor \delta^{-1/4}\rfloor\\[6pt]
\cD^\one_i & = \bigcup\limits_{1 < k < n_i}\, \bigcup\limits_{j =1}^{J_k} 
(a_{ikj}, b_{ikj}) \times\bigl([0, 1]\setminus B_{ikj}\bigr)
\end{align*}

\begin{lemma}\label{lem: 2-3} Let $E \in [\alpha_i + 2\delta^2, \beta_i - 
2\delta^2]$, $i \in R$.  Suppose $(x,\w) \in \cD^\one_i$ such that 
$|V(x) - E| \le \delta^2$.  Then $|V(x+j\w) - E| > \delta^2$ for 
$0 < |j| \le N^2_1$.
\end{lemma}

\begin{proof} $E \in [y_{i,k-1}, y_{ik}]$ for some $k$, $2 < k < n_i -1$.  
$|V(x) - E| \le \delta^2 \Longrightarrow x \in T_{i \ell}$, $|\ell - k| \le 1$.
  Say $x \in [a_{i \ell m}, b_{i \ell m}]$.  If $|V(x+j\w) - E| \le \delta^2$ 
then $x + j\w \in T_{i \tilde\ell}$, $|\tilde\ell - k| \le 1$.  Thus, we have 
$|\ell - \tilde\ell| \le 2 \Longrightarrow j \w \in \tb_{i \ell m} 
\Longrightarrow \w \in \tb_{i \ell m}(j)$.  This contradicts the construction 
of $\cD^\one_i$ if $0 < |j| \leq N^2_1$.  Hence $|V(x+j\w) - E| > \delta^2$ for $0 < |j| \le N^2_1$.
\end{proof}

Define $$ \cD^\one = \bigcup\limits_{i \in R}\, \cD^\one_i $$

\begin{corollary}\label{cor:2-4}  If $E \in \cE^\zero$, $(x, \w) \in \cD^\one$,
 $|V(x) - E| \le \delta^2$.  Then $|V(x+j\w) -E| > \delta^2$ for 
$0 < |j| \le N^2_1$.
\end{corollary}

For later sections, we also need an upper bound of the number of components in $\cD^\one$.

\definition A set $U_1 \subset \IR$ (or $U_2 \subset \IR^2$) is said to be 
$K$--simple if $U_1 = \bigcup\limits^K_{k=1}\, \cJ_k$ where $\cJ_k$ are 
intervals ($U_2 = \bigcup\limits^K_{k=1} D_k$ where $D_k$ are rectangles).  
If $U_1$ is $K$--simple, write $\compl U_1 \le K$.

\begin{lemma}\label{lem:2-5} For any $i \in R$, $\compl B_{ikj} 
\le 5 m_0 \delta^{-1}$, $\mes B_{ikj} < 40 m_0 \delta^{1/2}$.
\end{lemma}

\begin{proof} $\left\{[a_{i \ell m}, b_{i \ell m}] - [a_{ikj}, b_{ikj}]
\right\}$ is an interval.  Hence $\compl \tb_{ikj} \le 
\sum\limits_{|\ell - k| \le 2}\, J_\ell \le 5 m_0$
\begin{equation*}
\begin{split}
& \compl \tb_{ikj}(n)  \le n(5 m_0)\\[6pt]
& \compl B_{ikj}  \le 5 m_0 \sum_{0 < |n| \le N^2_1}\, n \le 5 m_0(N^2_1)^2 
  \le 5 m_0 \delta^{-1}\\[6pt]
& \mes \left\{[a_{i \ell m}, b_{i \ell m}] - [a_{ikj}, b_{ikj}]\right\} 
  \le (b_{i \ell m} - a_{i \ell m}) + (b_{ikj} - a_{ikj}) < 4\delta \\[6pt]
& \mes \tb_{ikj}(n) = \mes \tb_{ikj} < (5m_0)(4\delta) = 20m_0\delta\\[6pt]
& \mes B_{ikj} < (2N^2_1)(20 m_0 \delta) \le 40 m_0 \delta^{1/2}
\end{split}
\end{equation*}
\end{proof}

\begin{corollary} $\compl \cD^\one \le 5 m^2_0 C_0 \delta^{-3}$
\end{corollary}

\begin{proof}
\begin{align*}
\compl \cD^\one_i & \le n_i m_0(5 m_0 \delta^{-1})\\[6pt]
\compl \cD^\one & \le 5 m_0^2 \delta^{-1} \sum_{i \in R}\, n_i \le 5 m^2_0 \delta^{-1}(C_0 \delta^{-2})
\end{align*}
\end{proof}

\begin{corollary} 
Let $L = \sum\limits_{i \in R}\, \sum\limits_{1 < k < n_i}\,
\sum\limits_j (b_{ikj} - a_{ikj})$. Then
$$
\mes \Biggl(\bigcup_{i \in R}\, \bigcup_{1 < k < n_i}\,
\bigcup_j (a_{ikj}, b_{ikj})\times [0, 1]\setminus \cD^\one\Biggr) 
\le 40 m_0 \delta^{1/2} L
$$
\end{corollary}

To simplify notation, henceforth we will write $\cD^\one = \bigcup_k\, 
\bigcup_{\ell} (a^\one_k, b^\one_k)\times 
(\uw^\one_{k \ell}, \bw^\one_{k \ell})$.


\setcounter{section}{2}
\section{Eigenvalues and Eigenfunctions at first scale}
We now construct a function $E^{(1)}(x,w)$, which take values in $\rsp H_{[-N_1,N_1]}(x,\w)$, with corresponding eigenfunction $\vp^{(1)}(x,\w)$, which is exponentially decaying.  We will also derive some additional properties of $E^{(1)}$.

Take $\lambda = \delta^{-4}$,

$$
H_{[-N_1, N_1]}(x, \w) = \begin{pmatrix}
\lambda V(x - N_1 \w) & 1 &&\\
-1 & \lambda V\bigl(x-(N_1 -1)\w\bigr) & 1 &\\
& -1 & &\\
&& \ddots &\\
&&& -1\\
&& -1 & \lambda V(x + N_1 \w)
\end{pmatrix}
$$

\begin{lemma}\label{lem:3-1}
Let $(x,\w) \in \cD^\one$.  Then $H_{[-N_1, N_1]}(x,\w)$ has a unique eigenvalue in $\bigl[\lambda V(x) -2, \lambda V(x) +2\bigr]$.
\end{lemma}

\begin{proof} $(x,\w) \in \cD^\one \Longrightarrow V(x) \in \cE^\zero$, $|V(x+j\w) - V(x)| > \delta^2 = \lambda^{-1/2}$ for $0 < |j| \le N_1$. By Lemma B.4, $H_{[-N_1, N_1]}(x,\w)$ has a unique eigenvalue in $\bigl[\lambda V(x) -2, \lambda V(x) +2\bigr]$.
\end{proof}

Denote the eigenvalue of $H_{[-N_1, N_1]}(x,\w)$ in Lemma \ref{lem:3-1} by $E^\one(x,\w)$.

\begin{lemma}\label{lem:3-2}
$E^\one(x,\w)$ is $C^1$  in $\cD^\one$.  There are $C^1$ functions $\left\{\vp^\one (x, \w)(n)\right\}_{|n|\le N_1}$ defined in $\cD^\one$ such that $\sum\limits^{N_1}_{n = - N_1} \big |\vp^\one (x, \w)(n)\big |^2 = 1$,
$$
\hmap(x,\w) \vp^\one(x,\w) = E^\one(x,\w) \vp^\one(x,\w)\ .
$$
Also, 
\begin{align*}
\partial_x E^\one(x,\w) & = 
\sum^{N_1}_{n = -N_1}\, V'(x +n\w) |\vp^\one(x, \w)(n)|^2\\[6pt]
\partial_\w E^\one (x,\w) & = 
\sum^{N_1}_{n = -N_1}\, nV'(x + n\w) |\vp^\one(x, \w)(n)|^2
\end{align*}
\end{lemma}

\begin{proof} These follow from general results in Appendix A.
\end{proof}

\begin{lemma} \label{lem:3-3}
$\big |\vp^\one(x, \w)(n)\big | < \lambda^{-{1\over 3}|n|} \big | 
\vp^\one(x, \w)(0)\big |$.
\end{lemma}

\begin{proof} We prove the case for $n > 0$; the case for $n< 0$ is similar.

By Poisson's formula,
$$
\vp^{(1)}(x, \w)(n) = 
\left[H_{[1, N_1]} - E^\one (x, \w)\right]^{-1}(n, 1)\vp^\one (x, \w)(0)
$$
Since
\begin{align*}
|\lambda V(x + m\w) - E^\one(x,\w)| & \ge |\lambda V(x + m\w) - \lambda V(x)| - |\lambda V(x) - E^\one (x,\w)|\\
& > \lambda^{1/2} -2\qquad \text{for $1 \le m \le N_1$}\ ,
\end{align*}
and $N_1 \le \delta^{-1/4} = \lambda^{1/16}$, by Corollary C.8, one has
$$
|\vp^\one (x, \w)(n)| < \lambda^{-{1\over 3}|n|}| \vp^\one(x, \w)(0)|\ .
$$
\end{proof}

\begin{corollary}\label{cor:3-4}
\begin{align*}
{1\over 2}  \lambda^{39/40} & <|\partial_x E^{(1)}(x,\w)| \le C_0 \lambda\\
& |\partial_\w E^{(1)}(x,\w)| \le C_0 N_1 \lambda
\end{align*}
\end{corollary}

\begin{proof}
\begin{gather*}
\sum_{n\ne 0} |\vp^\one \xwn |^2 \le 2 \sum^\infty_{n=1}\, 
\lambda^{-{1\over 3}n} \le 4 \lambda^{-{1\over 3}}\\
(x,\w) \in \cD^\one \Longrightarrow |V'(x)| \ge \ve > \delta^{1\over 10} = 
\lambda^{-{1\over 40}}
\end{gather*}

\begin{align*}
|\partial_x E^\one (x,\w)| & \ge \lambda |V'(x)|\, |\vp^\one \xwo |^2\\
& \qquad - \lambda \sum_{n\ne 0} |V'(x + n\w)|\, |\vp^\one\xwn|^2\\
& \ge \lambda \ve(1- 4 \lambda^{-{1\over 3}}) - 
\lambda C_0(4\lambda^{-{1\over 3}})
 > {1\over 2} \lambda^{39\over 40}
\end{align*}
\end{proof}

We now construct a set of frequencies such that the image of $E^\one (., \w)$ 
contains most of the spectral values.  We will need this in Section 4 when 
we eliminate the situation when $\rsp \bigl( H_{[-N_1,N_1]}(x,\w) \bigr)$ and 
$\rsp \bigl( H_{[-N_1,N_1]}(x+j\w,\w) \bigr)$ are close to each other.  Recall that $L = \sum\limits_k (b^{(1)}_k - a^{(1)}_k )$.

\begin{lemma}\label{lem:3-5} There exists $\oone \subset [0, 1]$, 
$\mes \bigl([0,1]\setminus \oone\bigr) \le \delta^{1/4}$ such that for any 
$w \in \oone$, one has
\begin{align*}
\mes \bigl(V^{-1} (\cE^\zero) \setminus \cD^\one_\w\bigr) & 
\le 40 m_0 \delta^{1/4}L \\[6pt]
\compl \bigl(V^{-1}(\cE^\zero)\setminus \cD^\one_\w\bigr) & 
\le 40 m_0 C_0 \delta^{-7/4}L\ .
\end{align*}
\end{lemma}

\begin{proof} By Corollary 2.7,
$$
\int^1_0 \mes \bigl(V^{-1}(\cE^\zero) \setminus \cD^\one_\w \bigr) d\w = 
\mes \bigl[\bigcup_k (a^\one_k , b^\one_k )\times [0,1] \setminus 
\cD^\one\bigr]\le 40 m_0 \delta^{1/2}L\ .
$$
Let $B^\one = \left\{\w \in [0, 1]: \mes \bigl(V^{-1}(\cE^\zero) \setminus 
\cD^\one_w\bigr) \ge 40 m_0 \delta^{1/4}L\right\}$.  Then 
$\mes B^\one \le \delta^{1/4}$.

Take $\oone = [0, 1] \setminus B^\one$.  If $\w \in \oone$, then 
$\mes\bigl(V^{-1}(\cE^\zero)\setminus \cD^\one_\w\bigr) 
< 40 m_0 \delta^{1/4}L$.  Each interval $(a^\one_k , b^\one_k)$ is contained 
in either $\cD^\one_\w$ or $V^{-1}\bigl(\cE^\zero\bigr) \setminus \cD^\one_\w$.  Since $b^\one_k - a^\one_k \ge \delta^2 C_0^{-1}$
$$
\#\left\{ (a^\one_k , b^\one_k) \subset \bigl(V^{-1}(\cE^\zero)\setminus 
\cD^\one_\w \bigr)\right\}\le (40 m_0 \delta^{1/4}L)(\delta^2 C^{-1}_0)^{-1}\ .
$$
\end{proof}

For $\w \in \oone$, let $V\bigl(V^{-1}(\cE^\zero)\setminus \cD^\one_\w\bigr) = 
\bigcup\limits^{R_\w}_{r=1}(\ue_{\w r}, \be_{\w r})$, 
$R_\w \le 40 m_0 C_0 \delta^{-7/4}L$.  
Define 
$$
\cE^\one_\w = \cE^\zero\setminus \bigcup\limits_r 
(\ue_{\w r} - \delta^2, \be_{\w r} + \delta^2)
$$

\begin{lemma}\label{lem:3-6} For any $\w \in \oone$, one has 
$
\mes \bigl(\cE^\zero\setminus \cE^\one_\w\bigr) \leq120 m_0 C_0 \delta^{1/4}
$. 
Suppose $|V(x) - E| \le \delta^2$, $E \in \cE^\one_\w$.  Then 
$x \in \cD^\one_\w$.
\end{lemma}

\begin{proof}
\begin{equation*}
\begin{split}
\mes \bigl(\cE^\zero\setminus \cE^\one_\w\bigr) & 
\le (40 m_0 \delta^{1/4})C_0 + (40 m_0 C_0 \delta^{-7/4})(2 \delta^2)\\
& = 120 m_0 C_0 \delta^{1/4}
\end{split}
\end{equation*}

The second assertion follows from the construction of $\cE^\one_\w$.
\end{proof}

Next two lemmas provide us some flexibility concerning the size of the 
monodromy matrices we can use.  We will use them in Section 4 to obtain the 
hypothesis of the Avalanche Principle.

\begin{lemma}\label{lem:3-7} Let $\w \in \oone$,
$\lambda^{-1} E \in \cE^\one_\w$.  Suppose $N_1 \le b -a \le 5N_1$.  Then
$$
\log \big | \fmap(x, \w, E)| \ge (b-a) \log (\lambda^{1/2} - 2) + \log \dist\bigl(E, \rsp H_{[a, b]}(x,\w) \bigr)
$$
\end{lemma}

\begin{proof} Let $\rsp \hab (x,\w) = \bigl\{\mu_n\bigr\}^b_{n=a}$, $|\lambda V(x + n\w) - \mu_n| \le 2$.  Say
$$
|E - \mu_{n_0}| = \dist\bigl(E, \rsp \hab (x,\w)\bigr)\ .
$$
If $|E - \mu_{n_0}| > \lambda^{1/2} -2$ then $|E - \mu_n| > \lambda^{1/2} - 2$
for all $n$.  Hence
\begin{equation*}
\begin{split}
\log \big |\fmap (x,\w, E)\big | & = \sum_{n\ne n_0} \log |E - \mu_n| + |E - \mu_{n_0}|\\
& \ge (b-a) \log (\lambda^{1/2} - 2) + \log \dist\bigl(E, \rsp \hab (x,\w)\bigr)\ .
\end{split}
\end{equation*}
If $|E - \mu_{n_0}| \le \lambda^{1/2} - 2$ then $|E - \lambda V(x + n_{0}\w)| \le \lambda^{1/2}$ $\Longrightarrow x + n_0 \w \in \cD^\one_\w$.  Therefore,  
$|E - \lambda V(x + n\w)| > \lambda^{1/2}$ for $n \ne n_0$ $\Longrightarrow |E - \mu_n| > \lambda^{1/2} - 2$ for $n \ne n_0$.

Hence
\begin{equation*}
\log |\fmap (x,\w, E)|  \ge (b-a) \log (\lambda^{1/2} -2) +
 \log \dist\bigl(E, \rsp \hab (x, \w)\bigr)
\end{equation*}
\end{proof}

\begin{lemma}\label{lem:3-8}
Let $E \in \rsp \hab (x,\w)$, $N_1 \le b-a \le 5N_1$, $w \in \oone$, 
$\lambda^{-1}E \in \cE^\one_w$.  If \newline $|\lambda^{-1} E - V(x + j\w)| < 
\lambda^{-1/2}$ for some $j$, $a + N_1^\vt < j < b - N_1^\vt$ then 
$|E - E^\one (x + j\w, \w)| < 10 \lambda^{-{1\over 3}N_1^\vt}$.
\end{lemma}

\begin{proof} $|\lambda^{-1} E - V(x + j\w)| < \lambda^{-1/2} = \delta^2 
\Rightarrow x+j\w \in \cD^\one \Rightarrow E^\one(x+j\w,\w)$ is defined.  Also, similar to the construction of $E^\one$, $E$ is the only eigenvalue of 
$\hab (x,\w)$ in \newline $[V(x+j\w)-2,V(x+j\w)+2]$

Let
\begin{equation*}
\begin{split}
p & = \max \{a, j - N_1\}\ ,\quad q = \min\{b, j+N_1\},\\[6pt]
Y & = \Biggl[\sum^q_{\ell=p} \big | \vp^{(1)}(x + j\w, \w)(\ell - j)\big |^2
\Biggr]^{1/2}\\[6pt]
\tilde\vp(\ell) & = \begin{cases}
Y^{-1}\vp^{(1)}(x + j\w, \w)(\ell-j) & \text{if $\ell \in [p, q]$}\\[6pt]
0 & \ell \in [a, b]\setminus [p, q]
\end{cases}
\end{split}
\end{equation*}
Then $\|\tilde\vp\| = 1$.
\begin{equation*}
\begin{split}
p + N_1^\vt < j < q - N_1^\vt \Longrightarrow 1- Y^2 & \le 2 
\sum^\infty_{n = N_1^\vt} |\lambda^{-{1\over 3}n}|^2 \le {1\over 2}\\[6pt]
\big \| \bigl[\hab (x,\w) - E^{(1)}(x + j\w, \w)\bigr]\tilde\vp \big \| & 
\le Y^{-1}[ |\vp(x+j\w,\w)(p-1)| + |\vp(x+j\w,\w)(q+1)| ] \\
 & < 10 \lambda^{-{1\over 3}N_1^\vt}
\end{split}
\end{equation*}
Using Lemma A.5, since $E$ is the only eigenvalue of $\hab(x,\w)$ in 
$[V(x+j\w)-2,V(x+j\w)+2]$, we have $|E-E^\one(x+j\w,\w)|
< 10 \lambda^{-{1\over 3}N_1^\vt}$
\end{proof}


\setcounter{section}{3}
\section{Elimination of Resonances}
In this section, we use the result from Appendix E to avoid the situation when $\rsp \bigl( H_{[-N_1,N_1]}(x,\w) \bigr)$ and $\rsp \bigl( H_{[-N_1,N_1]}(x+j\w,\w) \bigr)$ are close to each other, by eliminate part of the domain in $\cD^\one$.  Then we construct a function $E^\two (x,\w)$ which takes value in $\rsp H_{[-\tilde{N}^\two, N^\two]}(x,\w)$, $\tilde{N}^\two, N^\two \asymp e^{N_1^\tau}$, and the corresponding eigenfunctions $\vp^\two(x,\w)$.  We will obtain some basic properties of $E^\two$ and $\vp^\two$ in this section; additional properties, including hypothesis for Avalanche Principle, will be derived in section 5, where the general inductive construction is done.

Recall that a set $U \subset \IR^2$ is called $K$--simple, denoted $\compl U \le K$, if $U = \bigcup\limits_{k = 1}^{K} D_k$ where $D_k$ are rectangles.

\begin{lemma}\label{lem:4-1} There is $m$--simple set 
$\tilde{\cD}^\two \subset \cD^\one$, $m \le e^{2N_1^\vt}$, which satisfies the 
following conditions:

\medskip
\noindent
(1) $\mes \bigl(\cD^\one \setminus \tilde{\cD}^\two\bigr) 
\le e^{-{1\over 2}N_1^\vt}L$

\medskip
\noindent
(2) If $(x,\w) \in \tilde{\cD}^\two$ and $(x + j\w,\w) \in \cD^\one$ for some $j$,
$0 < |j| \le N^2_2$, where $N_2 = \lfloor e^{N_1^\tau}\rfloor$, $\tau < \vt$, 
then $\big |E^\one(x, \w) - E^\one (x + j\w, \w)\big | > 4e^{-N_1^\vt}$. 
\end{lemma}

\begin{proof} For $0 < |j| \le N^2_1$, by construction of $\cD^\one$, $x \in \cD^\one_\w \Longrightarrow |V(x) - V(x + j\w)| > \lambda^{-{1\over 2}}$.  If $x + j\w \in \cD^\one_\w$, then
\begin{equation*}
\begin{split}
& \big |E^{(1)}(x, \w) - E^{(1)}(x + j\w,\w)\big | \\
&\quad \ge |\lambda V(x) - \lambda V(x + j\w)| - |\lambda V(x) - E^{(1)} (x, \w)| - |\lambda V(x+j\w) - E^{(N_1)}(x + j\w, \w)|\\
&\quad \ge \lambda^{1/2} - 4 > 4 e^{-N_1^\vt}\ .
\end{split}
\end{equation*}

Consider $D_k := \left(\bigl(a^\one_k, b^\one_k\bigr) \times [0, 1]\right)\cap 
\cD^\one$. Let $a^\one_k = \xmap = b^\one_{k}$, 
$x_i - x_{i-1} < e^{-N_1^\vt}$, $n \le 2e^{N_1^\vt}(b^\one_k - a^\one_k)$.
For any $|j| > N_1^2$, let $T_k(j) = \left\{(x, \w) \in D_k: x + j\w \in \cD^\one_\w \right\}$, $T_k(j)$ consists of at most $|j| C_0 \delta^{-2}$ components. (See Corollary 2.6)  On each of these components, the function
$$
F_j(x, \w) = E^{(1)} (x, \w) - E^{(1)}(x + j\w, \w)
$$
is $C^1$.
\begin{equation*}
\begin{split}
\partial_x F_j(x, \w) & = \partial_x E^{(1)}(x, \w) - \partial_x E^{(1)}(x + j\w, \w)\\
|\partial_x F_j(x, \w)| & \le 2C_0 \lambda\\[6pt]
\partial_\w F_j(x, \w) & = \partial_\w E^{(1)}(x, \w) - \partial_\w E^{(1)}(x + j\w, \w) - j \partial_x E^{(1)}(x + j\w, \w)\\
|\partial_\w F_j(x, \w)| & \ge |j|\bigl({1\over 2} \lambda^{39/40}\bigr) - 2C_0 N_1 \lambda > {1\over 4} |j| \lambda^{39/40} > 0
\end{split}
\end{equation*}
Apply Proposition E.1 to each of the components of $T_k(j)$. (Note that even though 
the components are not rectangles, the boundary consists of two verticle lines 
and two lines with slope ${-1\over j};  
{1\over |j|} \ll (2C_0 \lambda)\bigl({1\over 4}|j| 
\lambda^{39/40}\bigr)^{-1}$.)
We get rectangles $[x_{i-1}, x_i]\times [\uw_{i \ell}, \bw_{i \ell}]$ such that
\begin{equation*}
\bigl\{(x, \w) \in T_k(j) : |F_j(x, \w)| \le 2e^{-N_1^\vt}\bigr\}
\subset \bigcup_i\, \bigcup_\ell [x_{i-1}, x_i] \times 
[\uw_{i \ell}, \bw_{i \ell}] =: D_k(j)
\end{equation*}

\begin{equation*}
\mes D_k(j) \le |j| C_0 \delta^{-2}(b^\one_k - a^\one_k) 
\left[{2(4e^{-N_1^\vt)}\over {1\over 4}|j| \lambda^{39/40}} + 
2e^{-N_1^\vt} {2C_0 \lambda\over {1\over 4}|j| \lambda^{39/40}}\right]
 < (b^\one_k - a^\one_k) e^{-{3\over 4}N_1^\vt}
\end{equation*}

\begin{equation*}
\compl D_k(j) \le 2e^{N_1^\vt}(b^\one_k - a^\one_k) |j| C_0 \delta^{-2}
\end{equation*}

Set 
$$
\tilde{\cD}^\two = \cD^\one \setminus \Biggl(\bigcup\limits_k \, 
\bigcup\limits_{N_1^2 < |j| \le N_2^2} D_k(j)\Biggr)
$$  
$$
\mes \bigl(\cD^\one \setminus \tilde{\cD}^\two\bigr) \le \sum_k 
\sum_{N_1^2 < |j| \le N_2^2} 
(b^\one_k - a^\one_k) e^{-{3\over 4}N_1^\vt}< e^{-{1\over 2}N_1^\vt}L \ .
$$
$$
m= \compl \tilde{\cD}^\two \leq \sum_k \sum_{N_1^2 <|j| \le N_2^2} 2e^{N_1^\vt}
(b^\one_k - a^\one_k) |j| C_0 \delta^{-2} \le e^{2N_1^\vt}
$$
\end{proof}

By taking away a set of measure zero in $\tilde{\cD}^\two$, we obtain
$$
\cD^\two = \bigcup\limits_k\, \bigcup\limits_\ell (a^\two_k, b^\two_k) \times 
(\uw^\two_{k \ell}, \bw^\two_{k \ell}) \subset \tilde{\cD}^\two
$$
where $b^\two_k - a^\two_k < e^{-N_1^\vt}$, 
$\bw_{k \ell} - \uw_{k \ell}<e^{-N_1^\vt}$, 
$\compl \cD^\two \le e^{2N_1^\vt}$

We now show that $E^\one (x,\w)$ and $\rsp H_{[a,b]}(x,\w)$, $b-a \asymp N_1$, are separted, provide that the entries near the edge of the interval $[a,b]$ are separated from $V(x)$, i.e. $|V(x)-V(x+j\w)|>\lambda^{1/2}$ for $j \in [a,a+N_1^\vt] \cup [b-N_1^\vt,b]$.  The hypothesis of the Avalanche Principle will follow from the following lemma.

\begin{lemma}\label{lem:4-2}
Let $(x,\w) \in \cD^\two$, $\w \in \oone$, $\lambda^{-1}E \in \cE^\one_\w$, 
$\big | E - E^{(1)}(x, \w)\big | < 2e^{-N_1^\vt}$.  Suppose 
$|V(x +j\w) - \lambda^{-1} E| > \lambda^{-1/2}$ whenever
$a \le j \le a+ N_1^\vt$ or $b - N_1^\vt \le j \le b$ where 
$1 \le a \le a + N_1 \le b < a + 5N$.  Then 
$\dist\bigl(\rsp \hab(x, \w), E\bigr) \ge e^{-N_1^\vt}$.
\end{lemma}

\begin{proof} Let $\rsp \hab = \bigl\{\mu_j\bigr\}^b_{j=a}$, 
$|\mu_j - \lambda V(x + j\w)| \le 2$.  If 
$|E - \lambda V(x + j\w)| > \lambda^{1/2}$ for all $j \in [a, b]$ then 
$|E - \mu_j| > \lambda^{1/2} - 2 \Longrightarrow 
\dist \bigl(\rsp \hab (x, \w), E\bigr) > \lambda^{1/2} - 2$.

Suppose $|E - \lambda V(x + j_0\w)| \le \lambda^{1/2}$, 
$a + N_1^\vt < j_0 < b - N_1^\vt$.
Since $E \in \cE^\one_\w$, $x + j_0 \w \in \cD^\one_\w$.  So  
$|E - \lambda V(x + j\w)| > \lambda^{1/2}$ for all $j \ne j_0$, 
$a \le  j \le b$, $\dist \bigl(\rsp \hab (x, \w), E\bigr) = |E - \mu_{j_0}|$.
By Lemma 3.8, $\big |\mu_{j_0} - E^{(N_1)} (x + j_0 \w) \big | < 
10 \lambda^{-{1\over 3}N_1^\vt} < e^{-N_1^\vt}$.  Since $x \in \cD^\two_\w $, one has $\big |E^{(1)}(x, \w) - E^{(1)}(x + j_0 \w) \big | > 4e^{-N_1^\vt}$
Therefore, $$
\big | E - \mu_{j_0}\big |  \ge  \big |E^{(1)}(x, \w) - E^{(1)}(x + j_0 \w)\big |
- \big |E^{(1)}(x, \w) - E\big | - \big | E^{(1)}(x + j_0 \w) - \mu_{j_0} \ge  e^{-N_1^\vt} $$
\end{proof}

\begin{corollary}\label{cor:4-3} Let $(x,\w) \in \cD^\two$, $\w \in \oone$, $\lambda^{-1} E \in \cE^\one_\w$, $|E - E^{(1)}(x, \w)| < e^{-N_1^\vt}$.  Suppose $|V(x + j\w)- \lambda^{-1} E| > \lambda^{-1/2}$ whenever
$$
a \le j \le a+N_1^\vt\quad \text{or}\quad b - N_1^\vt \le j \le b+ N_1^\vt\quad \text{or}\quad c - N_1^\vt \le j \le c
$$
where
$$
1 \le a \le a + N_1 \le b \le b+1 + N_1 \le c \le a + 5N_1 \le N_2^2
$$
Then
\begin{align*}
\log \| \mab\xwe\| & \ge (b-a) \log (\lambda^{1/2} -2) - N_1^\vt\\
\log \| M_{(b, c]}\xwe\| & \ge (c - b -1) \log (\lambda^{1/2} -2) - N_1^\vt
\end{align*}
and
\begin{equation*}
 \log \| \mab\xwe\| +\log \| M_{(b, c]}\xwe\| - \log \|\mac \xwe\| 
\le 20 \log (\lambda C_0) + 20N_1^\vt\ .
\end{equation*}
\end{corollary}

\begin{proof} This follows from Lemma 3.7, Lemma C.10 and Lemma 4.2.
\end{proof}

Let $x \in \cD^\two_\w$, $\w \in \oone$, $\lambda^{-1} E \in \cE^\one_\w$, $|E - E^{(1)}(x, \w)| < 2e^{-N_1^\vt}$.  Suppose $0 < a < b \le N_2$ such that $|V(x + j\w) - \lambda^{-1} E| \ge \lambda^{-1/2}$ for $a \le j \le a+N_1^\vt$ and $b - N_1^\vt < j \le b$, $b -a \ge N_1$. 
Let $a < j_1 < j_2 < \dots < j_m < b$ be such that $|V(x + j_i \w) - \lambda^{-1} E| \le \lambda^{-1/2}$.  By construction of $\cD^\one$, $j_i - j_{i-1} > N^2_1$.  So choose $a = k_0 < k_1 < \dots < k_n = b$ such that $\min\limits_{i, \ell} |j_i - k_\ell| > N_1^\vt$, $N_1 < k_\ell - k_{\ell-1} \le 2N_1$.

Let 
\begin{align*}
A_1 & = M_{[a, k_1]}\xwe\begin{pmatrix} 1 & 0\\ 0 & 0\end{pmatrix}\\[6pt]
A_\ell & = M_{(k_{\ell-1}, k_\ell]}\xwe\ ,\quad \ell = 2, 3,\dots, n-1\\[6pt]
A_n & = \begin{pmatrix} 1 & 0\\ 0 & 0\end{pmatrix} M_{(k_{n-1}, b]}\xwe
\end{align*}

\begin{lemma}\label{lem:4-4}
With notation as above,
$$
\log |\fmap\xwe| = \sum^n_{\ell =2}\, \log \|A_\ell A_{\ell -1}\| - \sum^{n-1}_{\ell =2}\, \log \|A_\ell\| + O(\lambda^{-cN_1})\ .
$$
\end{lemma}

\begin{proof} If $[k_{\ell-1}, k_\ell]$ does not contain any $j_i$, then $|\lambda V(x + j\w) - E| \ge \lambda^{1/2}$ for all $j \in [k_{\ell-1}, k_\ell]$.  Hence $\log \|A_\ell\| \ge N_1 \log(\lambda^{1/2} - 2)$.

If $j_i \in [k_{\ell-1}, k_\ell]$, since $k_{\ell-1} + N_1^\vt < j < k_\ell - N_1^\vt$, $\dist\bigr(\rsp H_{[k_{\ell-1}, k_\ell]}\xwe \ge e^{-N_1^\vt}$ (Lemma 4.2).  By Lemma 3.7, $\log \|A_\ell\| \ge N_1 \log(\lambda^{1/2} - 2) - N_1^\vt$.

Also, applying Lemma C.10, we get
$$
\log \|A_\ell\| + \log \|A_{\ell-1}\| - \log\|A_\ell A_{\ell-1}\| \le 20\bigl(\log(\lambda C_0) + N_1^\vt\bigr)\ .
$$

The assertion follows from the Avalanche Principle.
\end{proof}

Next, we show that there is $N \approx N_2$ such that the Green's function 
$\bigl( H_{[1,N]}(x,\w)-E \bigr)^{-1}(k,1)$ decay exponentially.  In fact, the Green's function decay exponentailly if resonance does not occur near the edge.

\begin{lemma} Let $(x,\w) \in \cD^\two$, $\w\in\Omega^\one$, $\lambda^{-1} E \in \cE^\one_\w$, $|E - E^{(1)}(x,\w)| < e^{-N_1^\vt}$.   Suppose $|V(x + j\w)- \lambda^{-1} E| > \lambda^{-1/2}$ for $j\in[N-N_1^\vt,N]$ where $N_2 \le N < N^2_2$.  Then $E \notin \rsp H_{[1, N]}(x,\w)$ and 
$$
\log \big |\bigl[H_{[1, N]}(x,\w) - E\bigr]^{-1}(n, 1)\big| \le - {1\over 5}n \log \lambda
$$
for $n > N_1 - N_1^\vt$.
\end{lemma}

\begin{proof} We can choose $p_k, q_k$ such that
\begin{alignat*}{2}
& N_1 \le q_k - p_k \le 2N_1\ ,\qquad && p_k < q_{k-1} < p_{k+1} < q_k\\
& q_{k-1} - p_k > N_1^\vt\ , &&[1, N] = \bigcup\limits^K_{k=1}[p_k, q_k]\ ,\\
&|V(x + m\w) - \lambda^{-1}E| > \lambda^{-1/2}\quad &\text{for}\quad & |m- p_k|\le N_1^\vt\ \text{or}\ |m - q_k| \le N_1^\vt
\end{alignat*}

The first assertion follows from Lemma C.4 and C.9 if we can show that 
$\dist\bigl(H_{[p_k, q_k]} (x,\w), E\bigr) \ge e^{-N_1^\vt}$.  But this follows from Lemma 4.2.

Applying Lemma 4.4, and note that we can choose $k_i$ such that $\min\limits_i |n - k_i| =| n - k_{i_0}| \le 2N_1^\vt$, we get
\begin{equation*}
\begin{split}
\log \big |\bigl[H_{[1, N]}(x, \w) - E\bigr]^{-1}(k_{i_0}, 1)|  &= \log |f_{(k_{i_0}, N]}| - \log |f_{[1, N]}|\\
& = - \sum^{i_0 +1}_{\ell = 2} \bigl(\log \|A_\ell A_{\ell-1}\| - \|A_\ell\|\bigr) + O\bigl(\lambda^{-cN_1}\bigr)\\
& = - \sum^{i_0}_{\ell =1} \log \|A_\ell\| + \sum^{i_0 +1}_{\ell = 2}\bigl(\log \|A_\ell\| + \log \|A_{\ell-1}\| -\\
&\qquad \log \|A_{\ell} A_{\ell-1}\|\bigr) + O(\lambda^{-cN_1})\\
& \le - k_{i_0}\log (\lambda^{1/2} - 2) + i_0 N_1^\vt + i_0 \bigl(20(\log (\lambda C_0) + N_1^\vt)\bigr)\\
& \le - {1\over 4} k_{i_0} \log \lambda
\end{split}
\end{equation*}
and
\begin{equation*}
\begin{split}
\log \big |\bigl[H_{[1, N]}(x, \w) - E\bigr]^{-1} (n, 1)\big | & \le - {1\over 4}k_{i_0} \log \lambda + 2N_1^\vt \log (\lambda C_0 + 2)\\
& \le - {1\over 5} n \log \lambda
\end{split}
\end{equation*}
\end{proof}

In the next two lemmas, we find $N \approx N_2$ such that the entries near the edge of $[1,N]$ are separated from $V(x)$.

\begin{lemma}\label{lem:4-5}
Suppose $x \in \mathbb{T}$, $\w \in \oone$, $\lambda^{-1} E \in \cE^\one_\w$, $N \gg N_1$.  Then either
\begin{enumerate}
\item[(a)] $|\lambda V(x + j\w) - E| > {1\over 2}\lambda^{1/2}$ for $N - N_1 < j \le N$; or

\item[(b)] $|\lambda V(x + j\w) - E| > {1\over 2}\lambda^{1/2}$ for $N < j \le N + N_1$.
\end{enumerate}
\end{lemma}

\begin{proof} If $|\lambda V(x + j_0 \w) - E| < {1\over 2}\lambda^{1/2}$ for some $N - N_1 < j_0 \le N$ then $(x + j_0 \w, \w) \in \cD^\one$ since $E \in \cE^\one_\w$.  Hence $|\lambda V(x + j\w) - \lambda V(x + j_0 \w)| > \lambda^{1/2}$ for $|j- j_0| \le N_1^2$.  In particular, this is true for all $j, N < j \le {N + N_1}$.  Thus $|\lambda V(x + j\w) - E| > {1\over 2}\lambda^{1/2}$ for $N < j \le N + N_1$.
\end{proof}

Let $(x_1, \w_1), (x_2, \w_2) \in \bigl(a^\two_k, b^\two_k\bigr) \times \bigl(\uw^\two_{k \ell}, \bw^\two_{k \ell}\bigr) \subset \cD^\two$.  For $N_2 - N_1 < j \le N_2 + N_1$,  
\begin{align*}
|(x_1 + jw_1) - (x_2 + j\w_2)|  &\le |x_1 - x_2| + |j|\, |\w_1 - \w_2|\\
& \le \bigl(b^\two_k - a^\two_k\bigr) + (N_2 + N_1) \bigl(\bw^\two_{k \ell} - \uw^\two_{k \ell}\bigr)\\
& \lesssim e^{-N_1^\vt}\ .
\end{align*}

Suppose $|\lambda V(x_1 + j\w_1) - E_1| > {1\over 2}\lambda^{1/2}$ for $N_2 - N_1 < j \le N_2$.  The for $|E_2 - E_1| < {1\over 4}\lambda^{1/2}$,
$$
|\lambda V(x_2 + j\w_2) - E_2| > {\lambda^{1/2}\over 3}
$$
for $N_2 - N_1< j \le N_2$.  Hence, we have the following lemma.

\begin{lemma} For each $( a^\two_k , b^\two_k ) \times ( \uw^\two_{k\ell} , \bw^\two_{k\ell} )$, there exist $N^\two_{k\ell}$, $|N^\two_{k\ell}-N_2|\le N_1$, such that for any $(x,\w) \in ( a^\two_k , b^\two_k ) \times ( \uw^\two_{k\ell} , \bw^\two_{k\ell} ) \cap \Omega^\one$, $\lambda^{-1}E \in \cE^\one_\w$, $|E-E(x,\w)|<e^{-N_1^\vt}$, one has $$
|\lambda V(x+j\w)-E|>{\lambda^{1/2} \over 3} $$
for $N^\two_{k\ell}-N_1 < j \le N^\two_{k\ell}$.  Furthermore, the Green's function $[H_{[1,N^\two_{ij}]}(x,\w)-E]^{-1}(n,1)$ decay exponentially.
\end{lemma}

\begin{proof}
With no resonances near the edge, the second assertion follows from Lemma 4.5.
\end{proof}

For each component $(a^\two_k,b^\two_k) \times (\uw^\two_{k\ell}, \bw^\two_{k\ell})$, fix $\tilde{N}^\two_{k\ell}, N^\two_{k\ell}$ such that the Green's functions $\bigl( H_{[1,N^\two_{k\ell}]}-E \bigr)^{-1}(j,1)$ and 
$\bigl( H_{[-\tilde{N}^\two_{k\ell},-1]}-E \bigr)^{-1}(-j,-1)$ decay exponentially for all $(x,\w) \in (a^\two_k,b^\two_k) \times (\uw^\two_{k\ell}, \bw^\two_{k\ell})$.  We can now construct $C^1$ function $E^\two(x,\w)$ which take values in $\rsp H_{[-\tilde{N}^\two_{k\ell},N^\two_{k\ell}]}(x,\w)$. 

\begin{lemma}\label{lem:4-7} If $(x, \w) \in (a^\two_k, b^\two_k) \times 
(\uw^\two_{k \ell}, \bw^\two_{k \ell})$, then
$$
\rsp H_{[-\tilde{N}^\two_{k \ell}, N^\two_{k \ell}]}(x, \w)\cap 
\bigl(E^{(1)}(x, \w) - e^{-N_1}\ ,
 E^{(1)}(x, \w) + e^{-N_1}\bigr) \ne \emptyset\ .
$$
\end{lemma}

\begin{proof} Let
$$
\psi(j) = \begin{cases}
\vp^{(1)}(x, \w)(j) & |j| \le N_1\\[6pt]
0 & |j| > N_1
\end{cases}
$$
\begin{equation*}
\begin{split}
\big \|\bigl[H_{[-\tilde{N}^\two_{k \ell}, N^\two_{k \ell}]}(x, \w) - 
E^{(1)}(x, \w) \bigr]\psi\big\| & \le \big |\vp^{(1)}(x, \w)(1)\big | + 
\big | \vp^{(1)}(x, \w)(-N_1)\big |\\
& \le 2\bigl(\lambda^{-{1\over 3}N_1}\bigr) < e^{-N_1}\ .
\end{split}
\end{equation*}
The assertion now follows from Lemma A.4.
\end{proof}

\begin{lemma}\label{lem:4-8}
Suppose $E_1, E_2 \in \rsp H_{[-\tilde{N}^\two_{k \ell}, N^\two_{k \ell}]}(x, \w) \cap 
\bigl(E^{(1)}(x, \w) - e^{-N_1^\vt}, E^{(1)}(x, \w) + e^{-N_1^\vt}\bigr)$, 
$H_{[-\tilde{N}_{k \ell}, N^\two_{k \ell}]}(x, \w)\psi_i = 
E_i\psi_i$, $\|\psi_i\| = 1$.  If $|\psi_i(k)| \le e^{-c|k|}$ for 
$|k| \ge N_1$ then $E_1 = E_2$.
\end{lemma}

\begin{proof} Assume $E_1 \ne E_2$. Let
$$
\vp_i(k) = \begin{cases}
\psi_i(k) \Bigl[\sum_{|j| \le N_1}|\psi_i(j)|^2\Bigr]^{-1/2} & \text{for}\ |k| \le N_1\\[6pt]
0 & \hphantom{\text{for}}\ |k| > N_1
\end{cases}
$$
$
\|\vp_i\| = 1$, $\big \|\bigl[H_{[-N_1, N_1]}(x, w) - E_i\bigr]\vp_i \big \| \le 2 (2e^{-N_1})$.  Since
$$
\rsp H_{[-N_1, N_1]}(x, \w) \cap \bigl(E^{(N_1)} - {1\over 2} \lambda^{1/2}, E^{(N_1)} + {1\over 2} \lambda^{1/2}\bigr) = \left\{E^{(N_1)}(x, \w)\right\}\ ,
$$
by Lemma A.5
$$
\min_{|c| =1} \big \|\vp^{(N_1)}(x, \w) - c\vp_i \big \| < \sqrt 2(4e^{-N_1}) \bigl({1\over 2}\lambda^{1/2}\bigr)^{-1}\ .
$$
Hence, there is $|c| = 1$ such that
\begin{equation*}
\begin{split}
400 \lambda^{-1} e^{-2N_1} > \|\vp_1 - c\vp_2 \|^2 & = 2 + 2 \ree \bar c \la \vp_1, \vp_2\ra\\
& \Longrightarrow |\la \vp_1, \vp_2 \ra |> 1 - 200 \lambda^{-1}e^{-2N_1}
\end{split}
\end{equation*}

Since $H_{[-\bar{N}_{k \ell}, \tilde{N}_{k \ell}]}(x,\w)$ is hermitian,
\begin{equation*}
\begin{split}
0 & = |\la \psi_1, \psi_2\ra |\\
& \ge \Big | \sum_{|k| \le N_1}\, \psi_1(k) \overline{\psi_2(k)}\Big | - \sum_{|k| > N_1}\, \big |\psi_1(k)\overline{\psi_2(k)}\big |\\
& \gtrsim |\la \vp_1, \vp_2\ra| - \sum_{|k| > N_1}\, e^{-2|k|} > 0
\end{split}
\end{equation*}
\end{proof}

Lemma 4.8 and Lemma 4.9 show that 
$H_{[-\tilde{N}_{k \ell}, N^\two_{k \ell}]}(x, \w)$ has unique eigenvalue 
$E^\two(x, \w) \in \bigl(E^\one (x, \w) - {1\over 2} e^{-N_1^\vt}, 
E^\one (x,\w) + {1\over 2} e^{-N_1^\vt}\bigr)$ for $\w \in \oone$, 
$x \in \cD^\two_\w$, $\big | E^{(2)}(x, \w) - E^{(1)}(x, \w) \big | < e^{-N_1}$, the eigenfunction $\vp^\two (x, \w)$ corresponding to $E^\two (x, \w)$ decay 
exponentially, $\|\vp^{(2)}(x, \w) \| = 1$.  
Using Lemma A.5, by multiplying $c$ to $\vp^\two$, $|c| = 1$, we can choose
$$
\|\tilde\vp^\one (x, \w) - \vp^\two (x, \w) \| \le e^{-N_1}
$$
where
$$
\tilde\vp^{(1)}(x, \w)(k) = \begin{cases}
\vp^{(1)} (x, \w; k) & |k| \le N_1\\[6pt]
0 & |k| > N_1
\end{cases}
$$
Together with Lemma 4.7, this shows that $|\vp^\two(x,\w)(k)|\le 
e^{-{1\over 5}|k|}$ for all $k$, since $\vp^\one(x,\w)$ decay exponentially.


\setcounter{section}{4}
\section{Inductive construction of eigenvalue $E^{(s+1)}$ and eigenfunction $\vp^{(s+1)}$}
In this section, we will derive Theorem 1.1.  First, to construct $E^{(s+1)}$, we assume $\emap, \vpmap$ have been defined with the following properties:

\begin{enumerate}
\item The domain of $E^{(k)}$ and $\vp^{(k)}(.,.)(j)$ is $\cD^{(k)}$, $\cD^\one \supset \cD^\two \supset \dots \subset \cD^{(s)}$; $\compl \cD^{(k)} < e^{4N^\vt_{k-1}}$,  $\mes \bigl(\cD^{(k-1)} \setminus \cD^{(k)}\bigr)< e^{-N^\vt_{k-1}}$ for $1 \le k \le s$ where $N_k \asymp e^{N^\tau_{k-1}}\ (\tau < \vt)$.

\item $\cD^{(k)}=\bigcup\limits_i \, \bigcup\limits_j (a^{(k)}_i,b^{(k)}_i) \times(\uw^{(k)}_{ij},\bw^{(k)}_{ij})$, $b^{(k)}_j - a^{(k)}_j < e^{-N^\vt_k}$, $\bw^{(k)}_{j\ell} - \uw^{(k)}_{j\ell} < e^{-N^\vt_k}$.  For each $i,j$, there is $\tilde{N}^{(k)}_{ij},N^{(k)}_{ij}\approx N_k$ such that $H_{[-\tilde{N}^{(k)}_{ij}, N^{(k)}_{ij}]}(x,\w) \vp^{(k)}(x, \w) = E^{(k)}(x, \w)\vp^{(k)}(x, \w)$ for all $(x,\w)\in(a^{(k)}_i,b^{(k)}_i) \times(\uw^{(k)}_{ij},\bw^{(k)}_{ij})$.

\item $\sum\limits_j \big | \vp^{(k)}(x, \w)(j)\big |^2 = 1$; 
$\ \big | \vp^{(k)} (x, \w)(j) \big | \le \lambda^{-{1\over 5}|j|}$. \newline
${1\over 2} \lambda^{39/40} < |\partial_x E^\so(x,\w)| \le C_0 \lambda$; 
$\ |\partial_\w E^\so(x,\w)| \le C_0 N_{s+1}\lambda$.

\item $\big | E^{(k)} (x, \w) - E^{(k-1)}(x, \w) \big | 
< e^{-N_{k-1}}$, $\big \|\vp^{(k)}(x, \w) - \tilde\vp^{(k-1)}(x, \w)\big \| 
< e^{-N_{k-1}}$ where
$$
\tilde\vp^{(k-1)} (x, \w)(n) = \begin{cases} \vp^{(k-1)} (x, \w)(n) & 
\text{if}\ -\tilde{N}^{(k-1)}_{ij} \le n\le N^{(k-1)}_{ij} \\[6pt]
0 & \hphantom{\text{if}\ } \text{otherwise}
\end{cases}
$$

\item $\big | E^{(k-1)} (x, \w) - E^{(k-1)} (x + j\w, \w) \big | \ge 2e^{-N^\vt_{k-1}}$ if $(x,\w) \in \cD^{(k)}$ and $(x + j\w, \w) \in \cD^{(k-1)}$, $|j| \le N_k^2$

\item $\rsp H_{[-\tilde{N}^{(k)}_{ij}, N^{(k)}_{ij}]}(x, \w) \cap \bigl(E^{(k-1)} (x, \w) - e^{-N^\vt_{k-1}}, E^{(k-1)} (x, \w) + E^{-N^\vt_{k-1}}\bigr) = \bigl\{E^{(k)} (x, \w) \bigr\}$

\item There is $\oone \supset \Omega^\two \supset \dots \Omega^{(s-1)}$, $\mes \bigl(\Omega^{(k-1)} \setminus \Omega^{(k)}\bigr) < e^{-{1\over 2}N^\vt_{k-1}}$, such that $\mes \bigl(\cD^{(k-1)}_\w \setminus \cD^{(k)}_\w\bigr) < e^{-{1\over 2}N^\vt_{k-1}}$ for all $\w \in \Omega^{(k)}$

\item For $\w \in \Omega^{(k)}$, there is $\cE^{(k)}_\w \subset \cE^{(k-1)}_\w \subset \dots \subset \cE^\one_\w$, $\mes \bigl(\cE^{(k-1)}_\w \setminus \cE^{(k)}_\w \bigr) < e^{-{1\over 2}N^\vt_{k-1}}$, such that if $x \in \cD^{(k-1)}_\w$, $E \in \cE^{(k)}_\w$, $\big |E - E^{(k-1)} (x, \w) \big | < e^{-N^\vt_{k-1}}$ then $x \in \cD^{(k)}_\w$

\item Let $(x,\w)\in\cD^{(k)}$, $\w\in\O^{(k)}$, $\lambda^{-1}E\in\cE^{(k)}_\w$, $e^{-N^\vt_{k+1}} < |E^{(k)}(x +j\w,\w) -  E| < e^{-N^\vt_{k}}$.  Suppose 
$|E^{(k)}(x +j\w,\w) -  E| > e^{-N^\vt_{k}}$ whenever
$$
-a \le j \le -a+ N_k^\vt \quad\text{or}\quad b - N_k^\vt \le j \le b, \quad x+j\w\in\cD^{(k)}
$$
where $a,b\approx N_{k+1}$.  Then $$
\log |f_{[a,b]}(x,\w,E)| \ge {1\over 4}(a+b)\log \lambda $$

\end{enumerate}

For the first scale, these conditions were proven in Section 3.  Furthermore, we assume that for $(x,\w) \in \cD^{(k+1)}$, $\w \in \O^{(k)}$, $\lambda^{-1}E \in \cE^{(k)}_\w$, $\big | E - E^{(k)}(x, \w)\big | < 2e^{-N_{k}^\vt}$, the following conditions (and the corresponding conditions in [-N,-1]) hold.

\begin{list}{\arabic{num}.}{\usecounter{num}}
\setcounter{num}{9}

\item  Suppose 
$|E^{(k-1)}(x +j\w,\w) -  E| > e^{-N^\vt_{k-1}}$ whenever
$$
a \le j \le a+ N_k^\vt \quad\text{or}\quad b - N_k^\vt \le j \le b, \quad x+j\w\in\cD^{(k-1)}
$$
where $1 \le a \le a + N_k \le b < a + 5N$.  Then 
$\dist\bigl(\rsp \hab(x, \w), E\bigr) \ge e^{-N_k^\vt}$. 

\item   Suppose $|E^{(k-1)}(x +j\w,\w) -  E| > e^{-N^\vt_{k-1}}$ whenever
$$
a \le j \le a+N_k^\vt\quad \text{or}\quad b - N_k^\vt \le j \le b+ N_k^\vt\quad \text{or}\quad c - N_k^\vt \le j \le c, \quad x+j\w\in\cD^{(k-1)}
$$
where
$$
1 \le a \le a + N_k \le b \le b+1 + N_k \le c \le a + 5N_k \le N_{k+1}^2
$$
Then
\begin{align*}
\log \| \mab\xwe\| & \ge (b-a) \log (\lambda^{1/2} -2) - N_k^\vt\\
\log \| M_{(b, c]}\xwe\| & \ge (c - b -1) \log (\lambda^{1/2} -2) - N_k^\vt
\end{align*}
and
\begin{equation*}
 \log \| \mab\xwe\| +\log \| M_{(b, c]}\xwe\| - \log \|\mac \xwe\| 
\le 20 \log (\lambda C_0) + 20N_k^\vt\ .
\end{equation*}

\item  Suppose $|E^{(k-1)}(x + j\w,\w)- E| > e^{-N^\vt_{k-1}}$ for $j\in[N-N_1^\vt,N]$ where $N_{k+1} \le N < N_{k+1}^2$.  Then $E \notin \rsp H_{[1, N]}(x,\w)$ and 
$$
\log \big |\bigl[H_{[1, N]}(x,\w) - E\bigr]^{-1}(n, 1)\big| \le - {1\over 5}n \log \lambda
$$
for $n > N_k - N_k^\vt$.

\end{list}

With some abuse of notation, we use $E^{(0)}(.\w)$ to denote $\lambda V$ and $e^{-N_0^\vt}$ means $\lambda^{1/2}$.  Then the last three conditions were proven, for k=1, in Section 4.  We now begin to prove these condtions inductively for higher scale.  As in section 4, we need to eliminate the case when $\rsp H_{[-N_s,N_s]}(x,\w)$ and $\rsp H_{[-N_s,N_s]}(x+j\w,\w)$ are close.

\begin{lemma}\label{lem:5-1}
There is $\cD^{(s+1)} \subset \cD^{(s)}$, $\cD^{(s+1)} = \bigcup\limits^M_{j=1} D^{(s+1)}_j$ where $\cD^\so_j = \bigcup\limits_\ell \left(a^\so_j, b^\so_j\right)\times \left(\uw^\so_{j\ell}, \bw^\so_{j\ell}\right)$, $b^\so_j - a^\so_j < e^{-N^\vt_s}$, $\bw^\so_{j\ell} - \uw^\so_{j\ell} < e^{-N^\vt_s}$, $M \le e^{2N_s^\vt}$, which satisfies the following conditions:

\begin{enumerate}
\item $\mes \left(\cD^{(s)} \setminus \cD^\so\right) \le 
e^{-{1\over 2}N^\vt_s} L$

\item If $(x,\w) \in \cD^\so$ and $(x + j\w, \w) \in \cD^{(s)}$ for some $0 < |j| \le N_{s+1}^2$ where $N_{s+1} = \lfloor e^{N_s^\tau}\rfloor$, $\tau < \vt$, then $\big |E^{(s)}(x, \w) - E^{(s)}(x + j\w, \w)\big | > 4e^{-N_s^\vt}$.
\end{enumerate}
\end{lemma}

\begin{proof} For $|n| \le N_s^2$, this follows from conditions (4) and (5).  For $|n|>N_s^2$, the proof is essentially the same as proof of Lemma 4.1, given the estimate on derivatives of $E^{(s)}$ from condition (3).  (Also see remark after Lemma 4.1.)
\end{proof}

We will first prove conditions (10)--(12) for k=s.

\begin{lemma}
Let $(x,\w) \in \cD^\so$, $\w \in \O^{(s)}$, $\lambda^{-1}E \in \cE^{(s)}_\w$, 
$\big | E - E^{(s)}(x, \w)\big | < 2e^{-N_s^\vt}$.  Suppose 
$|E^{(s-1)}(x +j\w,\w) - E| >  e^{-N^\vt_{k-1}}$ whenever
$a \le j \le a+ N_s^\vt$ or $b - N_s^\vt \le j \le b$ where 
$1 \le a \le a + N_s \le b < a + 5N$.  Then 
$\dist\bigl(\rsp \hab(x, \w), E\bigr) \ge e^{-N_s^\vt}$.
\end{lemma}

\begin{proof} Let $a < j_1 < j_2 < \dots < j_m \le b$ be such that $x + j_i \w \in \cD^{(s)}_{\w}$ and $\big |E - E^{(s)}(x + j_i \w, \w) \big | < {1\over 2} e^{-N_s^\vt}$.  Then
\begin{equation*}
\begin{split}
\bigl | E^\smo (x+ j_i\w, \w) - E^\smo(x + j_{i+1} \w, \w)\big | &
     \le \bigl |E^\smo(x + j_i \w, \w) - E^{(s)}(x + j_i \w, \w)\big | \\
&\quad + \bigl| E^{(s)} (x + j_i, \w, \w) - E\big | + \bigl |E - E^{(s)}(x + j_{i+1} \w, \w)\big |\\
&\quad + \bigl| E^{(s)}(x + j_{i+1} \w, \w) - E^\smo (x + j_{i+1} \w, \w)\bigr|\\
& < e^{-N_s} + {1\over 2} e^{-N_s^\vt} + {1\over 2} e^{-N_s^\vt} + e^{-N_s} < 2e^{-N_s^\vt}
\end{split}
\end{equation*}
By condition (5), $|j_i - j_{i+1}| > N_s^2$.  Also, $j_1 > a+N_s^\vt$ and $j_m < b-N_s^\vt$.  The assertion follows from condition (12) and Lemma C.9.
\end{proof}

\begin{corollary} Let $(x,\w) \in \cD^{(s+1)}$, $\w \in \O^{(s)}$, $\lambda^{-1} E \in \cE^{(s)}_\w$, $|E - E^{(s)}(x, \w)| < e^{-N_s^\vt}$.  Suppose 
$|E^{(s)}(x +j\w,\w) -  E| > e^{-N^\vt_{s}}$ whenever
$$
a \le j \le a+N_s^\vt\quad \text{or}\quad b - N_s^\vt \le j \le b+ N_s^\vt\quad \text{or}\quad c - N_s^\vt \le j \le c, \quad x+j\w\in\cD^{(s)}_\w
$$
where
$$
1 \le a \le a + N_s \le b \le b+1 + N_1 \le c \le a + 5N_s \le N_{s+1}^2
$$
Then
\begin{align*}
\log \| \mab\xwe\| & \ge (b-a) \log (\lambda^{1/2} -2) - N_s^\vt\\
\log \| M_{(b, c]}\xwe\| & \ge (c - b -1) \log (\lambda^{1/2} -2) - N_s^\vt
\end{align*}
and
\begin{equation*}
 \log \| \mab\xwe\| +\log \| M_{(b, c]}\xwe\| - \log \|\mac \xwe\| 
\le 20 \log (\lambda C_0) + 20N_s^\vt\ .
\end{equation*}
\end{corollary}

\begin{proof}
This follows form Lemma 5.2, condition (9) and Lemma C.10.
\end{proof}

\begin{lemma} Let $(x,\w) \in \cD^{(s+1)}$, $\w \in \O^{(s)}$, $\lambda^{-1}E \in \cE^{(s)}_\w$, $\big | E - E^{(s)}(x, \w)\big | < 2e^{-N_{s}^\vt}$.  
Suppose $|E^{(s-1)}(x + j\w,\w)- E| > e^{-N^\vt_{s-1}}$ for $j\in[N-N_s^\vt,N]$ where $N_{s+1} \le N < N_{s+1}^2$.  Then $E \notin \rsp H_{[1, N]}(x,\w)$ and 
$$
\log \big |\bigl[H_{[1, N]}(x,\w) - E\bigr]^{-1}(n, 1)\big| \le - {1\over 5}n \log \lambda
$$
for $n > N_s - N_s^\vt$.
\end{lemma}

\begin{proof}
Let $1 < j_1 < j_2 < \dots < j_m \le N$ be such that $x + j_i \w \in \cD^{(s)}_{\w}$ and $\big |E - E^{(s)}(x + j_i \w, \w) \big | < {1\over 2} e^{-N_s^\vt}$.
By condition (5), $|j_i - j_{i+1}| > N_s^2$.  Also, $j_1 > N_s^2$ and $j_m < N-N_s^\vt$. (See the proof of Lemma 5.2)  By condition (11), we can partition $[1,N]$ so that the hypothesis of Avalanche Principle is satisfied, and the assertion follows.  (See Lemma 4.4 and Lemma 4.5)
\end{proof}

Next, we show that there is  $N \approx N_{s+1}$ such that the Green's function $\bigl( H_{[1,N]}(x,\w) - E \bigr)^{-1} (k,1)$ decays exponentially and then we will be able to define $E^{(s+1)}$ and $\vp^{(s+1)}$.

Let $(x,\w) \in \cD^{(s+1)}$, $\w \in \O^{(s)}$, $\lambda^{-1}E \in \cE^{(s)}_\w$, $\big | E - E^{(s)}(x, \w)\big | < 2e^{-N_{s}^\vt}$.  By condition (5), if there is $n_0\in [N_{s+1}-N_s,N_{s+1}]$ such that $x+n_0 \w \in \cD^{(s)}$ and 
$| E - E^{(s)}(x+n_0 \w, \w)\big | < 2e^{-N_{s}^\vt}$ then
$| E - E^{(s)}(x+n, \w)\big | \ge 2e^{-N_{s}^\vt}$ for all $n\in 
(N_{s+1},N{s+1}+N_s]$.  

For each $i,j$, we can find $N^\so_{ij}$, $|N^\so_{ij}-N_{s+1}|\le N_s$ such that $| E - E^{(s)}(x+n, \w)\big | \ge e^{-N_{s}^\vt}$ for all $n\in 
[N^\so_{ij}-N_s,N^\so_{ij}]$, $(x,\w)\in (a^\so_i,b^\so_i)\times(\uw^\so_{ij},\bw^\so_{ij})$.  Hence by Lemma 5.4, the Green's function $[H_{[1,N^\so_{ij}]}(x,\w)-E]^{-1}(n,1)$ decay exponentially.

Fix $N^\so_{ij},\tilde{N}^\so_{ij}$, such that $H_{[-\tilde{N}^\so_{ij},N^\so_{ij}]} (x,\w)$ has a unique eigenvalue $E^\so(x,\w)$ in $\bigl(E^{(s)}(x, \w) - e^{-N_s^\vt}, E^{(s)}(x, \w) + e^{-N_s^\vt}\bigr)$ for all $(x, \w) \in \left(a^\so_j, b^\so_j\right)\times \left(\uw^\so_{j\ell}, \bw^\so_{j\ell}\right)$, $\big |E^\so(x, \w) - E^{(s)}(x,\w)\big | < e^{-N_s^\vt}$.  (See Lemma 4.8 and Lemma 4.9)  We can define $\vp^\so(x , \w)(n)$ such that
\begin{equation*}
\begin{split}
\htmap (x,\w) \vp^\so(x,\w) & = E^\so(x,\w) \vp^\so(x,\w)\\
\sum_n |\vp^\so(x, \w)( n)|^2 & = 1 \\
\|\tilde{\vp}^{(s)}(x,\w) - \vp^\so(x,\w) \| & \lesssim e^{-N_s} e^{N_s^\vt} < e^{-{1\over 2}N_s}
\end{split} \end{equation*}
where $\tilde{\vp}^{(s)}$ is as defined in condition (4).  

\begin{lemma}
\begin{equation*} \begin{split}
|\vp^\so (x, \w)( n)| & < \lambda ^{-{1\over 5}\lambda}\\
{1\over 2} \lambda^{39/40} < |\partial_x E^\so(x,\w)| & \le C_0 \lambda\\
|\partial_\w E^\so(x,\w)| & \le C_0 N_{s+1}\lambda
\end{split}
\end{equation*}
\end{lemma}

\begin{proof}
The first assertion follows from Lemma 5.4 and the fact that $\|\tilde{\vp}^{(s)}(x,\w) - \vp^\so(x,\w) \| < e^{-{1\over 2}N_s}$ since $\vp^{(s)}$ decay exponentially.  This also gives the estimate of the derivatives of $E^{(s)}$.  (See Corollary 3.4) 
\end{proof}

Lemma 5.1 and Lemma 5.5 establish conditions (1)--(6) for the s+1 scale.

Let $B^\so = \left\{\w \in \O^{(s)}: 
\mes \bigl([E^{(s)}(.,\w)]^{-1}(\cE^{(s)}_\w) \setminus \cD^{(s+1)}_w\bigr) 
\ge e^{-{1\over 2}N_s^\vt} \right\}$, $\ \O^\so=\O^{(s)}\setminus B^\so$.
As in Lemma 3.5, $\mes(\O^{(s)} \setminus \O)\le e^{-{1\over 2}N_s^\vt}$.  Also, for any $\w \in \O^\so$, one has 
$$
\mes \Bigl( E^{(s)}(.,\w)^{-1}(\cE^{(s)} \setminus \cD^{(s+1)}_\w) \Bigr) 
\lesssim e^{-N_s^{\vt/2}}\ .
$$  
For $\w \in \O^\so$, suppose $\cD^{(s+1)}_\w=\bigcup\limits_r \bigl( a^{(s)}_r(\w),b^{(s)}_r(\w) \bigr)$.  Let $$ \tilde{\cE}^\so_{\w,r}= Ran E^{(s)}(.,\w) \Big|_{\cD^\so_\w \cap \bigl( a^{(s)}_r(\w),b^{(s)}_r(\w) \bigr)} = \bigcup\limits_k \bigl( \underline{E}_{r,k}, \bar{E}_{r,k} \bigr) $$
$$ \tilde{\cE}^\so_\w = \bigcup\limits_r \bigcup\limits_k 
\bigl( \underline{E}_{r,k}+e^{-N_s^\vt}, \bar{E}_{r,k}-e^{-N_s^\vt} \bigr) $$
$$ \cE^\so_\w = \bigcup\limits_r \bigcup\limits_k 
\bigl( \underline{E}_{r,k}+2e^{-N_s^\vt}, \bar{E}_{r,k}-2e^{-N_s^\vt} \bigr) $$
Then $\mes \bigl( \cE^{(s)} \setminus \cE^\so \bigr) < e^{-N_s^{\vt/4}}$.  
It is clear that for $\w \in \Omega^{(s+1)}$, if $x \in \cD^{(s)}_\w$, $E \in \cE^{(s+1)}_\w$, $\big |E - E^{(s)} (x, \w) \big | < e^{-N^\vt_{s}}$ then $x \in \cD^{(s+1)}_\w$.  This establishes conditions (7) and (8).

The next two lemma give us condition (9) and complete the inductive procedure.

\begin{lemma}\label{lem:5-3} Let $\w \in \O^\so$, $E \in \cE^\so_\w$.
Suppose $\big |E - E^\so(x + j\w, \w)\big | \ge e^{-2N_s^\vt}$ whenever $(x + j\w, \w) \in \cD^\so$ for $|j| \le 4N_{s+1}$.  Then 
$$
\log |\ftmap\xwe| \ge {N^{(s+1)}_{ij} + \tilde{N}^{(s+1)}_{ij} \over 4} \log \lambda
$$
\end{lemma}

\begin{proof}
For $\tx = x + j_0 \w$, $|j_0| \le 3N_\spo$, if $\bigl(\tx + j\w, \w) \in \cD^{(s)}$, $|j| \le 4N_s$, then either
\begin{enumerate}
\item $(\tx + j\w, \w) \notin \cD^\so$, or
\item $(\tx + j\w, \w) \in \cD^\so$, $|E - E^\so (x + j\w,\w)| > e^{-2N_s^\vt}$
\end{enumerate}

In (1), $E \in \cE^\so_\w \Longrightarrow |E^{(s)} (\tx + j\w, \w) - E| > {1\over 2}e^{-N_s^\vt}$.
In (2), \newline $|E - E^{(s)}(x,\w)| \ge |E - E^\so(x,\w)| - |E^\so (x,\w) - E^{(s)}(x,\w)| > {1\over 2} e^{-2N_s^\vt}$.

By condition (11), we can apply Avalanche Principle to obtain the assertion.
\end{proof}

\begin{lemma} Let $(x,\w)\in\cD^{(s+1)}$, $\w\in\O^{(s+1)}$, $\lambda^{-1}E\in\cE^{(s+1)}_\w$, $e^{-N^\vt_{s+1}} < |E^{(s+1)}(x +j\w,\w) -  E| < e^{-N^\vt_{s}}$.  Suppose 
$|E^{(s)}(x +j\w,\w) -  E| > e^{-N^\vt_{s}}$ whenever
$$
-a \le j \le -a+ N_k^\vt \quad\text{or}\quad b - N_s^\vt \le j \le b, \quad x+j\w\in\cD^{(k)}
$$
where $a,b\approx N_{k+1}$.  Then $$
\log |f_{[-a,b]}(x,\w,E)| \ge {1\over 4}(a+b)\log \lambda $$
\end{lemma}

\begin{proof} By construction
\begin{equation*}
\begin{split}
& \rsp H_{[-\tn, N]} (x,\w) \cap \bigl(E^{(s)}(x,\w) - e^{-N_s^\vt}, E^{(s)}(x,\w) + e^{-N_s^\vt}\bigr) = \bigl\{E^\so (x,\w)\bigr\}\\
& \big |E^{(s)}(x,\w) - E^\so(x,\w)\big | < e^{-N_s}\\
& \Longrightarrow \big |E - E^{(s)}(x + j\w, \w)\big | \ge e^{-N_s^\vt}\quad \text{for}\ (x + j\w, \w) \in \cD^{(s)}, \ 0 < |j| \le N_\spo^2\\
& \Longrightarrow \ \text{there exist $|\te - E| \le e^{-2N_s^\vt}$ such that}\ 
 \big |\te - E^{(s)}(x + j\w, \w)\big | \ge e^{-2N_s^\vt},\ |j| \le N^s_\spo
\end{split}
\end{equation*}

By Lemma 5.6, $\log |\ftmap (x, \w, \te)| > {N^{(s+1)}_{ij} + \tilde{N}^{(s+1)}_{ij} \over 4} \log \lambda$.  Applying Lemma 
B.5 gives
$$
\log | \ftmap \xwe| > {N^{(s+1)}_{ij} + \tilde{N}^{(s+1)}_{ij} \over 4} \log \lambda - 2N_\spo\ .
$$
\end{proof}

By taking the limit as $s$ goes to infinity, we can prove Theorem 1.1:

\begin{theorem}\label{th:5-5} There is $\O = \O (V,\lambda) \in [0, 1]$, 
$\mes \bigl([0,1]\setminus \O\bigr) \lesssim \lambda^{-{1\over 2}}$,
such that for any $\w \in \O$ there exists $\cE_\w \subset \lambda \cJ$, 
$\mes \bigl(\cJ \setminus \lambda^{-1}\cE_\w\bigr) \lesssim \lambda^{-{1\over 2}}$, 
with $\lambda(\w,E) \gtrsim \log \lambda$ for all $E \in \cE_\w$. 
Also, there is $x \in \tor$ and $\{\vp(n)\}$, 
$|\vp(n)| \lesssim e^{-c|n|}$ such that 
$$ -\vp(n+1) + \vp(n-1) + V(x + n\w)\vp(n) = E\vp(n). $$
\end{theorem}

\begin{proof} Take $\O = \bigcap \O^{(k)}$, $\cE_\w = \bigcap \cE_\w^{(k)}$.  For any $\w \in \O$ and $E \in \cE_\w$, one has
$$
{1\over N^{(k)}_{ij} + \tilde{N}^{(k)}_{ij}} \log \big \| M_{[-\tilde{N}^{(k)}_{ij}, N^{(k)}_{ij}]}(x_k, \w, E)\big \| \gtrsim \log \lambda,
$$ 
$k = 1, \dots$ where $e^{-N_k}< |E - E^{(k)}(x_k, \w)| < e^{-N_{k-1}}$, $\ x_k \in \cD^{(k)}_\w$.

A subsequent of $\{ x_k \}$ converges.  Without loss of generality, we may assume $x = \lim x_k$.  Take $\vp(n) = \lim\vp^{(k)}(x_k,\w)(n)$.
\end{proof}


\part{}
\setcounter{section}{5}
\section{Resonance at first scale}
In Part II, we introduce variations of potential (Section 8), and show that for typical variation, the resulting potential leads to positive Lyapunov exponent for all spectral value.  In this section, we show that by choosing appropriate length for the interval in the first scale, we can get an upper estimate of the number of terms where resonances can occur.

Let $V \in C^3(\tor)$ be such that $|V'(x)| + |V''(x)| \ge  2c > 0$ for all 
$x \in \tor$.  Note that the assumption implies $V$ has finitely many critical 
points.  Let $x_1,\dots, x_n$ be the critical points of $V$.  There exists $\delta>0$ 
such that
$$
|V'(y)| \ge c\ |y-x_i|\qquad\text{whenever}\ |y - x_i| < \delta
$$
for any $i=1,2,\dots,n$.

Suppose
\begin{align*}
C_0 & = \min\bigl\{|V'(y)|: y \in \tor \setminus \tcup_j(x_j - \delta, x_j + \delta)\bigr\} > 0\ ,\\[6pt]
C_1 & = \max_{y \in \tor}(|V'(y)|+|V''(y)|)\ .
\end{align*}
and $V$ has $m_0$ monotonicity intervals.

In the next two lemmas, we find an upper bound of the number of entries that can lead to resonance.

\begin{lemma} Take any $E \in \IR$, 
$0 < \ve < \min\{ {1\over 6} c\delta^2, C_0^2\}$.  Then
$$
V^{-1}(E - \ve, E + \ve) = \tcup^m_{i=1} (a_i, b_i),\quad m \le m_0,\quad 
b_i - a_i \lesssim \ve^{1/2}
$$
\end{lemma}

\begin{proof} Each monotonicity interval of $V$ can intersect with at most one 
component $(a_i, b_i)$ of $V^{-1}(E - \ve, E + \ve)$ so $m \le m_0$.
$$
\int^{x_j+\delta}_{x_j} |V'(y)| dy \ge \int^\delta_0 c t dt = {c\over 2} 
\delta^2
$$
so $(a_i, b_i)$ does not contain any $(x_j, x_j + \delta)$.  Similarly, 
$(a_i, b_i)$ does not contain any $(x_j-\delta, x_j)$.  Therefore, $(a_i, b_i)$
 intersects with at most one $(x_{j_1}, x_{j_1}+\delta)$ and one 
$(x_{j_2}-\delta, x_{j_2})$.  Suppose 
$(a_i, b_i)\cap (x_{j_1}, x_{j_1}+\delta) = (y_1, y_2)$
\begin{align*}
2\ve & \ge \int^{y_2}_{y_1} |V'(y)|dy\\
& \ge  c\int^{y_2}_{y_1} (y - x_{j_1}) dy\\
& = {c\over 2}\left[(y_2 - x_{j_1})^2 - (y_1 - x_{j_1})^2\right]\\
& \ge {c\over 2}(y_2 - y_1)^2
\end{align*}
Similarly, if $(a_i, b_i)\cap (x_{j_2} - \delta, x_{j_2}) = (y_3, y_4)$ then 
$2\ve \ge {c\over 2}(y_4 - y_3)^2$.
\begin{align*}
b_i - a_i & = (y_2 - y_1) + (y_4 - y_3) + \mes\Bigl[(a_i, b_i)\setminus 
\tcup_j(x_j - \delta, x_j + \delta)\Bigr]\\
& \le 2\Bigl({4\ve \over c}\Bigr)^{1/2} + {2\ve\over C_0}\
\end{align*}
\end{proof}

Fix $N \gg 1$, $\lambda = N^{8 \cdot 2^{m_0}}$, such that
$$
\lambda^{-1/2} < \min \{ {1\over 6} c\delta^2, C_0^2 \}\ .
$$

\begin{lemma} For any $x \in \tor$, $E \in \IR$, if $\w$ satisfies the 
Diophantine condition
\begin{equation}
\|\ell \w\| \gtrsim |\ell|^{-\beta} 
\end{equation}
$1 < \beta < 2\ $, then
$$
\#\left\{|\ell| \le N^{2^{m_0}}: x + \ell \w \in V^{-1}\bigl(E - \lambda^{-{1\over 2}}, E + \lambda^{-{1\over 2}}\bigr)\right\} \le m_0
$$
provided $\delta<\delta_0(m_0,\beta)$.
\end{lemma}

\begin{proof} Suppose, for contradiction, that
$$
x + \ell_1 \w, x + \ell_2 \w \in (a_i, b_i) \subset V^{-1} (E - \lambda^{-1/2}, E + \lambda^{-1/2})\ .
$$
Then, by Lemma 6.1,
$$
N^{-\beta 2^{m_0}} < |\ell_1 - \ell_2|^{-\beta} 
\lesssim \|(\ell_1 - \ell_2)\w\| 
< b_i - a_i \lesssim \Bigl(\lambda^{-1/2}\Bigr)^{1/2} =N^{-2\cdot 2^{m_0}}
$$
This is false for sufficiently large $N$ (or equivalently, for sufficiently 
small $\delta$).
\end{proof}

It is well known that almost all $\w\in\tor$ satisfied condition (6.1).  In particular, given any interval $(\uw,\bw)$, there is $\w\in(\uw,\bw)$ such that condition (6.1) is satisfied.

With the upper bound provided by Lemma 6.2, we can now find $N^\one$ such that all the resonance term is far away from the edge of $[-(N^\one)^2,(N^\one)^2]$.  In this case, the Green's function $H_{[1,(N^\one)^2]}(x,\w)(k,N^\one+1)$ will decay exponentially and we get exponentially decaying eigenfunctions.

\begin{lemma} For any $x \in \tor$, and any $\w$ satisfying condition $(6.1)$, 
there is $N^\one(x,\w) \le N^{2^{m_0 -1}}$ such that
$$
|V(x + j\w) - V(x)| \ge \lambda^{-1/2}
$$
for all
$$
j \in \left[-\bigl(N^\one(x,\w)\bigr)^2, -N^\one(x,\w) \right) \cup 
\left(N^\one(x,\w) , \bigl(N^\one(x,\w)\bigr)^2\right]\ .
$$
\end{lemma}

\begin{proof} If $|V(x + j\w) - V(x)| \ge \lambda^{-1/2}$ for all $j 
\in [-N^2, -N)\cup (N, N^2]$ then we can take $N^\one(x,\w)=N$.

So assume there is $j_1 \in [-N^2, -N)\cup(N, N^2]$ such that 
$|V(x + j_1 \w) - V(x)| < \lambda^{-1/2}$.  If 
$|V(x + j\w) - V(x)| \ge \lambda^{-1/2}$ for all
$j \in [-N^4, -N^2)\cup(N^2, N^4]$ then we can choose $N^\one(x,\w)=N^2$.

So assume there is $j_2 \in [-N^4, -N^2)\cup (N^2, N^4]$ such that 
$|V(x + j_2\w) - V(x)| < \lambda^{-1/2}$.  Continue this process if there is 
$j_i \in [-N^{2^i}, -N^{2^{i-1}}) \cup$ $(N^{2^{i-1}}, N^{2^i}]$ such that 
$|V(x+ j_i \w) - V(x)| < \lambda^{-1/2}$.  Since
$$
\#\left\{0 < |j| \le N^{2^{m_0}}: |V(x+j\w) -V(x)| < \lambda^{-1/2}\right\} \le m_0 -1
$$
we can find $N^\one(x,\w) \le N^{2^{m_0-1}}$ such that
$$
|V(x+j\w) - V(x)| \ge \lambda^{-1/2}
$$
for all
$$
j\in \left[-\bigl(N^\one(x,\w)\bigr)^2, -N^\one(x,\w)-1\right]\cup 
\left[N^\one(x,\w) +1, \bigl(N^\one(x,\w)\bigr)^2\right]\ .
$$
\end{proof}

Take $K,L \in \IN$ such that $20C_1 \lambda^{1/2} < K < 30C_1 \lambda^{1/2}$, 
$20C_1 \lambda^{5/8} < L < 30C_1 \lambda^{5/8}$.  Let $\tx_i = {i\over K}\ ,\ 
i=1,2, \ldots , K$ and $\tilde{\w}_j = {j\over L}\ , \ j=1,2, \ldots, L$.  
Define
$$
U_i = ( \tx_i - {1\over 32C_1} \lambda^{-1/2}, 
        \tx_i + {1\over 32C_1} \lambda^{-1/2} ), \qquad 
V_j = ( \tilde{\w}_j - {1\over 32C_1} \lambda^{-5/8}, 
        \tilde{\w}_j + {1\over 32C_1} \lambda^{-5/8} )
$$
Then $\tor \times \tor = \bigcup\limits_{i,j} U_i \times V_j$.  Moreover, there
 is $\hat{\w}_j \in V_j$ satisying condition (6.1) for each $j$.

\begin{lemma} For each $i,j$, there is $N^\one_{ij} \le N^{2^{m_0-1}}$ such 
that $|V(x+k\w) - V(x)| \ge {7\over 8}\, \lambda^{-1/2}$ for all 
$(x,\w) \in U_i \times V_j$, 
$k \in \left[-\bigl(N^\one_{ij}\bigr)^2, -N^\one_{ij} \right)\cup 
\left(N^\one_{ij}, \bigl(N^\one_{ij}\bigr)^2\right]$.
\end{lemma}

\begin{proof} Let $N^\one_{ij}=N^\one(\tx_i,\hat{\w}_j)$.  For any  
$(x,\w) \in U_i \times V_j$, 
$k \in \left[-\bigl(N^\one_{ij}\bigr)^2, -N^\one_{ij} \right)\cup 
\left(N^\one_{ij} , \bigl(N^\one_{ij}\bigr)^2\right]$, we have
\begin{eqnarray*}
|V(x + k\w) - V(x)| & \ge &   |V(\tx_i + k\hat{\w}_j) - V(\tx_i)|
                           - |V(\tx_i + k\hat{\w}_j) - V(\tx_i + k\w)| \\
& & \mbox{}              - |V(\tx_i + k\w) - V(x + k\w)| - |V(x) - V(\tx_i)| \\
& \ge & \lambda^{-1/2} - C_1|k(\hat{\w}_j - \w)| - 2C_1|\tx_i - x| \\
& \ge & {7\over 8}\, \lambda^{-1/2}
\end{eqnarray*}
\end{proof}

\noindent
{\bf Notation:}\quad $\Lambda^\one_{ij} = \left[-\bigl(N^\one_{ij}\bigr)^2, 
\bigl(N^\one_{ij}\bigr)^2\right]$.

\bigskip
Let
\begin{equation*}
\begin{split}
J^\one_{ij} &:= \#\left\{\ell \in \Lambda^\one_{ij}: 
|V(\tx_i + \ell \hat{\w}_j) - V(\tx_i)| < {1\over 2} \lambda^{-1/2}\right\}\\
& \le \#\left\{|\ell| \le N^{2^{m_0}}: |V(\tx_i + \ell \hat{\w}_j) - V(\tx_i)| < {1\over 2}\lambda^{-1/2}\right\} \le m_0\ .
\end{split}
\end{equation*}
We have distinct eigenvalues
$$
E^\one_{ijk}(\tx_i, \hat{\w}_j) \in \rsp H_{\Lambda^\one_{ij}}(\tx_i, \hat{\w}_j)\ ,\quad 
1 \le k \le J^\one_{ij}\ ,
$$
such that there is $\ell_k \in \bigl[-N^\one_{ij}, N^\one_{ij}]$, 
$\big |E^\one_{ijk}(\tx_i, \hat{\w}_j) 
- \lambda V(\tx_i + \ell_k \hat{\w}_j)\big | \le 2$, 
$|V(\tx_i + \ell_k \hat{\w}_j) - V(\tx_i)| < {1\over 2} \lambda^{-1/2}$.  
By perturbation theory and Lemma A.1, we can define $C^3$ functions 
$E^\one_{ijk}(x,\w)$, $\vp^\one_{ijk}(x,\w)(\ell)$, $\ell \in \Lambda^\one_{ij}$, for 
$(x,\w) \in U_i \times V_j$ such that
\begin{align*}
H_{\Lambda^\one_{ij}}(x,\w) \vp^\one_{ijk}(x,\w) & 
= E^\one_{ijk}(x,\w)\vp^\one_{ijk}(x,\w)\\[6pt]
\partial_x E^\one_{ijk}(x,\w) & = \lambda\, \sum_{\ell\in \Lambda^\one_{ij}}\, 
V'(x + \ell \w) \big |\vp^\one_{ijk}(x,\w) (\ell)\big |^2\\[6pt]
\partial_\w E^\one_{ijk}(x,\w) & = \lambda\, \sum_{\ell\in \Lambda^\one_{ij}}\, 
\ell \ V'(x + \ell \w) \big |\vp^\one_{ijk}(x,\w) (\ell)\big |^2\\[6pt]
\sum_{\ell\in \Lambda^\one_{ij}} \big|\vp^\one_{ijk}(x,\w)(\ell)\big|^2 & = 1
\end{align*}
Furthermore,
\begin{eqnarray*}
|E^\one_{ijk}(x,\w) - \lambda V(x)| 
& \le & |E^\one_{ijk}(x,\w) - E^\one_{ijk}(\tx_i,\w)|
      + |E^\one_{ijk}(\tx_i,\w) - E^\one_{ijk}(\tx_i,\hat{\w})|  \\
& & \mbox{} + |E^\one_{ijk}(\tx_i,\hat{\w}) - \lambda V(x+\ell_k \hat{\w})|
      + |\lambda V(x+\ell \hat{\w}) - \lambda V(x)| \\
& \le & (\lambda C_1)|x-\tx_i| + (\lambda N^{2^{m_0}} C_1)|\w-\hat{\w}| +2
      + {1\over 2} \lambda^{1/2} \\
& \le & {3\over 4} \lambda^{3/4}
\end{eqnarray*}

\begin{lemma} For any $(x,\w) \in U_i \times V_j$, 
$N^\one_{ij} < \ell \le (N^\one_{ij})^2$, one has
$$ 
\left(H_{\bigl(N^\one_{ij}, 
\bigl(N^\one_{ij}\bigr)^2\bigr]}(x,\w) - E^\one_{ijk}(x,\w)\right)^{-1} 
\bigl(N^\one_{ij} +1, \ell\bigr) \lesssim 
\lambda^{-{1\over 2}(\ell - N^\one_{ij})}\ .
$$
\end{lemma}

\begin{proof} 
\begin{align*}
|\lambda V(x + m\w) - E^\one_{ijk}(x,\w)| & \ge |\lambda V(x + m\w) 
- \lambda V(x)| - |\lambda V(x) - E^\one_{ijk}(x,\w)| \\
 & \ge {7\over 8} \lambda^{1/2} - \bigl({3\over 4} \lambda^{1/2} \bigr)
\ge {1\over 8} \lambda^{1/2}
\end{align*}
for any $m \in \bigl(N^\one_{ij} , (N^\one_{ij})^2\bigr]$.  Also 
$(N^\one_{ij})^2 \le N^{2^{m_0}} \le \lambda^{1/2}$.  
The assertion follows from Corollary C.8.
\end{proof}

\begin{corollary} $\big|\vp^\one_{ijk}(x,\w) (\ell)\big | \le 
\lambda^{-{1\over 2}(|\ell| - N^\one_{ij})}$, 
$N^\one_{ij} < |\ell| \le (N^\one_{ij})^2$.
\end{corollary}

In the next few sections, we will derive some properties of $E^\one_{ijk}$. But we need to know that these are the only eigenvalues of $H_{\Lambda^\one_{ij}}(x,\w)$ near $V(x)$.

\begin{lemma} For any $(x,\w) \in U_i \times V_j$,
$$
\rsp H_{\Lambda^\one_{ij}}(x,\w) \cap \left(\lambda V(x) - {1\over 4} \lambda^{1/2}, 
\lambda V(x) + {1\over 4} \lambda^{1/2}\right) \subset 
\tcup_k\left\{E^\one_{ijk}(x,\w)\right\}.
$$
\end{lemma}

\begin{proof} By construction, $\rsp H_{\Lambda^\one_{ij}}(\tx_i,\hat{\w}_j)\cap 
\bigl(\lambda V(\tx_i) - {1\over 2} \lambda^{1/2} +2, \lambda V(\tx_i) 
+ {1\over 2} \lambda^{1/2} -2\bigr) \subset \tcup_k 
\left\{ E^\one_{ijk} (\tx_i, \hat{\w}_j)\right\} $.  Suppose
$$
\mu(x_0,\w_0) \in \rsp H_{\Lambda^\one_{ij}}(x_0,\w_0) \cap \biggl(\lambda V(x_0,\w_0) - {1\over 4} \lambda^{1/2}, \lambda V(x_0,\w_0) + {1\over 4} \lambda^{1/2}\biggr)
$$
for some $(x_0,\w_0) \in U_i \times V_j$.  Then we can define $C^1$ function
$$
\mu(x,\w) \in \rsp H_{\Lambda^\one_{ij}}(x,\w)\ ,\quad 
|\partial_x \mu(x,\w)| \le \lambda C_1,\  
|\partial_\w \mu(x,\w)| \le \lambda N^{2^{m_0}} C_1
$$
on $U_i \times V_j$.  But
\begin{equation*}
\begin{split}
|\mu(\tx_i,\hat{\w}_j) - \lambda V(\tx_i)| & 
\le |\mu(\tx_i,\hat{\w}_j) - \mu(x_0,\hat{\w}_j)| 
+ |\mu(x_0,\hat{\w}_j) - \mu(x_0,\w_0)| \\
& \qquad + |\mu(x_0,\w_0) - \lambda V(x_0)| + |\lambda V(x_0) - \lambda V(\tx_i)|\\
& \le \lambda C_1 |\tx_i - x_0| + \lambda N^{2^{m_0}} C_1 |\hat{\w}_j-\w_0|
+ {1\over 4} \lambda^{1/2} + \lambda C_1 |x_0 - \tx_i| \\
& \le {1\over 2} \lambda^{1/2} -2
\end{split}
\end{equation*}
Hence $\mu(\tx_i,\hat{\w}_j) = E^\one_{ijk} (\tx_i,\hat{\w}_j)$ for some $k$.  
Therefore $\mu(x,\w) = E^\one_{ijk} (x,\w)$.
\end{proof}


\setcounter{section}{6}
\section{Separation of eigenvalues at first scale}
We develop a method to find a lower estimate for the separtion between eigenvalues of $H_{\Lambda_{ij}}(x,\w)$.  This method, first introduced in [GS2] for analytic potential, is based on the orthogonality of eigenfunctions corresponding to different eigenvalues of the self-adjoint matrix $H_{\Lambda_{ij}}(x,\w)$.

\begin{lemma} Let $(x,\w) \in U_i \times V_j$, $E_1, E_2 \in \bigl(\lambda V(x) - {3\over 4} \lambda^{1/2}, \lambda V(x) + {3\over 4} \lambda^{1/2}\bigr)$.  Then for any $n \in \bigl[-(N^\one_i)^2, -N^\one_i \bigr)$, one has
$$
\big |\log |f_{[-(N_{ij}^\one)^2, n]}(x,\w,E_1)| - \log |f_{[-(N^\one_{ij})^2, n]}(x,\w, E_2)| \big | \le |E_1 - E_2|\ .
$$
\end{lemma}

\begin{proof} Let $\bigl\{\mu_m\bigr\}^n_{m = -(N^\one_{ij})^2} = 
\rsp H_{[-(N^\one_{ij})^2, n]}(x,\w)$, $|\mu_m - \lambda V(x + m\w)| \le 2$.  Then for any $E \in \bigl(\lambda V(x) - {3\over 4} \lambda^{1/2}, \lambda V(x) + {3\over 4} \lambda^{1/2}\bigr)$
\begin{equation*}
\begin{split}
|\mu_m - E| & \ge |\lambda V(x + m\w) - \lambda V(x)| - |\lambda V(x+ m\w) - \mu_m| - |\lambda V(x) - E|\\
& \ge {7\over 8} \lambda^{1/2} - 2 - {3\over 4} \lambda^{1/2}\\
& \ge {1\over 16} \lambda^{1/2}\ .
\end{split}
\end{equation*}
For each $m$, there exists $E \in (E_1, E_2) \subset \bigl(\lambda V(x) - {1\over 4}\lambda^{1/2}, \lambda V(x) + {1\over 4} \lambda^{1/2}\bigr)$ such that
$$
\big |\log |\mu_m - E_1| - \log |\mu_m - E_2|\big | = {|E_1 - E_2|\over |\mu_m - E|} \le 2 \lambda^{-1/2} |E_1 - E_2|
$$
\begin{equation*}
\begin{split}
& \big |\log |f_{[-(N_{ij}^\one)^2, n]} (x,\w, E_1)| - \log |f_{[-(N^\one_{ij})^2, n]}(x,\w, E_2)|\big |\\
& \le \sum_m \big |\log |\mu_m - E_1| - \log |\mu_m - E_2|\big |\\
& \le (N^\one_{ij})^2 \bigl[16 \lambda^{-1/2}|E_1 - E_2|\bigr]\\
& \le \lambda^{1/8} \bigl[16 \lambda^{-1/2} |E_1 - E_2|\bigr]\\
& \le |E_1 - E_2|
\end{split}
\end{equation*}
\end{proof}

\begin{corollary} Let $(x,\w), E_1, E_2, n$ be as in Lemma 7.1.  Then
$$
\big | f_{[-(N^\one_{ij})^2, n]}(x,\w, E_1) - 
f_{[-(N^\one_{ij})^2, n]}(x,\w, E_2)\big |
\le |E_1 - E_2| e^{|E_1 - E_2|}\big |f_{[-(N^\one_{ij})^2, n]}(x,\w, E_1)\big |\ .
$$
\end{corollary}

\begin{proof} With $\bigl\{\mu_m\bigr\}$ as in proof of Lemma 7.1, note that $\sgn(\mu_j - E_1) = \sgn(\mu_j - E_2)$.  So 
$$
\sgn f_{[-(N^\one_{ij})^2, n]}(x,\w, E_1) = \sgn f_{[-(N^\one_{ij})^2, n]}(x,\w, E_2)\ .
$$
Hence, there is $\Gamma$ between $|f_{[-(N^\one_{ij})^2, n]}(x,\w,E_1)|$ and 
$|f_{[-(N^\one_{ij})^2, n]}(x,\w,E_2)|$ such that
\begin{equation*}
\begin{split}
& \big |f_{[-(N^\one_{ij})^2, n]}(x,\w, E_1) - f_{[-(N^\one_{ij})^2, n]}(x,\w, E_2)\big |\\
& \qquad = \big | |f_{[-(N^\one_{ij})^2, n]} (x,\w, E_1)| - |f_{[-(N^\one_{ij})^2 -n]}(x,\w, E_2)|\big |\\
& \qquad = \big |\log |f_{[-(N^\one_{ij})^2, n]}(x,\w, E_1)| - \log |f_{[-(N^\one_{ij})^2, n]}(x,\w, E_2)| \big | \cdot \Gamma \\
& \qquad \le |E_1 - E_2|\cdot \Gamma \ .
\end{split}
\end{equation*}
But
\begin{equation*}
\Gamma  \le \max \left\{|f_{[-(N^\one_{ij})^2, n]}(x,\w, E_1)|,\ 
|f_{[-(N^\one_{ij})^2, n]}(x,\w, E_2)|\right\}
 \le e^{|E_1 - E_2|} |f_{[-(N^\one_{ij})^2, n]}(x,\w, E_1)|\ .
\end{equation*}
\end{proof}

\begin{lemma} For any $(x,\w) \in U_i \times V_j$, $1 \le k \le J^\one_{ij}$,
\begin{equation*}
\sum_{N^\one_{ij} < |n| \le (N^\one_{ij})^2} \big |f_{[-(N^\one_{ij})^2, n]}(x,\w, E^\one_{ijk}(x,\w))\big |^2
 < 4 \lambda^{-1} \sum_{n \in \Lambda^\one_{ij}} \big |f_{[-(N^\one_{ij})^2, n]}(x,\w, E^\one_{ijk}(x,\w))\big |^2\ .
\end{equation*}
\end{lemma}

\begin{proof} $\left\{f_{[-(N^\one_{ij})^2, n]}(x,\w, E^\one_{ijk}(x,\w))\right\}_{n \in \Lambda^\one_{ij}}$ satisfies the Schr\"odinger equation with zero boundary condition.  Hence, there is $\mu \in\IR$ such that
$$
f_{[-(N^\one_{ij})^2, n]}(x,\w, E^\one_{ijk}(x,\w)) = \mu\, \vp^\one_{ijk}(x,\w)(n)\ .
$$
By Corollory 6.6,
\begin{equation*}
\begin{split}
\sum_{N^\one_{ij} < |n| \le (N^\one_{ij})^2}\big |f_{[-(N^\one_{ij})^2, n]}(x,\w, E^\one_{ijk}(x,\w))\big |^2 & = \mu^2 \sum_{N^\one_{ij} < |n| \le (N^\one_{ij})^2} \big |\vp^\one_{ijk}(x,\w)(n)\big |^2\\
& \le 2 \mu^2 \sum_{n > N^\one_{ij}}\, \lambda^{-(|n| - N^\one_{ij})}\\
& \le 2\mu^2 {\lambda^{-1}\over 1 - \lambda^{-1}}\\
& < 4 \lambda^{-1} \mu^2\\
& = 4 \lambda^{-1} \sum_{n \in \Lambda^\one_{ij}} \big |f_{[-(N^\one_{ij})^2, n]} (x,\w, E^\one_{ijk}(x,\w))\big |^2\ .
\end{split}
\end{equation*}
\end{proof}

\begin{lemma} For any $(x,\w) \in U_i \times V_j$, 
$\big |E^\one_{ijk}(x,\w) - E^\one_{ij\ell}(x,\w) \big | 
> e^{- |\Lambda^\one_{ij}|^{3/4}}$ if $k \ne \ell$.
\end{lemma}

\begin{proof} For any $n \in \bigl[-N^\one_{ij}, N^\one_{ij}\bigr]$,
\begin{equation*}
\begin{split}
& \big |f_{[-(N^\one_{ij})^2, n]} \bigl(x,\w, E^\one_{ijk}(x)\bigr) 
- f_{[-(N^\one_{ij})^2, n]} \bigl(x,\w, E^\one_{ij\ell}(x,\w)\bigr)\big |\\
& \le \Big \| \begin{pmatrix}
f_{[-(N^\one_{ij})^2, n]}\bigl(x,\w, E^\one_{ijk}(x,\w)\bigr)\\
f_{[-(N^\one_{ij})^2, n-1]}\bigl(x,\w, E^\one_{ijk}(x,\w)\bigr)
\end{pmatrix}
-
\begin{pmatrix}
f_{[-(N^\one_{ij})^2, n]}\bigl(x,\w, E^\one_{ij\ell}(x,\w)\bigr)\\
f_{[-(N^\one_{ij})^2, n-1]}\bigl(x,\w, E^\one_{ij\ell}(x,\w)\bigr)
\end{pmatrix}
\Big \|\\
& = \Big \|
M_{[-N^\one_{ij} -1, n-1]}\bigl(x,\w, E^\one_{ijk}(x,\w)\bigr)
\begin{pmatrix}
f_{[-(N^\one_{ij})^2, - N^\one_{ij} -1]} \bigl(x,\w, E^\one_{ijk}(x,\w)\bigr)\\
f_{[-(N^\one_{ij})^2, -N^\one_{ij} -2]} \bigl(x,\w, E^\one_{ijk}(x,\w)\bigr)
\end{pmatrix}\\
&\quad - 
M_{[-N^\one_{ij} -1, n-1]}\bigl(x,\w, E^\one_{ij\ell}(x,\w)\bigr)
\begin{pmatrix}
f_{[-(N^\one_{ij})^2, - N^\one_{ij} -1]} \bigl(x,\w, E^\one_{ij\ell}(x,\w)\bigr)\\
f_{[-(N^\one_{ij})^2, -N^\one_{ij} -2]} \bigl(x,\w, E^\one_{ij\ell}(x,\w)\bigr)
\end{pmatrix}
\Big \|\\
& \le
\Big \|
M_{[-N^\one_{ij} -1, n-1]}\bigl(x,\w, E^\one_{ijk}(x,\w)\bigr) -
M_{[-N^\one_{ij} -1, n-1]}\bigl(x,\w, E^\one_{ij\ell}(x,\w)\bigr) \Big \|\ \cdot\\
&\quad
\Big \|
\begin{pmatrix}
f_{[-(N^\one_{ij})^2, - N^\one_{ij} -1]} \bigl(x,\w, E^\one_{ijk}(x,\w)\bigr)\\
f_{[-(N^\one_{ij})^2, -N^\one_{ij} -2]} \bigl(x,\w, E^\one_{ijk}(x,\w)\bigr)
\end{pmatrix}
\Big \|
+ \Big \| M_{[-N^\one_{ij} -1, n-1]} \bigl(x,\w, E^\one_{ij\ell}(x,\w)\bigr)\Big \|\ \cdot\\
&\quad 
\Big \|
\begin{pmatrix}
f_{[-(N^\one_{ij})^2, - N^\one_{ij} -1]} \bigl(x,\w, E^\one_{ijk}(x,\w)\bigr)\\
f_{[-(N^\one_{ij})^2, -N^\one_{ij} -2]} \bigl(x,\w,E^\one_{ij\ell}(x,\w)\bigr)
\end{pmatrix} -
\begin{pmatrix}
f_{[-(N^\one_{ij})^2, -N^\one_{ij} -1]}\bigl(x,\w,E^\one_{ij\ell}(x,\w)\bigr)\\
f_{[-(N^\one_{ij})^2, -N^\one_{ij} -2]} \bigl(x,\w,E^\one_{ij\ell}(x,\w)\bigr)
\end{pmatrix} \Big \|\\
& \le
\Big |E^\one_{ijk}(x,\w) - E^\one_{ij\ell}(x,\w) \Big | (\lambda C_2)^{n + N^\one_{ij}} |\mu|\\
&\quad +
(\lambda C_2)^{n + N^\one_{ij} + 1} \Bigl(\big |E^\one_{ijk}(x,\w) - E^\one_{ij\ell}(x,\w)\big | e^{|E^\one_{ijk}(x,\w) - E^\one_{ij\ell}(x,\w)|} |\mu| \Bigr)
\end{split}
\end{equation*}
where 
$C_2 = \log \bigl(2 \max\limits_{y \in \tor}|V(y)| +1\bigr)$, 
$|\mu|^2 = \sum\limits_{m \in \Lambda^\one_{ij}}|f_{[-(N^\one_{ij})^2,n]}(x,\w,E^\one_{ijk}(x,\w))|^2$.  
($\mu $ is the same as in the proof of Lemma 7.3)

If $\big |E^\one_{ijk}(x,\w) - E^\one_{ij\ell}(x,\w)\big | 
\le e^{-|\Lambda^\one_{ij}|^{3/4}} = e^{-(N^\one_{ij})^{3/2}}$ then
$$
\big |f_{[-(N^\one_{ij})^2, n]}\bigl(x,\w, E^\one_{ijk}(x,\w)\bigr) 
- f_{[-(N^\one_{ij})^2, n]}\bigl(x,\w, E^\one_{ij\ell}(x,\w)\bigr)\big | \le  
e^{- {1\over 2} |\Lambda^\one_{ij}|^{-3/4}} |\mu|
$$
for all $n \in [-N^\one_{ij}, N^\one_{ij}]$.

Therefore,
\begin{equation*}
\begin{split}
& \sum_{n \in \Lambda^\one_{ij}} \Big | 
f_{[-(N^\one_{ij})^2, n]}\bigl(x,\w, E^\one_{ijk}(x,\w ) \bigr) -
f_{[-(N^\one_{ij})^2, n]}\bigl(x,\w, E^\one_{ij\ell}(x,\w) \bigr) \Big |^2\\
& = \sum_{N^\one_{ij} < |n| \le (N^\one_{ij})^2} \Big |
f_{[-(N^\one_{ij})^2, n]}\bigl(x,\w, E^\one_{ijk}(x,\w) \bigr) -
f_{[-(N^\one_{ij})^2, n]}\bigl(x,\w, E^\one_{ij\ell}(x,\w) \bigr) \Big |^2\\
& \quad + \sum_{|n| \le N^\one_{ij}} \Big |
f_{[-(N^\one_{ij})^2, n]}\bigl(x,\w, E^\one_{ijk}(x,\w) \bigr) -
f_{[-(N^\one_{ij})^2, n]}\bigl(x,\w, E^\one_{ij\ell}(x,\w) \bigr) \Big |^2\\
& \le \sum_{N^\one_{ij} < |n| \le (N^\one_{ij})^2} \biggl[\bigl | 
f_{[-(N^\one_{ij})^2, n]} \bigl(x,\w, E^\one_{ijk}(x,\w) \bigr) \bigl|^2 
+ \big |
f_{[-(N^\one_{ij})^2, n]}\bigl(x,\w, E^\one_{ij\ell}(x,\w) \bigr)\big |^2
\biggr]\\
& \quad + \bigl(2 N^\one_{ij} +1\bigr) e^{-|\Lambda^\one_{ij}|^{3/4}}|\mu|^2\\
& \le 4 \lambda^{-1} \sum_{n \in \Lambda^\one_{ij}} \biggl[ \big |
f_{[-(N^\one_{ij})^2, n]}\bigl(x,\w, E^\one_{ijk}(x,\w) \bigr)\big |^2 +
\big |
f_{[-(N^\one_{ij})^2,n]}\bigl(x,\w,E^\one_{ij\ell}(x,\w)\bigr)\big |^2\biggr]\\
& \quad + \bigl(2 N^\one_{ij} +1\bigr) e^{-|\Lambda^\one_{ij}|^{3/4}}|\mu|^2 \\
& \le \lambda^{-1/2} \sum_{n \in \Lambda^\one_{ij}} \biggl[ \big |
f_{[-(N^\one_{ij})^2, n]}\bigl(x,\w, E^\one_{ijk}(x,\w) \bigr)\big |^2 + \big |
f_{[-(N^\one_{ij})^2, n]}\bigl(x,\w, E^\one_{ij\ell}(x,\w)\bigr)\big |^2\biggr]
\end{split}
\end{equation*}
But $E^\one_{ijk}(x,\w) \ne E^\one_{ij\ell}(x,\w) \Longrightarrow \left\{
f_{[-(N^\one_{ij})^2, n]}\bigl(x,\w, E^\one_{ijk}(x,\w) \bigr)\right\}_{n \in \Lambda^\one_{ij}}$ and \newline
$\left\{
f_{[-(N^\one_{ij})^2, n]}\bigl(x,w, E^\one_{ij\ell}(x,\w) \bigr)\right\}_{n \in \Lambda^\one_{ij}}$ are orthogonal.  Hence
\begin{equation*}
\begin{split}
& \qquad
\sum_{n \in \Lambda^\one_{ij}} \big |
f_{[-(N^\one_{ij})^2, n]}\bigl(x,\w, E^\one_{ijk}(x,\w) \bigr) -
f_{[-(N^\one_{ij})^2, n]}\bigl(x,\w, E^\one_{ij\ell}(x,\w) \bigr) \big |^2\\
& = \sum_{n \in \Lambda^\one_{ij}} \biggl[ \big |
f_{[-(N^\one_{ij})^2, n]}\bigl(x,\w, E^\one_{ijk}(x,\w) \bigr) \big |^2 +
\big |
f_{[-(N^\one_{ij})^2, n]}\bigl(x,\w,E^\one_{ij\ell}(x,\w) \bigr) \big |^2\biggr]\ .
\end{split}
\end{equation*}
This contradiction shows that
$$
\big | E^\one_{ijk}(x,\w) - E^\one_{ij\ell}(x,\w) \big | 
> e^{-|\Lambda^\one_{ij}|^{3/4}}\ .
$$
\end{proof}


\setcounter{section}{7}
\section{Variations of potential}
In this section we introduce variation of potential and the notion of ``typical'' variations. We then derive some properties of $E^{(1)}_{ijk}$ for typical variations. 

\definition  Let $T$ be a large integer, $0< \delta \ll {1\over T^5}$.  Suppose $R_m(\eta_m,\xi_m,\theta_m;x)$ are $C^3$ functions, $m=1,2, \ldots, T$, $(\eta,\xi,\theta) \in \prod\limits_1^{3T} [-\delta,\delta]$, $x \in \tor$, satisfying the following conditions:
\begin{align}
|\partial_\alpha R_m(\eta_m,\xi_m,\theta_m;x)| \lesssim  {1\over T} & 
\qquad \text{for any index $|\alpha|\le 3$} \\
R_m(0,0,0;x) \equiv 0 & \\
R_m(\eta_m,\xi_m,\theta_m;x)=-x^{-3}\bigl( \eta_m + \xi_m x + {1\over 2}\theta_m x^2 \bigr)& 
\qquad \text{for $|x| \ge {1\over 2T}$}
\end{align} 
Define a $(T,\delta)$--variation of potential by
$$ 
W(\eta,\xi,\theta,\{ R_m \} ; x) = \sum\limits_{m=1}^{T} 
v_m\Bigl(\eta_m,\xi_m,\theta_m;x-{m\over T} \Bigr) 
$$
where
$$ v_m(\eta_m,\xi_m,\theta_m;x) = \eta_m + \xi_m x + {1\over 2} \theta_m x^2 + 
x^3 R_m (\eta_m,\xi_m,\theta_m;x) $$

By (8.2) and (8.3),
\begin{align*}
v_m(0, 0, 0, \{ R_m \} ; x) & \equiv 0 \\ 
\intertext{and}
v_m(\eta_m, \xi_m, \theta_m, \{ R_m \} ; x) & = 0\qquad \text{for $|x| \ge {1\over 2T}$}\ .
\end{align*}

Denote the collection of $(T,\delta)$--variations of potential by $\cS(T,\delta)$.  A set $S \subset \cS(T,\delta)$ is called $(1-\ve)$-typical if 
$$
|S| := \min\limits_{ \{ R_m \} } {1\over (2\delta)^{3T}}
\mes\bigl\{(\eta, \xi, \theta) \in [-\delta,\delta]^{3T}: 
v(\eta, \xi, \theta, \{ R_m \} ; . ) \in S \bigr\} \ge 1 - \ve
$$

\begin{remark} We assume $\delta \ll \lambda^{-1/2}$.  Since $\max\limits_{y \in \tor} \big |\tv(y) - V(y)| \lesssim \delta$, one has
$$
\big |\tv(x + j\w) - \tv(x) \big | \ge {7\over 8} \lambda^{-1/2}\qquad \text{for}\ j \in \Lambda^\one_{ij} \setminus \bigl[-N^\one_{ij}, N^\one_{ij}\bigr]
$$
for any $(x,\w) \in U_i \times V_j$.  Hence, we can define $\te^\one_{ijk}(x,\w)$
 and $\tvp^\one_{ijk}(x,\w)$ on 
$U_i \times V_j$, such that
$$
\th_{\Lambda^\one_{ij}}(x,\w) \tvp^\one_{ijk} (x,\w) = \te^\one_{ijk}(x,\w) \tvp^\one_{ijk}(x,\w)\ , \quad \|\tvp^\one_{ijk}(x,\w)\| = 1
$$
$\te^\one_{ijk}(x,\w)$ and $\tvp^\one_{ijk}(x,\w)$ have the same properties of $E^\one_{ijk}(x,\w)$ and $\vp^\one_{ijk}(x,\w)$, shown in Section 6 and Section 7.

\end{remark}

\begin{remark} These functions depend on $\{\eta_m\}, \{\xi_m\}, \{\theta_m\}$.  It is for the simplicity of notations that the dependence is not explicitly written.

\end{remark}

For the remaining of this section, we fix $(\tx,\tw) \in U_i \times V_j$, 
$\tw$ satisfies condition (6.1).

\begin{remark} We can define an one-to-one correspondence between $\Lambda^\one_{ij}$ and a subset of $[1, T]$ as follow: For any $\ell \in \Lambda^\one_{ij}$, there is unique $\ell' \in [1, T] \cap \IN$ such that $-{1\over 2T} < \{\tx + \ell \tw\} - {\ell'\over T} \le {1\over 2T}$.  If $m, \ell \in \Lambda^\one_{ij}$, $m \ne \ell$, then the Diophantine condition implies $m' \ne \ell'$.
\end{remark}

\begin{remark} If $p \ne q'$ for all $q \in \Lambda^\one_{ij}$, then $\partial_{\xi_p} \th_{\Lambda^\one_{ij}}(\tx,\tw) = 0$.  
\end{remark}

Let $\vr = {1\over 2} \dist \bigl(\te^\one_{ijk}(\tx,\tw), \rsp \th_{\Lambda^\one_{ij}}(\tx,\tw) \setminus \bigl\{\te^\one_{ijk}(\tx,\tw)\bigr\}\bigr)
> {1\over 2} e^{-| \Lambda^\one_{ij} |^{3/4}}$.  We now show that if the eigenvalues are separated, then we have control how much the eigenfunctions change when we perturbed the potential.

\begin{lemma} 
\begin{align*}
\big \|\partial_{\xi_p} \tvp^\one_{ijk}(\tx,\tw) \big \| & \le \begin{cases}
{\lambda \over \vr T} & \text{if $\exists \ q \in \Lambda^\one_{ij}$ such that $q' = p$}\\
0 & \text{otherwise}
\end{cases}\\[6pt]
\big \|\partial_{\theta_p}\tvp^\one_{ijk}(\tx,\tw) \big \| & \le \begin{cases}
{\lambda \over \vr T^2} & \text{if $\exists \ q \in \Lambda^\one_{ij}$ such that $q' = p$}\\
0 & \text{otherwise}
\end{cases}\\[6pt]
\big \| \partial_{\theta_p} \partial _x \tvp^\one_{ijk}(\tx,\tw) \big \| & \le \begin{cases} 
{\lambda \over \vr T} & \text{if $\exists \ q \in \Lambda^\one_{ij}$ such that $q' = p$}\\
0 & \text{otherwise}\end{cases}
\end{align*}
\end{lemma}

\begin{proof} By Lemma A.2,
\begin{equation*}
\begin{split}
\big \|\partial_{\xi_p} \tvp^\one_{ijk}(\tx,\tw) \big \|
&\le \Big\| {-1\over 2\pi i} \oint_{|z-\te^\one_{ijk}(\tx,\tw)| =\vr} \big[z - \th_{\Lambda^\one_{ij}}(\tx,\tw) \bigr]^{-1} \partial_{\xi_p} \th_{\Lambda^\one_{ij}}(\tx,\tw) \bigl[z - \th_{\Lambda^\one_{ij}}(\tx,\tw) \bigr]^{-1} dz\Big\|\\
& \le \begin{cases}
{1\over 2\pi} (2\pi \vr) {\lambda /T\over \vr^2} = {\lambda \over \vr T} & \text{if $\exists \ q \in \Lambda^\one_{ij}$ such that $q' = p$}\\
0 & \text{otherwise}
\end{cases}
\end{split}
\end{equation*}
The proofs for the other two inequalities are similar.
\end{proof}

Fix $\eta = (\eta_1, \dots, \eta_T) \subset (-\delta, \delta)^T$.  The next lemma shows that for ``typical'' $\xi,\theta$, the eigenvalues are Morse functions.

\begin{prop}
For $\lambda^2 {|\Lambda^\one_{ij}|^2\delta\over \vr \ T} \ll \ve \ll 1$, one has
\begin{equation*}
 \mes \left\{(\xi, \theta) \in (-\delta, \delta)^{2|\Lambda^\one_{ij}|}:
\big |\partial_x \te^\one_{ijk}(\tx,\tw)\big | \le \ve, \big |\partial_{xx} \te^\one_{ijk}(\tx,\tw)\big | \le \ve\right\}
 \le \Bigr(|\Lambda^\one_{ij}|{\ve\over \lambda \delta}\Bigr)^2
\end{equation*}
\end{prop}

\begin{proof} Let $\tvp_{00}(x,\w) = \tvp^\one_{ijk}(x,\w) \Big |_{(\xi, \theta) 
= (0,0)}$, $\te_{00}(x,\w) = \te^\one_{ijk}(x,\w) \Big |_{(\xi, \theta) = (0,0)}$
\begin{equation*}
\begin{split}
\lambda^{-1} \partial_x \te^\one_{ijk}(\tx,\tw) & = \sum_{\ell \in \Lambda^\one_{ij}}\, \partial_x \tv(\tx + \ell \tw) \big |\tvp^\one_{ijk}(\tx,\tw) (\ell)\big |^2\\
& = \sum_{\ell\in \Lambda^\one_{ij}}\left[V'(\tx + \ell \tw) + \xi_{\ell'} + O\bigl({\delta\over T}\bigr) \right] \left[ \big |\tvp_{00}(\tx,\tw) (\ell)\big |^2 + O\bigl({|\Lambda^\one_{ij}|\delta\lambda\over \vr \ T}\bigr) \right]\\
& = \lambda^{-1} \partial_x \te_{00}(\tx,\tw) + \sum_{\ell \in \Lambda^\one_{ij}} \xi_{\ell'} \big |\tvp_{00}(\tx,\tw)(\ell)\big |^2 + O\bigl({|\Lambda^\one_{ij}|^2\delta\lambda\over \vr \ T}\bigr)
\end{split}
\end{equation*}

By Lemma F.1,
\begin{equation*}
\begin{split}
& (2\delta)^{-T} \mes \left\{\xi: \bigg |\lambda^{-1} \partial_x\te_{00} (\tx,\tw) + \sum_{\ell \in \Lambda^\one_{ij}}\, \xi_{\ell'} \big |\tvp_{00}(\tx,\tw)(\ell)\big |^2 \bigg | \le \lambda^{-1} \ve\right\}\\
& = (2\delta)^{-|\Lambda^\one_{ij}|} \mes \left\{\bigl\{\xi_{m'}\bigr\}_{m \in \Lambda^\one_{ij}}: \big |\lambda^{-1} \partial_x \te_{00}(\tx,\tw) + \sum_{\ell \in \Lambda^\one_{ij}} \, \xi_{\ell'} \big |\tvp_{00}(\tx,\tw)(\ell) \big |^2 \le \lambda^{-1} \ve\right\}\\
& \le |\Lambda^\one_{ij}| {\lambda^{-1} \ve\over \delta}\ .
\end{split}
\end{equation*}

This expression does not depend on $\theta$.  So
$$
\left\{(\xi, \theta): \big |\partial_x \te^\one_{ijk}(\tx,\tw) \big | \le \ve\right\} \subset \Xi \times (-\delta, \delta)^{|\Lambda^\one_{ij}|}
$$
where $(2\delta)^{-T}\mes \Xi \le |\Lambda^\one_{ij}| {\lambda^{-1} \ve\over \delta}$.

Fix $\xi_0 \in \Xi$.  Let $\tvp_0(x,\w) = \tvp^\one_{ijk}(x,\w) \Big |_{\theta = 0}$, $\te_0(x,\w) = \te^\one_{ijk}(x,\w) \Big |_{\theta = 0}$
\begin{equation*}
\begin{split}
\lambda^{-1} \partial_{xx} \te^\one_{ijk}(\tx,\tw) & = \sum_{\ell \in \Lambda^\one_{ij}}\, \partial_{xx} \tv (\tx + \ell \tw) \big |\tvp^\one_{ijk}(\tx,\tw)(\ell)\big |^2 + 2 \sum_{\ell \in \Lambda^\one_{ij}}\, \partial_x \tv(\tx + \ell \tw) \tvp^\one_{ijk}(\tx,\tw)(\ell)\partial_x \tvp_{ijk}(\tx,\tw)(\ell)\\
& = \sum_{\ell\in\Lambda^\one_{ij}} \left[V''(\tx + \ell \tw) + \theta_{\ell'} + O\bigl({1\over T}\bigr)\right]\left[\big |\tvp_0(\tx,\tw)(\ell)\big |^2 + O\bigl({|\Lambda^\one_{ij}|\delta\lambda\over \vr T^2}\bigr)\right]\\
&\quad + 2 \sum_{\ell\in \Lambda^\one_{ij}} \left[V'(\tx + \ell \tw) + \xi_{\ell'} + O\bigl({\delta\over T}\bigr)\right]\left[\tvp_0(\tx,\tw)(\ell) + 
O\bigl({|\Lambda^\one_{ij}|\delta\lambda\over \vr T^2}\bigr) \right] \left[\partial_x \tvp_0(\tx,\tw)(\ell) + O\bigl({|\Lambda^\one_{ij}|\delta\lambda\over \vr T}\bigr)\right]\\
& = \lambda^{-1} \partial_{xx} \te_0(\tx,\tw) + \sum_{\ell\in \Lambda^\one_{ij}}\, \theta_{\ell'}\big |\tvp_0(\tx,\tw)(\ell)\big |^2 + O\bigl({|\Lambda^\one_{ij}|^2\delta\lambda\over \vr T}\bigr)
\end{split}
\end{equation*}
Similar calculation shows that
$$
(2\delta)^{-T} \mes \left\{\theta: \big |\partial_{xx} \te^\one_{ijk} (\tx,\tw) \big | \le \ve\quad \text{for fix}\ \xi_0 \in \Xi\right\}
\le |\Lambda^\one_{ij}|{\lambda^{-1}\ve\over \delta}
$$
\end{proof}

\begin{corollary}
Given $\w\in V_j$ satisfying condition (6.1), one has
\begin{equation*}
\begin{split}
  (2\delta)^{-3T} \mes \biggl\{(\eta, \xi, \theta): \min\limits_{x \in\tor} \Bigl[\big|\partial_x \te^\one_{ijk}(x,\w)\big | + \big |\partial_{xx}\te^\one_{ijk}(x,\w)\big |\Bigr]
\le \ve\ \text{for some $i,k$}\biggr\} \lesssim m_0 { |\Lambda_{ij}|^2 \over \vr} {\ve \over \delta^2}
\end{split}
\end{equation*}
provided $\ve > exp(-N^A/2)$.
\end{corollary}

\begin{proof} Let $K\asymp\bigl({\vr\ve\over\lambda|\Lambda^{(1)}_{ij}|}\bigr)^{-1}\in\IN$, $x_m={m\over K}$.  Since 
$$
\big |\partial_{xx} \te^\one_{ijk}(x,\w) \big | 
\lesssim {\lambda\over \vr}, \qquad 
\big |\partial_{xxx}\te^\one_{ijk}(x,\w) \big | \lesssim {\lambda\over \vr^2},
$$
if $\big | \partial_x \te^\one_{ijk} (x_m,\w) \big | > 2\ve$ then 
$\big | \partial_x \te^\one_{ijk} (x,\w) \big | > \ve$ for all 
$x \in \bigl( x_m - {1\over 2T},\ x_m + {1\over 2T} \bigr) \cap U_i$; and if 
$\big |\partial_{xx}\te^\one_{ijk}(x_m,\w) \big | > 2\ve$ then 
$\big |\partial_{xx}\te^\one_{ijk}(x,\w) \big | > \ve$ for all 
$x\in \bigl(x_m - {1\over 2T}, \ x_m + {1\over 2T}\bigr) \cap U_i$.  
The assertion follows from Proposition 8.6.
\end{proof}

\begin{prop}  There is a set of variations $S_1$, $1-|S_1| \le e^{-{1\over 2}N^A}$ such that for any $W\in S_1$, the potential $\tilde V = V+ W$ has the following property:  There is a set $\O_1 = \O_1(\tilde V) \subset \tor$, $\mes(\tor \setminus \O_1)\le e^{-{1\over 2}N^A}$ so that for any $i,j,k$ one has
$$
|\partial_x \tilde E_{ijk}(x,\w)| + |\partial_{xx} \tilde E_{ijk}(x,\w)| > e^{-N^A} 
$$
for all $x\in U_i$, $\w \in \O_1 \cap V_j$.
\end{prop}

\begin{proof}
Applying Corollary 8.7 with $\ve = e^{-N^A}$, the assertion follows from Fubini's theorem. 
\end{proof}


\setcounter{section}{8}
\section{Inductive construction using elimination of multiple resonances}

In this section, we will construct inductively the eigenvalues $E^{(s+1)}$, with exponentially decaying eigenfunctions $\vp^so$, of $H_{\Lambda^so}$.  We will also obtain an upper estimate on how many eigenvalues in an interval containing $\lambda V(x)$ and use it to show the norm of the monodromy matrix $M_{\Lambda^so}$ is large.

Suppose for each $r$, $1\le r\le s$, we have constructed a set of variations $S_r$, $1-|S_r|\le e^{N_r^A/2}$ where $N_r \asymp N_{r-1}^\tau$, such that for any potential
\begin{equation} 
\tilde V = V + \sum\limits_{r=1}^{s} W^{(r)}, \quad W^{(r)}\in S_r 
\end{equation} 
there is a set $\O_s=\O_s(\tilde V)\subset\O_{s-1}\subset\tor$, $\mes(\O_{s-1}\setminus\O_s\le e^{-N^A_s/2}$, satisfying the following inductive hypothesis:

\ \newline
For each $i,j$ there is $\Lambda^{(s)}_{ij} = \Bigl[ 
- \bigl( N^{(s)}_{ij} \bigr)^2, \bigl( N^{(s)}_{ij} \bigr)^2 \Bigr]$ and 
$C^3$ functions $\tilde E^{(s)}_{ijk}$ and $\vp^{(s)}_{ijk}$ defined on $U_i\times V_j$,
$$ \th_{\Lambda^{(s)}_{ij}}(x,\w) \tvp^{(s)}_{\ijk}(x,\w) = \te^{(s)}_{\ijk}(x,\w) \tvp^{(s)}_{\ijk}(x,\w) $$ such that, for any $(x,\w)\in U_i\times\O_s\cap V_j$,

\begin{enumerate}

\item $\sp \th_{ \Lambda_{ij}^{(s)} }(x,\w) \cap \Bigl(\lambda \tilde V(x)-{1\over 4}\lambda^{1/2}, \lambda+{1\over 4}\lambda^{1/2}\Bigr)=\{ \tilde E^{(s)}_{ijk}(x,\w)\}_{k=1}^{K^{(s)}_{ij}}$, $\ K^{(s)}_{ij}\lesssim N^{(s)}_{ij}$;

\item If $E\in\bigl(\lambda V(x)-{1\over 4}\lambda^{1/2},\lambda V(x)+{1\over 4}\lambda^{1/2}\bigr)$, $\dist \left( E, \rsp H_{\Lambda^{(s)}_{ij}}(x,\w)\right) < e^{-N_s^\beta}$ then for any $N \approx \bigl(N^{(s)}_{ij}\bigr)^2$, one has 
$$
\log \Big | \bigl[ H_{[1,N]}(x,\w)-E\bigr]^{-1} (n,1) \Big | 
\le -{1\over 4}n\log\lambda 
$$ 
for $N^{(s)}_{ij} < n \le N$.  In particular, 
$\big |\tvp^{(s)}_{\ijk}(x,\w)(n)\big | \le \lambda^{-{1\over 2}(|n| - N^\one_{ij})}$ if $N^{(s)}_{ij} < |n| \le \bigl(N^{(s)}_{ij}\bigr)^2$;

\item $\big | \te^{(s)}_{\ijk}(x,\w) - \te^{(s)}_{ij\ell}(x,\w)\big | 
> e^{-|\Lambda^{(s)}_{ij}|^{3/4}}$ if $k \ne \ell$;

\item $\big |\partial_x\te^{(s)}_{ijk}(x,\w)\big | + \big|\partial_{xx}\te^{(s)}_{\ijk}(x,\w)\big | \ge e^{-N_r^A /2}$.

\end{enumerate}

The above conditions, for s=1, are proven in Section 6--8.

Fix $V^{(s)}:=\tilde V$ in the form of (9.1).   To simplify notations, we will write $E_{\ijk}(x, \w)$ for $\te_{\ijk}(x,\w)$ and 
$\vp_{\ijk}(x,\w)$ for $\tvp_{\ijk}(x,\w)$.  Let $N_{s+1}\asymp e^{N_s^\tau}$.

To begin the inductive procedure, we use results from Appendix E to eliminate a set of frequencies to avoid the situation where the distance between $\rsp H_{\Lambda^{(s)}{ij}}(x,\w)$ and $\rsp H_{\Lambda^{(s)}_{pj}}(x+n\w,\w)$ is small.  After that, we will be able to apply the Avalanche Principle.

\begin{lemma} There exists $B^\so \subset O_s$, $\mes B^\so \le e^{-{1\over 2}N_s^\beta}$, such that for each $U_i\times V_j$, there is $N^\so_{ij}$, $\log N^\so_{ij} \asymp \log N_{s+1}$, so that $\dist \left(E, \rsp H_{\Lambda^{(s)}_{pj}}(x + n\w, \w)\right) > {1\over 2} e^{-(N_s^\beta} =: {\ve\over 2}$ whenever $\big |E - E^{(s)}_{\ijk}(x,\w)\big | < {\ve\over 2}$ for any $x \in U_i$, $x+n\w \in U_p$, $\w \in V_j \cap \O^{(s)}\setminus B^\so$, for all $n\in \Bigl[-(N^\so_{ij})^2,-N^\so_{ij}\Bigr)\cup\Bigl(N^\so_{ij},(N^\so_{ij})^2\Bigr]$
\end{lemma}

\begin{proof} If $\big |E^{(s)}_{\ijk}(x,\w) - E^{(s)}_{pjr}(x+n\w, \w)\big | > \ve$ for all $p, r$, $N_{s+1} < |n| \le \bigl(N_{s+1}\bigr)^2$, $(x,w) \in U_i\times V_j\cap\O^{(s)}$ then choose $N^\so_{ij} = N_{s+1}$.  Otherwise, let
\begin{align*}
F_1(x,\w) & = E^{(s)}_{\ijk}(x,\w) - E^{(s)}_{pjr}(x+n_1 \w, \w)\\[6pt]
F_2(x,\w) & = E^{(s)}_{pjr}(x+n_1 \w, \w) - E^{(s)}_{ajc}(x+n_2 \w, \w)
\end{align*}
where $N_{s+1} \le |n_1| \le \bigl(N_{s+1}\bigr)^2$, $\bigl(N_{s+1}\bigr)^4 \le |n_2| \le \bigl(N_{s+1}\bigr)^8$.

Since $\big|\partial_x E^{(s)}(x,\w)\big | + \big |\partial_{xx}E^{(s)}(x,\w)\big | \ge e^{-N^A /2}$ by Theorem E.6,
$$
\mes B^\so_j := \mes \left\{ \w \in V_j\cap\O_s: \begin{array}{l}
\exists x \in U_i,\ x + n_1 \w \in U_k\ \text{such that}\\
\big |F_1(x,w)\big | \le \ve/2\ ,\ \big |F_2(x+n_1 \w, \w)\big | \le \ve/2\end{array}\right\} \le \ve^\vt |U_i|\, |V_j|
$$
Let $B^\so = \sum_j B^\so_j$,
\begin{equation*}
\begin{split}
\mes(O_s\setminus B^\so) & \le \sum_{i,j}\, \ve^\vt |U_i|\, |V_j|\\
& \le \lambda \ve^\vt \le \ve^{\vt/2}\ .
\end{split}
\end{equation*}
\end{proof}

Denote $\Lambda^\so_{ij} = \left[-\bigl(N^\so_{ij}\bigr)^2, \bigl(N^\so_{ij}\bigr)^2\right]$.

The following lemma and its corollary provide the hypotheis for Avalanche Principle.  They are similar to Lemma 5.2 and Corollory 5.3 in Part I.  (Note that because of Lemma 9.1, we already know that there is no resonance entries near the edge.)

\begin{lemma} Let $(x,\w)\in U_i\times V_j\cap\O_s\setminus B^\so$, $E\in \bigl(\lambda V(x)-{1\over 4}\lambda^{1/2},\lambda V(x)+{1\over 4}\lambda^{1/2}\bigr)$.  Suppose $\dist \left(E, \rsp H_{\Lambda^{(s)}_{ij}}(x,\w)\right) < {1\over 2} e^{-N_s^\beta}$.  Then $\dist \left(E, \rsp H_{[-a,b]}(x + n\w, \w)\right) > {1\over 2} e^{-N_s^\beta}$ where $N^\so_{ij} \lesssim |n| \le \bigl(N^\so_{ij}\bigr)^2$, $a, b \asymp N_s$.
\end{lemma}

\begin{proof} The assertion follows from condition (2) and Lemma C.9.
\end{proof}

\begin{corollary} Let $x, \w, E$ as in Lemma 9.2, $-\bigl(N^\so_{ij}\bigr)^2 \le c \le d < N^\so_{ij}$, $d-c > 10N_s$ then we can choose $c = a_0 < a_1 < \dots < a_K = d$, $a_k - a_{k-1} \asymp N_s$,
\begin{align*}
A_k & = M_{(a_{k-1}, a_k]}(x,\w,E)\quad k = 2, \dots, K-1\\[6pt]
A_1 & = M_{[a_0, a_1]}(x,\w,E)\begin{pmatrix}
1 & 0\\ 0 & 0\end{pmatrix}\\[6pt]
a_K & = \begin{pmatrix}
1 & 0\\ 0 & 0\end{pmatrix} M_{(a_{K-1}, a_K]}(x,\w,E)
\end{align*}
such that
$$
\log \big |f_{[c, d]}(x, \w, E)\big | = \sum^K_{k=1} \log \big \|A_k A_{k-1}\big \| - \sum^{K-1}_{k=2} \log \|A_k\| + O\left(N_{s+1} \lambda^{-N}\right)\ .
$$
\end{corollary}

\begin{proof} The hypothesis of Avalanche principle is verified by Lemmas 9.2 and C.10.
\end{proof}

Similar to Section 7, we need now to show that the spectrum of 
$H_{ \Lambda^\so_{ij} }(x,\w)$ near $E^{(s)}_{ijk}(x,\w)$ is separated.

\begin{lemma} If $E_1, E_2 \in \left(E^{(s)}_{\ijk} (x,\w) - {1\over 2} e^{-\bigl(N^{(s)}\bigr)^\beta}, E^{(s)}_{\ijk}(x,\w) + {1\over 2} e^{-\bigl(N^{(s)}\bigr)^\beta}\right)$ then for $-\bigl(N^\so_{ij}\bigr)^2 \le a < b < - N^\so_{ij}$, $|\Lambda^{(s)}_{pq}| < b - a \le 5|\Lambda^{(s)}_{pq}|$, one has
$$
\Big | \log \big| f_{[a,b]}(x,\w,E_1)\big | - \log \big |f_{[a,b]}(x, \w, E_2)\big |\, \Big | \lesssim  e^{{1\over 2}N_s^\beta} |E_1 - E_2|\ .
$$
\end{lemma}

\begin{proof} Let $J = \left\{j \in [a, b]:\ |E^{(s)}_{pj\ell}(x + j\w,\w) - E^{(s)}_{ijk}(x,\w)| < e^{-N_s^\beta}\right\}$.  Then $\#J \le N^\so_{ij}$.

Let $\{\mu_j\}^b_{j=a} = \rsp H_{[a, b]}(x,\w)$, $\ |\mu_j - \lambda V(x + j\w)| \le 2$
\begin{equation*}
\begin{split}
& \Big |\log \big |f_{[a, b]}(x, \w, E_1)\big | - \log \big |f_{[a,b]}(x,\w, E_2)\big |\, \Big |\\
& \le \sum_{j \in J} \big | \log |\mu_j - E_1| - \log |\mu_j - E_2|\, \big | + \sum_{j \notin J} \big |\log |\mu_j - E_1| - \log |\mu_j - E_2|\, \big |\\
& \le N^\so_{ij}\Bigl({1\over 2} e^{-\bigl(N^{(s)}\bigr)^\beta}\Bigr)^{-1} |E_1 - E_2| + (b-a) \bigl({1\over 2} \lambda^{1/2}\bigr)^{-1} |E_1 - E_2|\ .
\end{split}
\end{equation*}
\end{proof}

\begin{corollary} $E_1, E_2$ as in Lemma 9.4, then
\begin{equation*}
\begin{split}
& \Big | \log \big |f_{[-(N^\so_{ij})^2, n]} (x, \w, E_1)\big | - \log \big |f_{[-(N^\so_{ij})^2, n]}(x, \w, E_2)\big |\, \Big |\\
& \lesssim  e^{{1\over 2}N_s^\beta} |E_1 - E_2|
\end{split}
\end{equation*}
\end{corollary}

\begin{corollary} If $E_1, E_2 \in \left(E^{(s)}_{\ijk}(x, \w) - {1\over 2} e^{-N^\beta}, E^{(s)}_{\ijk}(x,\w) + {1\over 2} e^{-N_s^\beta}\right) \cap \rsp H_{\Lambda^\so_{ij}}(x,\w)$, $E_1 \ne E_2$, then $|E_1 - E_2| > e^{-(N^\so_{ij})^\vt}$.
\end{corollary}

\begin{proof} The proof is essentially the same as in Lemmas 7.2, 7.3, 7.4.
\end{proof}

Applying the same method as in Section 8, we can show the following lemma, which gives conditon (4).

\begin{lemma} There is a set of variations $S_{s+1}$, $1-|S_{s+1}|\le e^{-N_{s+1}^A/2}$, such that for any $W\in S_{s+1}$, the potential $\tilde V = V^{(s)} + W$ has the following property: There is a set $\O_{s+1}=\O_{s+1}(\tilde V)\subset\O_s$, $\mes (\O_s\setminus\O_{s+1})\lesssim e^{-N_{s+1}^A/2}$ so that for any $i,j,k$ one has $$
|\partial_x \tilde E^{(s+1)} (x,\w)|+|\partial_{xx} \tilde E^{(s+1)}(x,\w)|>e^{-N_{s+1}^A} $$
for all $x\in U_i$, $\w\in \O_{s+1}\cap V_j$.
\end{lemma}

For the first scale, we know that the number of eigenvalues close to $V(x)$ is bounded by $m_0 \ll N^\one_{ij}$ since there are at most $m_0$ resonance entries.  For the general inductive step, we will count the number of eignevalues of 
$ H_{ \Lambda^\so_{ij} }(x,\w)$ near $E^{(s)}_{ijk}(x,\w)$ in the following lemma.

\begin{lemma} If $\w \in \O_{s+1}$, then
$$
\#\left\{E \in \rsp H_{\Lambda^\so_{ij}}(x, \w): \dist\bigl(E, H_{\Lambda^{(s)}_{ij}}(x,\w)\bigr) < {1\over 2} e^{-N_s^\beta}\right\} \lesssim \big |\Lambda^\so_{ij}\big |^{1/2}
$$
\end{lemma}

\begin{proof} By Corollary 9.3, if $H_{\Lambda^\so_{ij}} \vp = E\vp$, $\|\vp\| =1$, $\dist\bigl(E, H_{\Lambda^{(s)}_{ij}}(x,\w) \bigr) < {1\over 2} e^{-N_s^\beta}$ then
$|\vp(n)| \lesssim e^{-|n| - N^\so_{ij}}$ if $|n| > N^\so_{ij}$.

Let $\cL = \bigoplus\limits_{\dist\bigl(E, H_{\Lambda^{(s)}_{ij}}(x,\w)\bigr) <
{1\over 2} e^{-N_s^\beta}} 
\ker \bigl(H_{\Lambda^\so_{ij}} - EI\bigr)$.  Then $|\vp(n)| \lesssim e^{-|n|- N^\so_{ij}}$ if $|n| > N^\two_{ij}$ for all $\vp \in \cL$, $\|\vp\| =1$.  Hence there is $\psi \in \IR^{[-2N^\so_{ij}, 2N^\so_{ij}]}$ such that $\vp \cdot \psi \ne 0$.

But by Lemma A.6, $\dim \cL \lesssim N^\so_{ij} \lesssim \big |\Lambda^\so_{ij}\big |^{1/2}$.
\end{proof}

\begin{lemma} If $w \in \O_{s+1}$, $e^{-(N_{s+1})^\gamma} < \dist\bigl(E, \rsp H_{\Lambda^\so_{ij}}(x,\w) \bigr)$ then $\log \big | f_{\Lambda^\so_{ij}}(x,\w) \big | > {|\Lambda^\so_{ij}|\over 4} \log \lambda$.
\end{lemma}

\begin{proof} If $\big |E - E^{(s)}_{\pjk}(x + n\w, \w)\big | \ge {1\over 2} e^{-N_s^\beta}$ for $n, p, k$ then the assertion follows immediately from the Avalanche Principle.

So suppose $\big |E - E^{(s)}_{p_0 j k_0}(x+ n_0 \w, \w)\big | < {1\over 2} e^{-N_s^\beta}$, $\w \in \O_{s+1}$.  Then $\big |E - E^{(s)}_{\pjk}\bigl(x + (n_0 + n) \w, \w\bigr)\big | \ge {1\over 2} e^{-(N^\one)^\beta}$, $N^\so_{pj} < |n| \le \bigl(N^\so_{pj}\bigr)^2$.

Take $\te = |E - \te| \lesssim e^{-N_s^\beta}$ such that $\big|\te - E^{(s)}_{\pjk}(x + n\w, \w)\big | \ge {1\over 4} e^{-N_s^\beta}$.  By Lemma 9.8 and Lemma B.5,
$$
\log \big |f_{\Lambda^\so_{ij}}(E)\big | > \log \big |f_{\Lambda^\so_{ij}}(\te)\big | - \big |\Lambda^\so_{ij}\big |^{1/2} \bigl(N^\so\bigr)^\gamma\ .
$$
\end{proof}


\setcounter{section}{9}
\section{Proof of Main Theorem}

We have defined $C^3$ functions $V^\one, V^\two,\dots$, 
$\big \|V^{(\ell+1)} - V^{(\ell)} \big \|\lesssim e^{-(N^{(\ell)})^\nu}$,
and \newline $\tor \supset \O^\two \supset \O^\three\supset \dots$,
$\mes\bigl(\O^{(\ell+1)} \setminus \O^{(\ell)}\bigr) \lesssim e^{-(N^{(\ell)})}$.  Let $\hv = \lim V^{(\ell)} \in C^3(\tor)$, $\O = \tcap \O^{(\ell)}$,
$$
\hh_{\Lambda_{ij}^{(\ell)}} (x,\w) = \begin{pmatrix}
\hv\bigl(x - |\Lambda_{ij}^{(\ell)}| \w\bigr) & -1 & \\
-1 & \hv\bigl(x - \bigl(|\Lambda_{ij}^{(\ell)}| -1)\w\bigr) & -1\\
& -1 & \ddots\\
& \ddots & -1\\
& -1 & \hv\bigl(x + |\Lambda_{ij}^{(\ell)}| \w\bigr)
\end{pmatrix}
$$
Then $\big \|\hh_{\Lambda_{ij}^{(\ell)}}(x,\w) - H_{\Lambda_{ij}^{(\ell)}}(x,\w) \big \| \lesssim e^{-(N^{(\ell)})^\nu}$.  

To establish Theorem 1.2, we need to show that the eigenvalues and eigenfunctions of $\hh_{\Lambda_{ij}^{(\ell)}} (x,\w)$ have the same properties shared by the eigenvalues and eigenfunctions of the potential $V^{(\ell)}$.  We do this by comparing the corresponding eigenvalues and eigenfunctions.

By Lemma B.4, we can define
$$
\he_{\ijk}^{(\ell)}(x,\w) \in \rsp \hh_{\Lambda_{ij}^{(\ell)}}(x,\w)\ ,\quad \w \in \O\ ,\ x \in U_{ij}
$$
such that $\big |\he_{\ijk}^{(\ell)} (x,\w) - E_{\ijk}^{(\ell)}(x,\w) \big | \lesssim e^{-(N^{(\ell)})^\nu}$.

\begin{lemma} $\dist \left(\he_{\ijk}^{(\ell)} (x,\w),\rsp \hh_{\Lambda_{ij}^{(\ell)}}(x,\w) \setminus \bigl\{\he_{\ijk}^{(\ell)}(x,\w)\bigr\} \right) > e^{-(N_{ij}^{(\ell)})^\nu}$.
\end{lemma}

\begin{proof}
This follows from Corollary 9.4, Lemma B.4 and the definition of $\he_{\ijk}^{(\ell)}(x,\w)$ above.
\end{proof}

\begin{lemma} There is $\big \|\hat\varphi_{\ijk}^{(\ell)}(x,\w) \big \| = 1$ such that 
\begin{equation*}
\begin{aligned}
& \hh_{\Lambda_{ij}^{(\ell)}}(x,\w) \hat\varphi_{\ijk}^{(\ell)}(x,\w) =  \he_{\ijk}^{(\ell)}(x,\w) \hat\varphi_{\ijk}^{(\ell)}(x,\w),\\
& \big \|\hat\varphi_{\ijk}^{(\ell)}(x,\w) - \varphi_{\ijk}^{(\ell)}(x,\w)\big \| \lesssim e^{-(N^{(\ell)})^\nu}
\end{aligned}
\end{equation*}
\end{lemma}

\begin{proof} 
\begin{align*}
\big \|\bigl[\hh_{\Lambda_{ij}^{(\ell)}}(x,\w)- \he_{\ijk}^{(\ell)}(x,\w)\bigr] \varphi_{\ijk}^{(\ell)}(x,\w)\big \|
& \le \big \|\hh_{\Lambda_{ij}^{(\ell)}}(x,\w)- H_{\Lambda_{ij}^{(\ell)}}(x,\w)\big \| + \big \| \bigl[H_{\Lambda_{ij}^{(\ell)}}(x,\w) -
E_{\ijk}^{(\ell)}(x,\w)\bigr] \varphi_{\ijk}^{(\ell)}(x,\w)\big \|\\
& \quad + \big | E_{\ijk}^{(\ell)}(x,\w) - \he_{\ijk}^{(\ell)}(x,\w) \big | \\
&  \lesssim e^{-(N^{(\ell)})^\nu}
\end{align*}
The assertions follows from Lemma A.5 and Lemma 10.1.
\end{proof}

\begin{corollary} $w \in \O$.  If $\big | E - \he_{\ijk}^{(\ell)} (x,\w)\big | < {1\over 2}  e^{-(N^{(\ell)})^\beta}$ then
$$
\dist \left(E, \rsp \hh_{\Lambda_{pj}^{(\ell)}}(x + n\w, \w)\right) > {1\over 2}  e^{-(N^{(\ell)})^\beta}
$$
for all $N^{(\ell+1)} < |n| \le \bigl(N^{(\ell+1)}\bigr)^2$.  Also,
$$
\# \left(E \in \rsp \hh_{\Lambda_{ij}^{(\ell+1)}}(x,\w): \dist 
\left(E, \rsp \hh_{\Lambda_{ij}^{(\ell)}}(x,\w)\right) < {1\over 2} e^{-(N^{(\ell)})^\beta}\right) \lesssim \big |\Lambda_{ij}^{(\ell+1)}\big|^{1/2}
$$
\end{corollary}

\begin{proof} Same as in Lemma 9.1 and Lemma 9.6.
\end{proof}

Let $f_{\Lambda_{ij}^{(\ell)}}(x,\w,E) = \det \bigl(\he_{\Lambda_{ij}^{(\ell)}}(x,\w) - E\bigr)$.

\begin{lemma} If $\w \in \O \cap V^{(\ell)}_j$, $x \in U_{i}$
$$
\dist\left(E, \rsp H_{\Lambda_{ij}^{(\ell)}}(x , \w)\right) > {1\over 2}  e^{-(N^{(\ell)})^\gamma}
$$
Then $\log \big | \hat f_{\Lambda_{ij}^{(\ell)}}(x,\w,E)\big | > {|\Lambda_{ij}^{(\ell)}\over 4} \log \lambda$.
\end{lemma}

\begin{proof} For $\ell =1$, let $\bigl\{\mu_k\bigr\} = \rsp H_{\Lambda_{ij}^{(2)}}(x,\w)$, $\big | \lambda\hv (x + k\w) - \mu_k\big | \le 2$.

Let $K = \left\{k: \big |E - \mu_k\big | \le \lambda^{1/2}\right\}$.  If $k \in K$, then
\begin{align*}
\big |V(x + k\w) - \lambda^{-1}E\big | & \le \big | V(x + k\w) - \hv(x + k\w)\big | + \big |\hv(x + k\w) - \lambda^{-1} E\big |\\
& \lesssim \lambda^{-1/2}\ .
\end{align*}
By Lemma 6.2, $\#K \le m_0$.  Hence
\begin{equation*}
\begin{aligned}
\log \big |\hat f_{\Lambda_{ij}^{(1)}} (x,\w, E)\big | & >
\bigl[\Lambda^\one_{ij}(x,\w) - m_0\bigr] \log \lambda^{1/2} - m_0\bigl(N^\one\bigr)^\gamma\\
& > {|\Lambda_{ij}^\one|\over 4} \log \lambda\ .
\end{aligned}
\end{equation*}
For $\ell > 1$, there is $|\te - E| \lesssim e^{-(N^{(\ell-1)})^\beta}$ such that $\dist \left(\te, \rsp H_{\Lambda_{pj}^{(\ell-1)}}(x + n\w)\right) > e^{-N^{(\ell-1)})^\beta}$. By the Avalanche Principle
$$
\log \big |\hat f_{\Lambda_{ij}^{(\ell)}}(x, \w, \te)\big | > {\Lambda_{ij}^{(\ell)}|\over 4} \log \lambda\ .
$$
In view of Corollary 10.3, we get
$$
\log \big |\hat f_{\Lambda_{ij}^{(\ell)}}(x, \w, E)\big| > {|\Lambda_{ij}^{(\ell)}|\over 4} \log \lambda
$$
(see Lemma 9.7).
\end{proof}

\begin{lemma} $\big|\partial_x \he_{\ijk}^{(\ell)}(x, \w) - \partial_x E_{\ijk}^{(\ell)}(x,\w) \big | \lesssim e^{-(N^{(\ell)})^\nu}$.
\end{lemma}

\begin{proof}
\begin{equation*}
\begin{aligned}
\big|\partial_x \he_{\ijk}^{(\ell)}(x, \w) - \partial_x E_{\ijk}^{(\ell)}(x,\w) \big |
& \le \lambda \sum_{n \in \Lambda_{ij}^{(\ell)}} \big |\hv'(x+n\w)\big |\, \big |\hat\varphi_{\ijk}^{(\ell)} \xw(n)\big |^2 - V'(x + n\w)  \big |\varphi_{\ijk}^{(\ell)} \xw(n)\big |^2\\
&\quad \le \lambda \sum_n \hv'(x + n\w) - V^{\two'}(x + n\w) \big |\, \big|\hat\varphi_{\ijk}^{(\ell)}\xw (n)\big |^2\\
&\qquad + \lambda\sum_n \big | V^\two (x+n\w)\big |\, \big |\hat\varphi_{\ijk}^{(\ell)}\xw (n)\big |^2 - \big |\varphi_{\ijk}^{(\ell)}\xw (n)\big |^2\\
&\quad \lesssim e^{-(N^{(\ell)})^2}
\end{aligned}
\end{equation*}
\end{proof}

\begin{lemma} $\big \|\partial_x \hat\varphi_{\ijk}^{(\ell)}\xw - \partial_x \varphi_{\ijk}^{(\ell)}\xw \big \| \lesssim e^{-{1\over 2}N_\ell^\nu}$.
\end{lemma}

\begin{proof}
$$
\partial_x \varphi_{\ijk}^{(\ell)}\xw =\left[
\oint_{{|z - E^{(\ell)}_{\ijk}|=\varrho}} \bigl(H_{\Lambda_{ij}^{(\ell)}}\xw - z\bigr)^{-1}
\bigl[\partial_x H_{\Lambda_{ij}^{(\ell)}}\xw\bigr]
\bigl(H_{\Lambda_{ij}^{(\ell)}}- z\bigr)^{-1} dz \right]
\varphi_{\ijk}^{(\ell)}\xw
$$
where $\varrho = {1\over 2} e^{-(N^{(\ell)})^\nu}$.  Hence 
\begin{equation*}
\begin{aligned}
\big \|\partial_x \hat\varphi_{\ijk}^{(\ell)}\xw - \partial_x \varphi_{\ijk}^{(\ell)}\xw \big \|
& \le \oint \big \|\bigl(\hh - z\bigr)^{-1} \bigl(\partial_x \hh\bigr)\bigl(\hh - z\bigr)^{-1} - \bigl(H - z\bigr)^{-1}\bigl(\partial_x H\bigr)\bigl(H - z\bigr)^{-1}\big \| dz\\
&\qquad + \Bigl[\oint \big \|\bigl(H - z\bigr)^{-1} \bigl(\partial_x H\bigr) \bigl(H - z\bigr)^{-1} \big \| dz\Bigr]
\big \| \hat\varphi_{\ijk}^{(\ell)}\xw -  \varphi_{\ijk}^{(\ell)}\xw\big \|
\end{aligned}
\end{equation*}
\begin{equation*}
\begin{aligned}
& \big \|(\hh - z\bigr)^{-1} \bigl(\partial_x \hh\bigr)        \bigl(\hh - z\bigr)^{-1} - \bigl(H - z\bigr)^{-1}\bigl(\partial_x H\        bigr)\bigl(H - z\bigr)^{-1}\big \|\\
&\quad \le \big \|\bigl(\hh - z\bigr)^{-1} - \bigl(H - z\bigr)^{-1}\big \|\, \big \|\bigl(\partial_x \hh\bigr)\bigl(\hh - z\bigr)^{-1}\big \|\\
&\qquad + \big \|\bigl(H - z\bigr)^{-1}\big \|\, \big \|\partial_x H - \partial_x \hh \big \|\, \big \|\bigl(\hh - z\bigr)^{-1}\big \|\\
&\qquad + \big \|\bigl(H - z\bigr)^{-1}\big \|\, \big \|\partial_x H\big \|\, \big \|\bigl(\hh - z\bigr)^{-1} - \bigl(H - z\bigr)^{-1}\big \|\\
&\quad \lesssim e^{-{1\over 2}(N^{(\ell)})^\nu}
\end{aligned}
\end{equation*}
since $\bigl(\hh - z\bigr)^{-1} - \bigl(H - z\bigr)^{-1} = \bigl(\hh - z\bigr)^{-1} \left[\bigl(H - z\bigr) - \bigl(\hh - z\bigr)\right]\bigl(H - z\bigr)^{-1}$ and $\big \|\bigl(\hh - z\big)^{-1} \big \| \lesssim e^{N_\ell^\nu}$, $\big \|\bigl(H - z\bigr)^{-1}\big \| \lesssim e^{N_\ell^\nu}$.
\end{proof}

\begin{lemma} $\big |\partial_{xx}\he_{\ijk}^{(\ell)}\xw - \partial_{xx} E_{\ijk}^{(\ell)} \xw \big | \lesssim e^{-{1\over 2}(N^{(\ell)})^\nu}$.
\end{lemma}

\begin{proof}
\begin{equation*}
\begin{aligned}
\big |\partial_{xx} \he_{\ijk}^{(\ell)}\xw - \partial_{xx} E_{\ijk}^{(\ell)}\xw\big \|
& \le \lambda \sum_{n \in \Lambda^{(\ell)}_{ij}} \Big | \hv''(x + n\w) \big |\hat\varphi_{\ijk}^{(\ell)} \xw (n) \big |^2 - V^{(\ell)''}(x + n\w) \big | \varphi_{\ijk}^{(\ell)}\xw (n)\big |^2 \Big |\\
&\qquad + 2 \lambda \sum_{n \in \Lambda_{ij}^{(\ell)}} \Big |\hv'(x + n\w) \hat\varphi_{\ijk}^{(\ell)} \xw \partial_x \hat\varphi_{\ijk}^{(\ell)} - V^{(\ell)'}(x + n\w) \varphi_{\ijk}^{(\ell)}\xw \partial_x \varphi_{\ijk}^{(\ell)}\Big |\\
&\quad \lesssim e^{-{1\over 2}(N^{(\ell)})^\nu}
\end{aligned}
\end{equation*}
\end{proof}

\begin{corollary} $\big |\partial_x \he_{\ijk}^{(\ell)} \xw \big | +
\partial_{xx} \he_{\ijk}^{(\ell)} \xw \big | \ge e^{-{1\over 2}N_\ell^\nu}$.
\end{corollary}

\begin{proof} Since $\big |\partial_x E_{\ijk}^{(\ell)} \xw \big | +
\big |\partial_{xx} E_{\ijk}^{(\ell)} \xw \big | \ge e^{-(N^{(\ell)})^\nu}$, the assertion follows from Lemma 10.5 and Lemma 10.7.
\end{proof}


\begin{theorem} Given any $V \in C^3(\tor)$, $|V'(x)|+|V''(x)|\ge c>0$, there is $\lambda_0=\lambda_0(V)$ such that for $|\lambda| > \lambda_0$, one has a collection of pertubed potentials $\{ S_\ell=S_\ell(V,\lambda) \}_{\ell=1}^\infty$, $S_\ell \subset \cS(T^{(\ell)},\delta_\ell)$, $\log T^{(\ell+1)} \asymp \bigl( T^{(\ell)} \bigr)^\alpha$, $0< \alpha \ll 1$, $\sum\limits_{\ell=1}^{\infty} \bigl(1-|S_\ell|\bigr) \le \lambda^{-\beta}$, so that for any potential 
$$ \tv(x) = V(x) + \sum\limits_{\ell=1}^\infty 
 W^{(\ell)}\bigl(\eta^{(\ell)}, \xi^{(\ell)}, \theta^{(\ell)}, \{ R^{(\ell)}_m \} ; x \bigr) $$ 
where $W^{(\ell)} \in S_\ell$, there exists $\Omega=\Omega(\lambda,\tilde V)$, $\mes \Omega \le \lambda^{-\beta}$, so that the Lyapunov exponent $L(\w,E) \ge {1 \over 4} \log \lambda$ for any $\w \in \Omega,\ E\in \IR$.
\end{theorem}

\begin{proof} Fix $E, \w$.  For each $\ell$, from Corollary 11.8, we have
$$
\mes \left\{ x \in \tor: \dist\bigl(E, \rsp H_{\Lambda_{ij}^{(\ell)}} \xw\bigr) \le e^{-(N^{(\ell)})^r}\right\} \le e^{-c(N^{(\ell)})^\nu}\ .
$$

Let $\cJ = \bigcup\limits^\infty_{\ell =1} \left\{x \in \tor: \dist\bigl(E, \rsp H_{\Lambda_{ij}^{(\ell)}} \xw \bigr) \le e^{-(N^{(\ell)})^r} \right\}$.  If $x \in \tor \setminus \cJ$ then
\begin{align*}
{1\over |\Lambda_{ij}|} \log \big \|M_{\Lambda_{ij}}(x, \w, E) \big \| & \ge
{1\over |\Lambda_{ij}|}\log \big |f_{\Lambda_{ij}}(x, \w, E) \big |\\
& \ge {1\over 4} \log \lambda\ .
\end{align*}

From above, $\mes \tor \setminus \cJ > 0$.  But ${1\over |\Lambda_{ij}|} \log \big \| M_{ij}(x, \w, E) \big \| \to L(\w,E)$ for almost all $x$ since $\w \in \IR \setminus \IQ$.  Therefore $L(\w,E) \ge {1\over 4} \log \lambda$.
\end{proof}


\appendix
\section{Matrix Functions}

Let $A: (a,b) \to M_{m\times m}(\IC)$ be $C^1$, $A(x)$ self-adjoint for every $x \in \ab$.  Suppose $E: \ab \to \IR$ is such that $E(x)$ is a simple eigenvalue of $A(x)$, and
$$
\rsp A(x) \cap \left[E_0 - {3\delta\over 2},\ E_0 + {3\delta\over 2}\right] = \{E(x)\} \subset \left(E_0 - {\delta\over 2},\ E_0 + {\delta\over 2}\right)\ ,\quad \delta > 0\ .
$$
Let $P(x)$ be the orthogonal projection on the eigenspace of $A(x)$ with eigenvalue $E(x)$.  By Riesz formula for orthogonal projection of self-adjoint matrix
\begin{align*}
P(x)  &= {1\over 2\pi i} \oint_{|z - E_0| = \delta} \bigl(zI - A(x)\bigr)^{-1} dz\\[6pt]
P'(x) & = {1\over 2\pi i } \oint_{|z - E_0| = \delta} \bigl(zI - A(x)\bigr)^{-1}A'(x) \bigl(zI - A(x)\bigr)^{-1} dz
\end{align*}
Hence $P:\ab \to M_{m\times m}(\IC)$ is $C^1$
$$
\|P'(x)\| \le {1\over 2\pi}\bigl({\delta\over 2}\bigr)^{-1} \|A'(x)\| \bigl({\delta\over 2}\bigr)^{-1} (2\pi \delta) = {1\over 4\delta}\|A'(x)\|\ .
$$

\begin{lemma}\label{lem:A-1}
$A: \ab \to M_{m\times m}(\IC)$, $E: \ab \to \IR$ as above.  Then $E$ is $C^1$ and \newline $E'(x) = \bigl(A'(x) \vp(x), \vp(x)\bigr)$ where $\vp(x)$ is a normalized eigenvector of $A(x)$ with eigenvalue $E(x)$.

\end{lemma}

\begin{proof} For any $x_0 \in \ab$, choose $\vp_0 \in \ran P(x_0)$, $\|\vp_0\| = 1$.  Then $\bigl(P(x_0)\vp_0, \vp_0\bigr) = 1$.  There exists neighborhood $U$ of $x_0$ such that $\bigl(P(x)\vp_0, \vp_0\bigr) > 0$ for every $x \in U$.
$$
E(x) = {\bigl(A(x) P(x) \vp_0, \vp_0\bigr)\over \bigl(P(x) \vp_0, \vp_0\bigr)}\quad \text{for $x \in U$}
$$
$E$ is $C^1$ since $A$ and $P$ are $C^1$.
\begin{equation*}
\begin{split}
E'(x_0) & = {\bigl(A'(x_0) P(x_0)\vp_0, \vp_0\bigr) + \bigl(A(x_0) P'(x_0)\vp_0, \vp_0\bigr)\over \bigl(P(x_0)\vp_0, \vp_0\bigr)}
 - \bigl(A(x_0) P(x_0)\vp_0, \vp_0\bigr)\ {\bigl(P'(x_0) \vp_0, \vp_0\bigr)\over \bigl(P(x_0)\vp_0, \vp_0\bigr)^2}\\[6pt]
& = \pamap + \bigl(P'(x_0)\vp_0, A(x_0)\vp_0\bigr) - \amap \ppmap\\[6pt]
& = \pamap + E(x_0) \ppmap - E(x_0)\ppmap\\[6pt]
& = \pamap\ .
\end{split}
\end{equation*}
\end{proof}

\begin{lemma}\label{lem:A-2}
Let $x_0,  \vp_0, U$ be as in proof of Lemma A.1.  Define $\vp(x) = \bigl(P(x) \vp_0, \vp_0\bigr)^{-1/2}\bigl(P(x) \vp_0\bigr)$ for $x \in U$.  Then $\vp(x)$ is a normalized eigenvector of $A(x)$ with eigenvalue $E(x)$.  
$\vp: U \to \mathbb{R}^n$ is $C^1$ and $\vp'(x_0) = P'(x_0)\vp_0$.
\end{lemma}

\begin{proof} Since $P(x)$ is an orthogonal projection, 
$$
\bigl(P(x) \vp_0, \vp_0\bigr) = \bigl(P(x)^2 \vp_0, \vp_0 \bigr) 
= \bigl(P(x)\vp_0, P(x)\vp_0 \bigr) = \|P(x_0)\vp_0\|^2\ .
$$  
Therefore, $\|\vp(x)\| = 1$.

Since $P$ is $C^1$, $\vp$ is $C^1$,
\begin{align*}
\vp'(x_0) & = - {1\over 2}\pmap^{-{3\over 2}} \ppmap \bigl(P(x_0)\vp_0\bigr)
 + \pmap^{-{1\over 2}} \bigl(P'(x_0)\vp_0\bigr)\\[6pt]
& = - {1\over 2}\ppmap \vp_0 + P'(x_0)\vp\ .
\end{align*}
It remains to show $\ppmap = 0$.
\begin{align*}
P(x) & = P(x)^2\\
P'(x_0) & = P'(x_0) P(x_0) + P(x_0) P'(x_0) \\
\ppmap & = \bigl(P'(x_0) P(x_0)\vp_0, \vp_0\bigr) + \bigl(P(x_0) P'(x_0)\vp_0, \vp_0\bigr)\\
\ppmap & = 2 \ppmap\ .
\end{align*}
Hence $\ppmap = 0$.
\end{proof}

\begin{lemma}\label{lem:A-3}
$A: \ab \to M_{m\times m}(\IC)$, $E: \ab \to \IR$ as above.  If $A$ is $C^2$, then $E$ is $C^2$ and 
$$
|E''(x)| \le \|A''(x)\| + {1\over 2\delta} \|A'(x)\|^2\ .
$$
\end{lemma}

\begin{proof} For any $x_0 \in \ab$, define $\vp(x)$ in a neighborhood $U$ of $x_0$ as in Lemma A.2, $\|\vp(x)\| = 1$, 
\begin{alignat*}{2}
A(x)\vp(x)  &= E(x)\vp(x) &\quad\text{for every} & x \in U\\
E'(x) & = \bigl(A'(x)\vp(x), \vp(x)\bigr) &\text{for every} & x \in U\ .
\end{alignat*}
Since $\vp(x)$ is $C^1$, if $A$ is $C^2$ then $E'$ is $C^1$.  So $E$ is $C^2$.
\begin{align*}
E''(x_0) & = \bigl(A''(x_0)\vp(x_0), \vp(x_0)\bigr) + \bigl(A'(x_0)P'(x_0)\vp(x_0), \vp(x_0)\bigr)
 + \bigl(A'(x_0)\vp(x_0), P'(x_0)\vp(x_0)\bigr)\\
& = \bigl(A''(x_0)\vp(x_0), \vp(x_0)\bigr) + 2\bigl(A'(x_0) P'(x_0) \vp(x_0),\vp(x_0)\bigr)\\[6pt]
|E''(x_0)| & \le \|A''(x_0)\| + 2\|A'(x_0)\|\, \|P'(x_0)\|\\
& \le \|A''(x_0)\| + {1\over 2\delta} \|A'(x_0)\|^2
\end{align*}

Let $V: \IR \to \IR$ be continuous, piecewise smooth, $1$--periodic.  Define
$$
H_{[m, n]}(x,w) = \begin{pmatrix}
V(x+mw) & -1 &&\\[6pt]
-1 & V\bigl(x+ (m+1)w\bigr) & -1 &\\[6pt]
& -1 &&\ddots\\[6pt]
&\ddots && -1\\[6pt]
&& -1 & V(x + nw)\end{pmatrix}
$$
Then $H_{[m,n]}(x,w)$ is continuous; $H_{[m, n]}(., w)$ and $H_{[m, n]}(x, .)$ are piecewise smooth.  It is well-known that the eigenvalues of $H_{[m, n]}(x, w)$ are simple.  So the eigenvalues $E^{(m)}_{[m, n]}(x,w) < \dots < E^{(n)}_{[m, n]}(x,w)$ of $H_{[m, n]}(x, w)$ are continuous.

Let $\left\{\vp^{(\ell)}_{[m, n]}(x,\w)(j)\right\}_{m\le j \le n}$ be a normalized eigenvector of $H_{[m,n]}(x,\w)$ with eigenvalue  
$E^{(\ell)}_{[m, n]}(x,\w)$.  Suppose $\partial_x H_{[m, n]}(x_0, \w_0)$ 
exists.  For fixed $k$, let
\begin{align*}
E_0 & = E^{(k)}_{[m, n]}(x_0, \w_0),\\
\delta & = {1\over 2} \dist\Bigl(E^{(k)}_{[m, n]}(x_0, \w_0), \bigl\{E^{(\ell)}_{[m, n]}(x_0, \w_0)\bigr\}_{\ell \ne k}\Bigr) > 0\ .
\end{align*}
Since $E^{(\ell)}_{[m, n]}$ are continuous, there exists neighborhood $U$ of $x_0$ such that for every $x \in U$
$$
\rsp H_{[m, n]}(x, \w_0) \cap \Bigl[E_0 - {3\delta\over 2}, E_0 + {3\delta\over 2}\Bigr] = \bigl\{E^{(k)}_{[m, n]}(x, \w_0)\bigr\} \subset \Bigl(E_0 - {\delta\over 2}, E_0 + {\delta\over 2}\Bigr)\ .
$$
By Lemma A.1,
$$
\partial_x E^{(k)}_{[m, n]}(x_0, \w_0) = \sum^n_{j=m}\, V'(x_0 + j\w_0) \big |\vp^{(k)}_{[m, n]}(x_0, \w_0)(j)\big |^2\ .
$$
Similarly,
$$
\partial_\w E^{(k)}_{[m, n]}(x_0, \w_0) = \sum^n_{j = m}\, jV'(x_0 + j\w_0) \big |\vp^{(k)}_{[m, n]}(x_0, \w_0)(j)\big |^2\ .
$$

\end{proof}

\begin{lemma}\label{lem:A-4}
If $A$ is $n\times n$, self-adjoint, and $\|A\vp\| < \ve$ for some $\vp$, $\|\vp\| = 1$.  Then there exists $\lambda \in \rsp A \cap (-\ve , \ve )$.
\end{lemma}

\begin{proof} If $A$ is not invertible, then take $\lambda = 0$.  If $A^{-1}$ exists, then $1 = \|\vp\| = \|A^{-1}A\vp\| < \|A^{-1}\|\ve$.  Let $\lambda_1, \dots, \lambda_n$ be the eigenvalues of $A$,
$$
{1\over \ve} < \|A^{-1}\| = {1\over \min |\lambda_i|} \Longrightarrow \min |\lambda_i | < \ve\ .
$$
\end{proof}

\begin{lemma}\label{lem:A-5} Let $A$ be $n\times n$, self-adjoint.  Suppose $\|A\vp \| < \ve$, $A\psi = 0$, $\|\vp\| = \|\psi\| =1$.  If $|\mu| > \delta$ for every $\mu \in \rsp A \setminus \{0\}$ then $\min\limits_{|c| =1} \|\vp - c\psi\| \lesssim \ve/\delta$ provided 0 is simple eigenvalue.
\end{lemma}

\begin{proof} Let $\psi_\mu$ be normalized eigenvectors of $A$ with eigenvalues $\mu \ne 0$,
\begin{align*}
\ve^2 > \|A\vp\|^2 & = \lambda^2 |(\vp, \psi)|^2 + \sum_{\mu \ne 0}\, \mu^2 |(\vp, \psi_\mu)|^2 > \delta^2 \sum_{\mu \ne 0} |(\vp, \psi_\mu)|^2\\[6pt]
\|\vp - c\psi\|^2 & = \|\vp\|^2 - \bar c(\vp, \psi) - c(\psi, \vp) + |c|^2\|\psi\|^2\\
& = 2\left[1 - \ree \bar c(\vp, \psi)\right]\qquad [|c| = 1]\\[6pt]
\min_{|c| =1} \|\vp - c\psi\|^2 & = 2\left[1 - |(\vp, \psi)|\right]\\
& \le 2\left[1 - |(\vp, \psi)|^2\right]\qquad [|(\psi, \vp)| \le 1]\\
& = 2\|\vp - (\vp, \psi)\psi\|^2\\
& = 2 \sum_{\mu \ne 0} |(\vp, \psi_\mu)|^2\\
& < 2\ve^2/\delta^2
\end{align*}
\end{proof}

\begin{lemma} Let $\cL, \cM$ be subspaces of $\IR^n$.  If $\dim \cL > \dim \cM$
, then there is $\psi \in \cL$, $\| \psi \| = 1$, such that $(\psi,\vp)=0$ for 
all $\vp \in \cM$.
\end{lemma}

\begin{proof}  Suppose, for contradiction, that $\cL \cap \cM^\bot = \{ 0 \}$. 
Since $\IR = \cM \oplus \cM^\bot$, we have $\cL \subset \cM$.  But this is 
impossible because $\dim \cL > \dim \cM$.  Hence, there is 
$\psi \in \cL \cap \cM^\bot$, $\| \psi \| = 1$.
\end{proof}


\appendix
\setcounter{section}{1}
\section{Mini-max principle}

\noindent
Mini-max Principle: Let $A$ be $n\times n$ hermitian matrix, with eigenvalues $\lambda_1 \le \lambda_2 \le \dots \le \lambda_n$.  Then
\begin{align*}
\lambda_j & = \min_{\dim M = j}\ \max_{\substack{x \in M\\ \|x\| =1}} (Ax, x)\\[6pt]
& = \max_{\dim M = n-j +1}\ \min_{\substack{x \in M\\ \|x\| =1}} (Ax, x)\ .
\end{align*}

\begin{lemma}\label{lem:B-1}
Let $A, B$ be $n\times n$ hermitian matrices, with eigenvalues $\lambda_1 \le \dots \lambda_n$ and $\mu_1 \le \dots \le \mu_n$ respectively.  Suppose there is $\alpha > 0$, $y \in \IR^n$ such that $Ax = Bx + \alpha(x, y) y$ for all $x \in \IR^n$.  Then the eigenvalues of $A$ and $B$ interlace, i.e. $\mu_1 \le \lambda_1 \le \mu_2 \le \lambda_2 \le \dots \le \mu_n \le \lambda_n$.
\end{lemma}

\begin{proof}
\begin{align*}
\mu_j & = \min_{\substack{x \in M_j\\ \|x\| =1}}\ (Bx, x)\qquad \text{for some $M_j$, $\dim M_j = n - j +1$}\\[6pt]
\lambda_j & = \max_{\dim M = n-j+1}\, \min_{\substack{x \in M\\ \|x\| =1}} (Ax, x)\\
& \ge \min_{\substack{x \in M_j\\ \|x\| =1}}(Ax, x) = \min_{\substack{x \in \tm_j\\ \|x\| =1}} \left[(Bx, x) + \alpha |(x, y)|^2\right] \ge \mu_j\\[6pt]
\end{align*}
On the other hand, \begin{align*}
\lambda_j & = \min_{\substack{x \in M_j\\ \|x\| =1}}(Ax, x)\qquad \text{for some $\tm_j,\ \dim \tm_j = n-j +1$}
\end{align*}
Let $U_j \subset \tm_j \cap (\IR y)^\bot$, $\dim U_j = n - j$ for $j < n$.
\begin{align*}
\mu_{j+1} & = \max_{\dim M = n-j}\ \min_{\substack{x \in M\\ \|x\| =1}} (Bx, x)\\
& \ge \min_{\substack{x \in U_j\\ \|x\| =1}} (Bx, x) = \min_{\substack{x \in U_j\\ \|x\| =1}}(Ax, x) \ge \lambda_j\ .
\end{align*}
\end{proof}

\begin{corollary}\label{cor:B-2}
Let $A, B$ be hermitian matrices, with eigenvalues $\lambda_1 \le \dots \le \lambda_n$ and $\mu_1 \le \dots \le \mu_n$ respectively.  Suppose ${\rm{rank}}(A - B) = k$.  Then
\begin{align*}
\mu_j  &\le \lambda_{j+k}\\
\lambda_j  & \le \mu_{j+k}\ .
\end{align*}
\end{corollary}

\begin{lemma}\label{lem:B-3}
Let $A, B$ be $n\times n$ hermitian matrices, $\rank(A-B) = k > 0$, $B \ne 0$.  Then
$$
\log |\det B| - \log |\det A| \le 2k \log \|B\| - 2k \log \dist(\rsp A, 0)\ .
$$
\end{lemma}

\begin{proof} $\det A = 0 \Longleftrightarrow \dist(\rsp A, 0) = 0$.  In this case, the inequality is trivial.  So assume $\det A \ne 0$.

Let $\lmap$ and $\mmap$ be the eigenvalues of $A$ and $B$ respectively.  Suppose $\mu \le 0 < \mu_{i+1}$.  By Lemma B.2,
$\lambda_{j-k}  \le \mu_j < 0$ for $k < j \le i\ $; $\ 0 < \mu_\ell \le \lambda_{\ell +k}$ for $i < \ell \le n-k$.
\begin{equation*}
\begin{split}
\log |\det B| - \log |\det A| & = \sum_{k < j \le i}\bigl(\log |\mu_{j}| - \log |\lambda_{j-k}|\bigr) + \sum_{i < j \le n-k} \bigl(\log |\mu_{j}| 
- \log |\lambda_{j+k}|\bigr)  \\
&\quad + \sum_{j \le k}\, \log |\mu_j| + \sum_{j > n - k}\, \log |\mu_j|
- \sum_{i - k < j \le i+k}\, \log |\lambda_j|\\
& \le 2k \log \|B\| - 2k \log \dist(\rsp A, 0)\ .
\end{split}
\end{equation*}
\end{proof}

\begin{lemma}\label{lem:B-4}
Let $A, B$ be $n\times n$ hermitian matrices, with eigenvalues $\lmap$ and 
 $\mmap$ respectively.  Then $|\lambda_j - \mu_j| \le \|A - B \|$, 
$j = 1, \dots, n$.
\end{lemma}

\begin{proof}
\begin{align*}
\mu_j & = \max_{\substack{x \in M_j\\ \|x\| =1}}(Bx, x)\quad \text{for some $\overline M_j, \dim \overline M_j = j$}\\[6pt]
\lambda_j & = \min_{\dim M = j}\ \max_{\substack{x \in M\\ \|x\| =1}}(Ax, x) \le \max_{\substack{x \in \overline M_j\\ \|x\| =1}}(Ax, x)\\
&\le \max_{\substack{x \in \overline M_j\\ \|x\| =1}}(Bx, x) + \max_{\substack{x \in \overline M_j\\ \|x\| =1}}\bigl((A-B)x, x\bigr)\\
& \le \mu_j + \|A - B\|\ .
\end{align*}
Similarly, $\mu_j \le \lambda_j + \|A - B\|$.
\end{proof}

\begin{lemma}\label{lem:B-5}
Let $A$ be an $n \times n$ hermitian matrix.  Given an interval $(E', E'')$, 
let $\ue = E' - n(E''- E')$, $\be = E'' + n(E'' - E')$.  If 
$m = \#\bigl(\rsp A\cap (\ue, \be)\bigr)$, then for any 
$E_1, E_2 \in (E', E'')$ one has
$$
\log |\det (A-E_2)| - \log |\det (A-E_1)|
\le 1 + m\log \| A-E_2 \| - m\log \dist(\rsp A, E_1)
$$
\end{lemma}

\begin{proof} We prove the case for $E_1 < E_2$.  (The case for $E_2 < E_1$ is 
similar.)  Let $\lmap$ be the eigenvalues of $A$, 
$\lambda_r \le \ue < \lambda_{r+1}$, 
$\lambda_s < \be \le \lambda_{s+1}$.
\begin{alignat*}{3}
\text{For $j > s$},   & \quad |E_2 - \lambda_j| & < |E_1 - \lambda_j|\\[6pt]
\text{For $j \le r$}, & \quad {|E_2 - \lambda_j|\over |E_1 - \lambda_j|}  \le 
{|E_2 - E_1| + |E_1 - \lambda_j|\over |E_1 - \lambda_j|} 
\le {E'' - E'\over n(E'' - E')} + 1 = 1 + {1\over n}\\[6pt]
\text{For $r < j \le s$}, & \quad {|E_2 - \lambda_j|\over |E_1 - \lambda_j|}  
\le {\| A - E_2\|\over \dist ( \rsp A, E_1)}
\end{alignat*}

\begin{equation*}
\begin{split}
\log |\det (A-E_2)| & = \log \prod^n_{j=1} |E_2 - \lambda_j|\\
& = \log \prod^n_{j=1} |E_1 - \lambda_j| + \sum_{j \le r}\, 
\log {|E_2 - \lambda_j|\over |E_1 - \lambda_j|} + \sum_{r < j \le s}\, \log {|E_2 - \lambda_j|\over |E_1 - \lambda_j|} 
+ \sum_{j > s}\, \log {|E_2 - \lambda_j|\over |E_1 - \lambda_j|}\\
& \le \log |\det (A-E_1)| + n \log \bigl(1 + {1\over n}\bigr) + m \Bigl( \log \|A - E_2\|\big | - \log \dist(\rsp A, E_1) \Bigr)\\[6pt]
\end{split}
\end{equation*}
The assertion now follows since $n \log \bigl(1 + {1\over n}\bigr) \le 1 $.
\end{proof}


\appendix
\setcounter{section}{2}
\section{1 dimensional difference Schr\"odinger equation}

Consider the 1 dimensional difference Schr\"odinger equation:
$$
-\vp(k-1) - \vp(k+1) + \lambda v(k) \vp(k)= E\vp (k)\ ,\qquad k \in \IZ
$$
where $v(k) \in \IR$, $|v(k)| \le C$.

The eigenvalues of this equation with Dirichlet boundary conditions, $\vp(a-1) = \vp(b+1) = 0$, are the same as the eigenvalues of
$$
H_{[a, b]} = \begin{pmatrix}
\lambda v(a) & -1 &&\\
-1 & \lambda v(a+1) & -1 &\\
& -1 & \ddots &\\
& \ddots && -1\\
&& -1 & \lambda v(b)
\end{pmatrix}
$$
Given initial condition $\vp(a-1), \vp(a)$, the solution of this equation, for $b \ge a$, can be written as
$$
\begin{pmatrix}
\vp(b+1)\\
\vp(b)\end{pmatrix} = M_{[a, b]} (E) \begin{pmatrix}
\vp(a)\\
\vp(a-1)\end{pmatrix}
$$
where $M_{[a, b]}(E)$ is the monodromy matrix
$$
M_{[a, b]}(E) = \prod^a_{k=b}\, A_k, \qquad A_k = \begin{pmatrix}
\lambda v(k) - E & -1\\
1 & 0\end{pmatrix}
$$
The entries of the monodromy $M_n$ are
$$
M_{[a, b]}(E) = \begin{pmatrix}
f_{[a, b]}(E) & -f_{[a+1, b]}(E)\\
f_{[a, b-1]}(E) & -f_{[a+1, b-1]}(E)\end{pmatrix}
$$
where $f_{[a, b]}(E) = \det \bigl(H_{[a, b]} - E\bigr)$.

If $E \notin \rsp H_{[a, b]}$, then by Cramer's rule,
\begin{equation*}
\begin{split}
\bigl(H_{[a, b]} - E\bigr)^{-1}(k, \ell) & = \bigl(H_{[a, b]} - E\bigr)^{-1}(\ell, k)\qquad [a \le k \le \ell \le b]\\
& = {f_{[a, k-1]}(E) f_{[\ell+1, b]}(E)\over f_{[a, b]}(E)}\ .
\end{split}
\end{equation*}
By convention, $f_{[a, a-1]}(E) = f_{[b+1, b]}(E) = 1$.

\bigskip
\noindent
{\bf Poisson's Formula:} If $-\vp(n-1) - \vp(n+1) + v(n)\vp(n) = E\vp(n)$, $E\notin \rsp H_{[a, b]}$.  Then for $a \le m \le b$
$$
\vp(m) = \bigl(H_{[a, b]} -E\bigr)^{-1}(m, a)\vp(a-1) + \bigl(H_{[a, b]} - E\bigr)^{-1}(m, b)\vp(b+1)\ .
$$

\bigskip
\noindent
{\it Notation:}
\begin{align*}
T_{[a, b]} & = H_{[a, b]} - \diag\bigl(\lambda v(a), \dots, \lambda v(b)\bigr)\\
& = \begin{pmatrix}
0 & -1 &&&\\
-1 & 0 & -1 & &\\
& -1 & 0 & \ddots &\\
&& & \ddots & -1\\
&&&-1 & 0
\end{pmatrix}
\end{align*}
Note that $\big \|T_{[a, b]}\big\| \le 2$.

Fix $a, b$.  Let $n = b - a + 1$.

\begin{lemma}\label{lem:C-1}
If $\mu = {1\over 2} \min\limits_{a \le j \le b} |\lambda v(j) - E| > 2$, then $\dist (\rsp H_{[a, b]}, E) > \mu$ and  $\log |f_{[a, b]}(E)| > n \log \mu$.
\end{lemma}

\begin{proof} Let $\mmap$ be the eigenvalues of $\bigl(H_{[a, b]} -E\bigr)$,
$\tilde\mu_j = \lambda v(i_j) - E\ $, 
$\ \tilde\mu_1 \le \dots \le \tilde\mu_n\ $.
By Lemma B.4, $|\mu_j - \tilde\mu_j| \le \big \|T_{[a, b]}\big \| \le 2$.  Hence $|\mu_j| \ge 2 \mu - 2 > \mu$.  Therefore
$$
\log |f_{[a, b]}(E)| = \Sigma \log |\mu_j| > n \log \mu\ .
$$
\end{proof}

\begin{lemma}\label{lem:C-2}
If $\mu = {1\over 2} \min\limits_{j \ne j_0}| \lambda v(j) -E| > 2$, then $\#\bigl(\rsp(H_{[a, b]}-E)\cap [-\mu, \mu]\bigr) \le 1$ and
$$
\log |f_{[a, b]}(E)| > (n-1) \log \mu + \log \dist(\rsp H_{[a, b]}, E)\ .
$$
\end{lemma}

\begin{proof} Let $\mmap$ be the eigenvalues of $\bigl(H_{[a, b]} -E\bigr)$, 
$$
\tilde \mu_j = \lambda v(i_j) - E\ ,\quad \tilde\mu_1 \le \dots \le \tilde\mu_n\ ,\quad \tilde\mu_{\bar j} = \lambda v(j_0) - E\ .
$$
Then $|\mu_j| > \mu$ for $j \ne \bar j$
\begin{align*}
\log |f_{[a, b]}(E)| & = \sum_{j \ne \bar j}\, \log |\mu_j| + \log |\mu_{\bar j}|\\
& > (n-1) \log \mu + \log \dist \bigl(\rsp H_{[a, b]}, E\bigr)
\end{align*}
\end{proof}

\begin{lemma}\label{lem:C-3}
If $\mu = {1\over 2} \min |\lambda v(j) - E| > 2$, $|E| \le \max |\lambda v(j)| = \lambda C$ then $$
\log \big |(H_{[a, b]} -E)^{-1}(k, \ell) \big | < - {1\over 2} |\ell - k| \log \mu $$ 
provided $|\ell -k| > \lambda^\vt$, $\lambda > \lambda_0 (\vt, C)$.
\end{lemma}

\begin{proof} Without loss of generality, assume $\ell > k$.  Let
\begin{align*}
H_{[a, b; k, \ell]} & = \begin{pmatrix}
H_{[a, k-1]} &&\\
& H_{[k, \ell]} &\\
&& H_{[\ell +1, b]}
\end{pmatrix}\\
f_{[a, b; k, \ell]}(E) & = \det \bigl(H_{[a, b; k, \ell]} - E\bigr)\ 
\end{align*}
$\rank\bigl(H_{[a, b]} - H_{[a, b; k, \ell]}\bigr) \le 4$.  Since $\log\dist(\rsp H_{[a, b]}, E) > \log \mu > 0$, by Lemma B.3
$$
\log |f_{[a, b; k, \ell]}(E)| - \log |f_{[a, b]}(E)| \le 8 \log \|H_{[a, b; k, \ell]} - E\|
$$
\begin{align*}
\log \big |(H_{[a, b]}-E )^{-1} (k, \ell) \big | & = \log \big |f_{[a, k-1]}(E)\big | + \log \big | f_{[\ell +1, b]}(E)\big | - \log\big |f_{[a, b]}(E)\big |\\
& = \log \big |f_{[a, b; k, \ell]}(E)\big | - \log \big |f_{[a, b]}(E) \big | - \log \big |f_{[k, \ell]}(E)\big |\\
& \le 8 \log \big \|H_{[a, b; k, \ell]} - E\big \| - \log \big |f_{[k, \ell]}(E)\big |\\
& \le 8 \log (2 \lambda C + 2) - (\ell - k +1) \log \mu\\
& < - {1\over 2} (\ell -k)\log \mu\qquad \text{provided}\ (\ell - k) > \lambda^\vt
\end{align*}
\end{proof}

\begin{lemma}\label{lem:C-4}
If $\mu = {1\over 2} \min\limits_{j \ne j_0} |\lambda v(j) - E| > 2$, $|E| \le \max |\lambda v(j) | = \lambda C$, $E \notin {\rsp} H_{[a, b]}$. Then
$$
\log \big |(H_{[a, b]} -E)^{-1} (k, \ell)\big | < - {1\over 2} |k - \ell| \log \mu + 8 \big | \log \dist (\rsp H _{[a, b]}, E)\big |
$$
provided $|k - \ell| > \lambda^\vt$, $\lambda > \lambda_0(\vt, C)$.
\end{lemma}

\begin{proof} Without loss of generality, assume $k < \ell$.  If $j_0 < k < \ell$ or $k < \ell < j_0$, then
\begin{align*}
\log \big | (H_{[a, b]} - E)^{-1} (k, \ell)\big | & = \log \big |f_{[a, k-1]}(E) \big | + \log \big |f_{[\ell +1, b]}(E) \big | - \log \big | f_{[a, b]}(E)\big |\\
& = \log \big | f_{[a, b; k, \ell]}(E)\big | - \log \big | f_{[a, b]}(E)\big | - \log \big | f_{[k, \ell]}(E)\big |\\
& \le 8 \log \big \|H_{[a, b; k, \ell]} -E\big \| + 8 \big |\log \dist(\rsp H_{[a, b]},E)\big | - (\ell -k +1) \log \mu\\
& <  - {1\over 2} |k - \ell| \log \mu + 8 |\log \dist(\rsp H_{[a, b]}, E)|
\end{align*}

Suppose $k \le j_0 \le \ell$.  Let
\begin{align*}
\th_{[a, b]}(E)& = \diag\bigl(\lambda v(a), \dots, \lambda v(j_0 -1), 2\mu + E, \lambda v(j_0 +1), \dots, \lambda v(b)\bigr) + T_{[a, b]}\\
\tilde f_{[a, b]}(E) & = \det \bigl(\th_{[a, b]}(E) - E\bigr)\ .
\end{align*}
By Lemma C.3,
$$
\log \big |(\th_{[a, b]}(E) - E)^{-1} (k, \ell)\big | \le 8 \log (2 \lambda C + 2) - (\ell -k +1) \log \mu
$$
$\rank\bigl(\th_{[a, b]}(E) - H_{[a, b]}\bigr) = 1$.  By Lemma B.3
\begin{align*}
\log \big | \tf_{[a, b]}(E)\big | - \log \big |f_{[a, b]}(E)\big | & \le 2 \log \big \|\th_{[a, b]}(E) - E\big \| - 2 \log \dist(\rsp H_{[a, b]}, E)\\
& \le 2 \log (2\lambda C + 2) + 2 \big |\log\dist(\rsp H_{[a, b]}, E)\big |\\[6pt]
\log \big |(H_{[a, b]} -E)^{-1} (k, \ell) \big | & = \log \big |f_{[a, k-1]}(E)\big | + \log \big |f_{[\ell+1, b]}(E) \big | - \log \big |f_{[a, b]}(E)\big |\\
& = \log \big |(\th_{[a, b]}(E) - E)^{-1} (k, \ell)\big | + \log \big |\tf_{[a, b]} (E) \big | - \log \big |f_{[a, b]}(E) \big |\\
& \le 10 \log (2 \lambda C + 2) - (\ell -k +1) \log \mu + 2 \big |\log \dist(\rsp H_{[a, b]}, E)\big |\\
& < - {1\over 2} |k - \ell| \log \mu + 8 \big | \log \dist(\rsp H_{[a, b]}, E)\big |
\end{align*}
\end{proof}

\begin{lemma}\label{lem:C-5}
Let $K \in M_{n\times n}(\IC)$.  Then
$\big |\log |\det(I + K)| \big | \lesssim n \|K\|$ provided $\|K\| < {1\over 2}$.
\end{lemma}

\begin{proof} $\rsp (I + K) \subset \left\{z \in \IC: |z-1| \le \|K\|\right\}$, $\det (I+K) = \Pi \lambda$ where the product runs over the eigenvalues of $I+K$ with corresponding algebraic multiplicity.
So
$$
\bigl({1\over 2}\bigr)^n < \bigl(1 - \|K\|\bigr)^n \le |\det(I + K)| \le \bigl(1 + \|K\|\bigr)^n
$$
if $\|K\| < {1\over 2}$.  Therefore
$$
\big | \log |\det(I + K)| \big | \lesssim n \|K\| \ .
$$
\end{proof}

\begin{lemma}\label{lem:C-6}
$$
\Bigg | \log \bigg|\det \begin{pmatrix}
a_1 & 1 & &&\\
1 & a_2 & 1 &&\\
& 1 & \ddots &&\\
&&&&1\\
&&&1 & a_n\end{pmatrix}
\bigg | - \sum^n_{j=1} \log |a_j| \Bigg | \lesssim {n\over \min |a_j|}
$$
provided $\min |a_j| > 2$.
\end{lemma}

\begin{proof}

$$
\det \begin{pmatrix}
a_1 & 1 & &&\\
1 & a_2 & 1 &&\\
& 1 & \ddots &&\\
&&&&1\\
&&&1 & a_n\end{pmatrix} =
\Biggl(\prod^n_{j=1}\, a_j\Biggr)
\begin{pmatrix}
1 & 1/a_1 & &\\
1/a_2 & 1 & 1/a_2 &\\
& 1/a_3 & & \ddots \\
&& \ddots & \end{pmatrix}
$$

$$
\Bigg | \log \bigg | \det
\begin{pmatrix}
a_1 & 1 & &&\\
1 & a_2 & 1 &&\\
& 1 & \ddots &&\\
&&&&1\\
&&&1 & a_n\end{pmatrix}
\bigg | - \sum^n_{j=1} \log |a_j| \Bigg | =
\Bigg | \log \bigg | \det \bigg[I + 
\begin{pmatrix}
0 & 1/a_1 & &\\
1/a_2 & 0 & 1/a_2 &\\
& 1/a_3 & & \ddots \\
&&\ddots & \end{pmatrix}
\bigg]\bigg |\Bigg | \lesssim {n\over \min |a_j|}
$$
provided $\min |a_j| > 2$.
\end{proof}

\begin{lemma}\label{lem:C-7}
If $\min\limits_{a \le k \le b} |\lambda v(k) - E| \ge \lambda^A$, $a \le i < j < b$.  Then
$$
\Big | {f_{[a, i]}(E) f_{[j, b]}(E)\over f_{[a, b]}(E)}\Big | \le \lambda^{-A|j - i -1|}
$$
provided $n = b - a +1 \lesssim \lambda^A$.
\end{lemma}

\begin{proof}
\begin{align*}
\log \big |f_{[a, i]}(E) \big | & \le \sum^i_{k=a} \log |\lambda v(k) - E| + n \lambda^{-A}\\
\log \big | f_{[j, b]}(E)\big | & \le \sum^b_{k=j} \log |\lambda v(k) - E| + n \lambda^{-A}\\
\log \big | f_{[a, b]}(E)\big | & \ge \sum^b_{k=a} \log |\lambda v(k) - E| - n \lambda^{-A}
\end{align*}

\begin{align*}
\log \Big |{f_{[a, i]}(E) f_{[j, b]}(E)\over f_{[a, b]}(E)} \Big | & \le - \sum_{i < k < j} \log |\lambda v(k) - E| + 3 n\lambda^{-A}\\
& \le - |j - i -1| \log \lambda^A + 3n \lambda^{-A}
\end{align*}
\end{proof}

\begin{corollary}\label{cor:C-8}
$\big | (H_{[a, b]} - E)^{-1} (i, j)\big | \le \lambda^{-A(|i - j| +1)}$ if $\min\limits_{a \le k \le b} |\lambda v(k) - E| \ge \lambda^A$, $n \lesssim \lambda^A$.
\end{corollary}

\begin{lemma}\label{lem: C-9}
Suppose $[a, b] = \bigcup\limits^K_{k=1} [a'_k, b'_k]$, $a'_1 = a$, $b'_K = b$, $a'_k < b'_{k-1} < a'_{k+1} < b'_k$ for \newline $k = 2, 3, \dots, K-1$, $b'_{k-1} - a'_k > n^\tau$ for $k = 2, 3, \dots, K$.  Assume that
$$
\big | (H_{[a', b']} - E)^{-1} (i, j) \big | \le \exp(-\gamma|i - j|)\ 
$$
when $|i - j| \ge {1\over 2} n^\tau$ for all $k$.  Then $E \notin \rsp H_{[a, b]}$ provided $n > n_0 (\tau, \gamma)$.
\end{lemma}

\begin{proof} Suppose, for contradiction, that $H_{[a, b]} \vp = E\vp$, $\|\vp\| = 1$.  Let
$$
x_k = \lfloor {1\over 2} (a'_k + b'_{k-1})\rfloor\ ,\quad k = 2, \dots, K\ .
$$
Then $a'_k < x_k < x_{k+1} < b_k$.  For any $j \in [x_k, x_{k+1}]$, by Poisson's formula
\begin{equation*}
\begin{split}
\vp(j) & = \bigl(H_{[a'_k, b'_k]} - E\bigr)^{-1} (j, a'_k)\vp (a'_k -1) + 
\bigl(H_{[a'_k, b'_k]} - E\bigr)^{-1} (j, b'_k) \vp(b'_k +1)\\
|\vp(j)| & \le 2 \exp \bigl[- {\gamma\over 2} n^\tau\bigr]
\end{split}
\end{equation*}
since $|j - a'_k| \ge {1\over 2} n^\tau$, $|j - b'_k| \ge {1\over 2} n^\tau$.

Similarly, for $j \in [a, x_2] \cup [x_k, b]$,
$$
|\vp (j)| \le \exp \bigl[ - {\gamma\over 2} n^\tau\bigr]
$$
Since 
\begin{align*}
[a, b]  &= \Bigl(\bigcup\limits^{K-1}_{k = 2} [x_k, x_{k+1}]\Bigr) \cup [a, x_2] \cup [x_k, b]\\[6pt]
1 & = \sum^b_{j=a} |\vp(j)|^2 \le n \bigl[4 \exp(-\gamma n^\tau)\bigr]
\end{align*}
But this is false for sufficiently large $n$.
\end{proof}

\begin{lemma}\label{lem:3-9} Suppose 
$|E| \le \lambda \max\limits_{a \le n \le c}  |v(n)| = : \lambda C_0$, 
$\dist \bigl(\rsp \hac, E\bigr) = \ka$.  If $a < b < c$ then
$$
\log \big \| M_{[a, b]}(E)\big \| + \log \big \| M_{[b+1, c]}(E)\big \|
- \log \big \|M_{[a, c]}(E) \big \| \le 20 \bigl[\log(\lambda C_0) - \log \ka\bigr]
$$
provided $\lambda C_0 \gg 1$.
\end{lemma}

\begin{proof} 
\begin{equation*}
\begin{split}
\| M_{[a, c]} (E) \| & \ge |\facmap(E)|\\[6pt]
\| M_{[a, b]}(E)\| & \le |\fmap(E)| + |f_{[a, b-1]}(E)| + |f_{[a+1, b]}(E)| + |f_{[a+1, b-1]}(E)|\\[6pt]
\|M_{[b+1, c]}(E)\| & \le |f_{[b+1, c]}(E)| + |f_{[b+1, c-1]}(E)| + |f_{[b+2, c]}(E)| + |f_{[b+1, c-1]}(E)|
\end{split}
\end{equation*}

\begin{equation}
{\|M_{[a, b]}(E)\|\, \|M_{[b+1, c]}(E)\|\over
\|M_{[a, c]} (E)\|}  \le
{\bigl(|\fmap(E) + \cdots \bigr)\bigl(|f_{[b +1, c]}(E)| + 
\cdots\bigr)\over |\facmap (E)|} 
\end{equation}
Expand the numerator, we get 16 terms.  One of the terms is
$$
|f_{[a+1, b-1]}(E)|\, |f_{[b+2, c-1]}(E)| = \Bigg | \det \begin{pmatrix}
I_1 &&&&\\
& H_{[a+1, b-1]}&&&\\
&& I_2 &&\\
&&& H_{[b+2, c-1]} &\\
&&&& I_1\end{pmatrix}  \Bigg | =: |\det \tilde H|\\
$$
$\rank\bigl(\hac - \tilde H\bigr) = 8$.  By Lemma B.3 $$
\log |f_{[a+1, b-1]}(E) f_{[b+1, c-1]}(E)| - \log | \facmap (E)
 \le 16 \log \|\tilde H\| - 16\log \dist (\rsp \hac, E) $$
Therefore $$
{|f_{[a+1, b-1]}(E)|\, |f_{[b+1, c-1]}(E)|\over |\facmap (E)|} \le \left({2\lambda C_0 + 2\over \ka}\right)^{16} $$
Same estimate holds for other terms in (C.1).  So
\begin{equation*}
\begin{split}
 & {\|M_{[a, b]}(E)\|\, \|M_{[b+1, c]}(E)\|\over
\|M_{[a, c]} (E)\|} \le 16 \left({2 \lambda C_0+2\over \ka}\right)^{16}\\
\end{split}
\end{equation*}
Hence
$$
\log \|M_{[a, b]}(E)\| + \log \|M_{[b+1, c]}(E)\|
- \log \|M_{[a, c]}(E)\| \le \log 16 + 16\log \left({2 \lambda C_0 + 2\over \ka}\right)
$$
\end{proof}


\appendix
\setcounter{section}{3}
\section{Avalanche Principle}

\begin{prop} \textbf{(Goldstein, Schlag)}[GS1] Let $A_1, \dots, A_n$ be a sequence of $2\times 2$ matrices whose determinants satisfy
$$
\max_{1 \le j \le n} |\det A_j| \le 1\ .
$$
Suppose that
$$
\min_{1 \le j \le n} \|A_j\| \ge \mu > n
$$
and
$$
\max_{1 \le j < n} \bigl[\log \|A_{j+1}\| + \log \|A_j\| - \log \|A_{j+1} A_j\| \bigr] < {1\over 2} \log \mu\ .
$$
Then
$$
\big | \log \|A_n\dots A_1\| + \sum^{n-1}_{j=2} \log \|A_j\| - \sum^{n-1}_{j=1} \log \|A_{j+1} A_j\| \big | \lesssim {n\over \mu}\ .
$$
\end{prop}

\begin{prop} \textbf{(Goldstein, Schlag)}[GS2] For $a_0 < a_1 < \dots < a_n$, if the monodromy matrices $M_{(a_{j-1}, a_j]} =: A_j$ satisfy the conditions of Proposition 1, then
$$
\Big | \log \|M_{(a_0, a_n]}\| + \sum^{n-1}_{j=2} \log \|M_{(a_{j-1}, a_j]}\| - \sum^n_{j=2} \log \| M_{[a_{j-2}, a_j]}\| \Big | \lesssim {n\over \mu}\ .
$$
Moreover, if $A_j := M_{(a_{j-1}, a_j]}$ for $1 < j < n$,
$$
A_1 := M_{[a_0, a_1]}\begin{pmatrix}
1 & 0\\ 0 & 0\end{pmatrix} \quad \text{and}\quad A_n := \begin{pmatrix}
1 & 0\\ 0 & 0\end{pmatrix} M_{(a_{n-1}, a_n]}
$$
satisfy the conditions of the proposition, then
\begin{equation*}
\begin{split}
 & \Bigg | \log f_{[a_0, a_n]}(E) + \sum^{n-1}_{j=2} \log \|M_{(a_{j-1}, a_j]}\| - \sum^{n-1}_{j=3} \log \|M_{(a_{j-2}, a_j]}\|\\
&\qquad - \log \Big \|M_{[a_0, a_2]}\begin{pmatrix}
1 & 0\\ 0 & 0\end{pmatrix}\Big \| - \log \Big \| \begin{pmatrix}
1 & 0\\ 0 & 0\end{pmatrix} M_{(a_{n-2}, a_n]}\Big\|\Bigg | \lesssim{n\over \mu}
\end{split}
\end{equation*}
\end{prop}


\section{Implicit Function}
\definition $D_0 \subset \IR^2$ is $n$--simple if $D_0 = \bigcup\limits^{J_0}_{j=1} D_j$, where $D_j$ are rectangles, $J_0 \le n$.

\begin{prop}\label{lem:E-1}
Let $f(x,y)$ be $C^1$ on $\ab \times (c, d) = : D$.  Assume $\partial_y f(x,y) \ge \mu > 0$, $\big |\partial_x f(x,y) \big | \le \ka\ for all\ (x,y) \in D$.  Given $\rho > 0$, there exists $n$--simple set $D_0 \subset D$, $n = 2m$, $|D \setminus D_0| \le (b-a) \bigl({2\rho\over \mu} + {1\over m}\, {\ka\over \mu}\bigr)$ such that $|f(x,y)| \ge \rho$ on $D_0$.
\end{prop}

\begin{proof} For $x \in \ab$, define
$s^+(x) = \{y \in (c, d): f(x,y) \ge \rho\}, \qquad 
s^-(x) = \{y \in (c,d): f(x,y) \le -\rho\}$ and

\begin{align*}
y^+(x) & = \begin{cases}
d & \text{if}\ s^+ = \emptyset\\
\inf s^+ &\hphantom{\text{if}\ }s^+ \ne \emptyset\end{cases}\\[6pt]
y^-(x) & = \begin{cases}
c & \text{if}\ s^- = \emptyset\\
\sup s^- & \hphantom{\text{if}\ } s^- \ne \emptyset
\end{cases}
\end{align*}
Let
$$
J_k  = \left(a + {k-1\over m} (b-a),\ a + {k\over m}(b-a)\right), \qquad
x_k  = a + {2k-1\over 2m} (b-a), \qquad k = 1, 2, \dots, m,
$$
\begin{align*}
y_{k, 1} & = \min \left\{y^+(x_k) + {1\over 2m}\, {\ka \over \mu}\, , d\right\}\\[6pt]
y_{k, 2} & = \max \left\{y^-(x_k) - {1\over 2m}\, {\ka\over \mu}\, , c\right\}
\end{align*}
Take
\begin{align*}
D_{k, 1} & = J_k \times (y_{k, 1}, d)\\[6pt]
D_{k, 2} & = J_k \times (c, y_{k, 2})\\[6pt]
D_0 & = \bigcup_k (D_{k, 1}\cup D_{k,2})
\end{align*}
Then $|f(x,y)| \ge \rho$ on $D_0$ and
$$
0 < y_{k, 1} - y_{k, 2} \le {2\rho\over \mu} + {1\over m}\, {\ka\over \mu}
\Longrightarrow
|D \setminus D_0| \le (b-a)\bigl({2\rho\over \mu} + {1\over m}\, {\ka\over \mu}
\bigr)
$$
\end{proof}

\begin{lemma} Suppose $f:[a,b] \to \IR$ is $C^2$, monotone.  If $|f''(x)| \ge C$ for all $x \in (a,b)$, then
$$
\mes \left\{x \in [a, b]: |f(x)| \le \ve\right\} \lesssim \bigl({\ve\over C}\bigr)^{1/2}\ .
$$
\end{lemma}

\begin{proof} Without loss of generality, assume $f$ is increasing. 
(Otherwise, consider $-f$.)

Let $(\tilde a, \tilde b) \subset [\bar a, \bar b] = \bigl\{x \in [a, b]: |f(x)| \le \ve\bigr\}$.  If $f''(x) \ge C$ then there is 
$\tilde c \in (\tilde a, \tilde b)$ such that
\begin{align*}
f(\tilde b) - f(\tilde a) & = f'(\tilde a)(\tilde b - \tilde a) + {1\over 2} f''(\tilde c)(\tilde b - \tilde a)^2\\[6pt]
2 \ve & \ge {1\over 2}C(\tilde b - \tilde a)^2\\[6pt]
\tilde b - \tilde a & \lesssim \bigl({\ve\over C}\bigr)^{1/2}\ .
\end{align*}
If $f''(x) \le -C$ then
\begin{align*}
f(\tilde a) - f(\tilde b) & = f'(\tilde b)(\tilde a - \tilde b) + {1\over 2} f''(\tilde c)(\tilde b - \tilde a)^2\\[6pt]
f(\tilde b) - f(\tilde a) & = f'(\tilde b)(\tilde b - \tilde a) + {1\over 2}\bigl[-f''(\tilde c)\bigr](\tilde b - \tilde a)^2\\[6pt]
2\ve & \ge {1\over 2}C(\tilde b - \tilde a)^2\\[6pt]
\tilde b - \tilde a & \lesssim\bigl({\ve\over C}\bigr)^{1/2} \\[6pt]
\mes\bigl\{x \in [a, b]: |f(x)| \le \ve\bigr\} & = \sup(\tilde b - \tilde a) \lesssim \bigl({\ve\over C}\bigr)^{1/2}
\end{align*}
\end{proof}

\begin{corollary} Suppose $f:[a, b] \to \IR$ is $C^3$, such that 
$|f'(x)| + |f''(x)| \ge C$, $|f''(x)| + |f'''(x)| \le \K$.  Then for 
$\ve \lesssim C \lesssim \K(b-a)$, one has
$$
\mes \left\{x \in [a, b]: |f(x)| \le \ve\right\} 
\lesssim {\K\over C}\bigl({\ve\over C}\bigr)^{1/2}(b-a).
$$
\end{corollary}

 \begin{proof} Let $x_i = a + {b-a\over N}i$, $i = 0, 1, 2, \dots, N$ where 
${C\over 8\K} \le {b-a\over N} \le{C\over 4\K}$.  On each $x_i$, either 
$|f'(x_i)| \ge {C\over 2}$ or $|f''(x_i)| \ge {C\over 2}$.  

If $|f''(x_i)| \ge {C\over 2}$ then $|f''(x)| \ge {C\over 4}$ for all $x \in [x_{i-1}, x_i]$ since $\sgn f''$ does not change in $[x_{i-1}, x_i]$, $f$ has at most two monotonicity intervals in $[x_{i-1}, x_i]$.  Hence  
$\mes\left\{x \in [x_{i-1}, x_i]: |f(x)| \le \ve\right\} 
\lesssim \bigl({\ve\over C}\bigr)^{1/2}$.

If $|f'(x_i)| \ge {C\over 2}$ then $|f'(x)| \ge {C\over 4}$ for all $x \in [x_{i-1}, x_i]$.  So  
$\mes\left\{x \in [x_{i-1}, x_i]: |f(x)| \le \ve\right\} \le {2\ve\over C/4} 
\lesssim \bigl({\ve\over C}\bigr)^{1/2}$.
Therefore
$$
\mes \left\{x \in [a, b]: |f(x)| \le \ve\right\} \lesssim N\bigl({\ve\over C}\bigr)^{1/2} \lesssim {\K\over C}\bigl({\ve\over C}\bigr)^{1/2}(b-a)\ .
$$
\end{proof}

Let $F_\ell(x,y)$ be $C^3$ on $(a_\ell, b_\ell)\times (c, d)$, where  
$(a_\ell, b_\ell),\  (c,d)\subset \mathbb{T}$, $\ell \in \{1, 2\}$.  Suppose
$$
0 < \mu \le  |\partial_y F_1 (x,y) |
$$
$$
|\partial_y F_\ell|, |\partial_x F_\ell|, |\partial_{xx} F_\ell|, 
|\partial_{yy} F_\ell|, |\partial_{xy} F_\ell| \le \K
$$
Since $\partial_y F_1 \ne 0$, we can define $y(., t)$ such that $F_1\bigl(x, y(x, t)\bigr) = t$ for $x \in \pi_1\bigl(F^{-1}_1(t)\bigr)$.  Furthermore, 
$$
|\partial_x y(x,t)| \le {\ka\over \mu}, \qquad 
|\partial_{xx}y(x, t)| \le {4\K^3\over \mu^2}
$$
Let $x_i = a_1 + {i\over N}$, $i = 0, 1, 2, \dots, \lfloor N(b_1 - a_1)\rfloor =: N_1$.  Define
$$
\xi_i = \min \left\{ |\partial_x y(x, t)| : 
(x,t) \in [x_{i-1}, x_i] \times [-\ve, \ve] \right\} \ .
$$
Suppose $\xi_i \le \vr$.  Then there is $t \in [-\ve, \ve]$ such that
$$
\mes \left\{y \left(\bigl[x_{i-1}, x_i\bigr], t\right)\right\} \le {\vr\over N} + {4\K^3\over \mu^3 N^2}\ .
$$
Since $|y(x, s) - y(x, t)| \le {|s-t|\over \mu}$,
$$
\tcup_{t \in [-\ve, \ve]} \left\{y \left(\bigl[x_{i-1}, x_i\bigr], t\right)\right\} \subset \bigl[\tilde c_i, \tilde d_i\bigr]
$$
$\tilde d_i - \tilde c_i \le 
{\vr\over N} + {4\K^3\over \mu^2 N^2} + {2\ve\over \mu}$.  Write 
$(c, d) \setminus \tcup_{i: \xi_i \le \vr} [\tilde c_i, \tilde d_i] 
= \tcup^J_{j=1}(c_j, d_j)$, $J \le N_1 + 1$.  On each $(c_j, d_j)$, define 
the inverse functions of $y$, i.e. $F_1\bigl(\phi_{jk}(y, t), y\bigr) = t$ for 
$y \in (c_j, d_j)\cap \pi_2\bigl(F^{-1}_1(t)\bigr)$, $k = 1, 2, \dots, K_j$, 
$K_j \le N_1 + {b_1-a_1\over d - c}\, {\ka\over \mu} \le 2N_1$ provided 
$N \ge {1\over d - c}\ {\ka\over \mu}$.

\begin{lemma} Let $y_0 \in (c_j + {2 \ve \over \mu}, d_j - {2 \ve \over \mu})$.
  Then $|\phi_{jk}(y_0, s) - \phi_{jk}(y_0, t)| \le {2\ve\over \mu \vr}$ for 
any $s,t \in [-\ve,\ve]$ such that $\phi_{jk}(y_0,s)$ and $\phi_{jk}(y_0,t)$ 
are defined.
\end{lemma}

\begin{proof} Let $a = \phi_{jk}(y_0, s),\ b = \phi_{jk}(y_0, t)$.  Then 
$F_1(a,y_0)=s$. Since $|\partial_y F_1 (x,y)|>\mu$ and $|s-t|\le 2\ve$, 
there is
$y_1 \in [y_0 - {2\ve \over \mu}, y_0 + {2\ve \over \mu}] \subset (c_j,d_j)$ 
such that $F_1(a,y_1)=t$.  Hence, $y(x,t)$ is defined for all $x$ between 
$a$ and $b$.  Therefore, there exists $\tilde x$ between $a$ and $b$ such that 
$$
{2\ve / \mu \over |\phi_{jk}(y_0, s) - \phi_{jk}(y_0, t)|} \ge
{|y(a, s) - y(a, t)|\over |\phi_{jk}(y_0, s) - \phi_{jk}(y_0, t)|}
= |\partial_x y(\tilde x,t)| > \vr
$$
\end{proof}

\begin{lemma} Suppose $b_1-a_1>\ve^{1/100}$ and $d-c>\ve^{1/100}$.  If 
$$ |\partial_x F_2(x,y)| + |\partial_{xx}F_2(x,y)| \ge C > \ve^{1/100} $$ 
for all $(x,y) \in (a_2, b_2)\times (c, d)$, then for
$\K \le \ve^{-{1/100}}$, $\mu \ge \ve^{1/100}$, one has
$$
\mes\left\{y \in (c, d):
\begin{array}{l}
\exists\ x \in (a_1, b_1),\ n > \ve^{-1/10}\\
\text{such that $|F_1(x, y)| \le \ve,\ |F_2(x+ ny, y)| \le \ve$}
\end{array}
\right\} \lesssim \ve^\vt(d-c)(b_1-a_1)\ .
$$
\end{lemma}

\begin{proof} Choose $\vr = \ve^{1/25},\ N \asymp \ve^{-1/10}$.  
Fix $jk, t, n$. Let $f(y) = F_2\bigl(\phi_{jk}(y, t) + ny, y\bigr)$.  Then
\begin{eqnarray*}
f'(y) & = &(\partial_x F_2) (\partial_y \phi_{jk} + n) + \partial_y F_2\\
f''(y) & = &(\partial_{xx} F_2)(\partial_y\phi_{jk} + n)^2 
+ (\partial_x F_2)(\partial_{yy}\phi_{jk})
+ 2(\partial_{xy} F_2)(\partial_y \phi_{jk}+n) + \partial_{yy} F_2\\
\end{eqnarray*}
for all $y$ where $f$ is defined.

\begin{eqnarray*}
|f'(y)| + |f''(y)| & \ge & \begin{cases}
{C \over 2}(n - \vr^{-1}) - \ka & \text{if $|\partial_x F_2|\ge {C\over 2}$} 
\\[6pt]
{C \over 2}(n - \vr^{-1})^2  - \ka {4 \ka^3 \over \mu^2 \vr^3} 
        - 2 \ka (\ka + n) - \ka & \text{if $|\partial_{xx}F_2|\ge {C\over2}$}
\end{cases} \\[6pt]
& \ge & {nC \over 4}
\end{eqnarray*}
By Corollary E.3,
\begin{align*}
& \mes \left\{y \in (c_j + {2\ve \over \mu}, d_j - {2\ve \over \mu}): 
|f(y)| \le \ve^{1/2}\right\} \lesssim 
{\K\over nC}\bigl({\ve^{1/2}\over nC}\bigr)^{1/2} (d_j - c_j)
\end{align*}
Also, by Lemma E.4,
$$
|F_2(\phi_{jk}(y, s) + ny, y) - F_2(\phi_{jk}(y, t) + ny, y)|
\le \K|\phi_{jk}(y, s) - \phi_{jk}(y, t)| \le 2{\K\ve\over \mu\vr}
$$
for all $s$ such that $\phi_{jk}(y,s)$ is defined.  
If $|f(y)| > \ve^{1/2}$ then
$$
|F_2\bigl( \phi_{jk}(y, s) + ny, y \bigr)| 
> |f(y)| - 2{\K\ve\over \mu\vr} > \ve
$$
Hence,
\begin{equation*}
\begin{split}
& \mes\left\{y \in (c, d): \begin{array}{l}
\exists\ x \in (a_1, b_1),\ n > \ve^{-1/10}\\
\text{such that $|F_1(x, y)| \le \ve,\ |F_2(x+ ny, y)| \le \ve$}
\end{array} \right\} \\
& \lesssim (N_1 +1)(2N_1){\K\over C}\Bigl({\ve^{1/2}\over C}\Bigr)^{1/2}(d-c)
\sum\limits_{n > \ve^{-1/10}} {1\over n^{3/2}} + N\Bigl({\vr\over N} + {4\K^3\over \mu^2 N^2} 
+ {6\ve\over \mu}\Bigr)\\
& \lesssim \ve^\vt (d-c)(b_1-a_1)
\end{split}
\end{equation*}
\end{proof}

Fix $n_1 \ll  n_2$.  Let
\begin{align*}
F_1(x,\w) & = E_1(x,\w) - E_2(x + n_1\w,\w)\qquad \text{for}\ (x,\w) 
\in (a_1, b_1)\times (c,d)\\[6pt]
F_2(\tx,\w) & = E_2(\tx,\w) - E_3(\tx + n_2\w,\w)\qquad \text{for}\ (\tx,\w) 
\in (a_2, b_2)\times (c,d)
\end{align*}
where $|\partial_x E_\ell(x,\w)| + |\partial_{xx} E_\ell(x,\w)| > \delta$.  
Also, assume
$$
|\partial_x E_\ell|,\ |\partial_\w E_\ell|,\ |\partial_{xx}E_\ell|,\ |\partial_{x\w}E_\ell|,\ |\partial_{\w\w}E_\ell|,\ |\partial_{xxx}E_\ell|,\ |\partial_{xx\w}E_\ell| \le \K
$$

\begin{theorem} For $\ve \ll n_1 \delta$,
$$
\mes\left\{\w \in (c, d):
\begin{array}{l}
\exists\ x \in (a_1, b_1),\ x + n_1\w \in (a_2, b_2)\\
\text{such that $|F_1(x,\w)| \le \ve,\ |F_2(x+ n_1 \w,\w)| \le \ve$}
\end{array}
\right\} \lesssim \ve^\vt(d-c)(b_1 - a_1)\ .
$$
\end{theorem}

\begin{proof} Let $\tx_i = a_1 + {n_1\delta\over 4\K}i$, $\ \tw_j = c + {n_1\delta\over 4\K}j$, $\ D_{ij} = [\tx_{i-1}, \tx_i] \times [\tw_{j-1}, \tw_j]$.  Either
\begin{enumerate}
\item[(1)] $|\partial_\w F_1| > {n_1\delta\over 4}$ for all $(x,\w) \in D_{ij}$; or
\item[(2)] $|\partial_{\w\w}F_1| > {n_1\delta\over 4}$ for all $(x,\w) \in D_{ij}$.
\end{enumerate}
Choose $\mu$ so that $\bigl({n_1\delta\over 4}\bigr)^{100}> \mu > \ve^{1/100}$.
  In case (1), 
$$
\mes \left\{ \w \in (\tw_{j-1}, \tw_j): 
\begin{array}{l}
\exists x \in (\tx_{i-1}, \tx_i)\\
|F_1(x,\w)| \le \ve,\ |F_2(x+n_1 \w, \w)| \le \ve
\end{array}\right\} \lesssim \ve^\vt (\tw_j-\tw_{j-1})(\tx_i-\tx_{i-1})
$$
In case (2), consider $G_1 = \partial_\w F_1$, $G_2 = F_2$.  Applying Lemma E.5 to $G_1, G_2$ gives
\begin{align*}
\mes\O_{ij} & := \mes\left\{\w \in (\tw_{j-1}, \tw_j):
\begin{array}{l}
\exists x \in (\tx_{i-1}, \tx_i)\\
|G_1(x, \w)| \le \mu\ ,\ |G_2(x + n_1 \w, \w)| \le \mu
\end{array}\right\} \\
 & \lesssim \mu^\vt (\tw_j-\tw_{j-1})(x_i-x_{i-1})
\end{align*}

Also,
$$
\mes \left\{
\begin{array}{l}
\w\in (\tw_{j-1}, \tw_j) \setminus \O_{ij}: \exists x \in (\tx_{i-1}, \tx_i)\\
|F_1(x,\w)| \le \ve\ ,\ |F_2(x+ n_1\w, \w)| \le \ve\end{array}
\right\} \lesssim \ve^\vt (\tw_j-\tw_{j-1})(\tx_i-\tx_{i-1})\ .
$$
Hence,
\begin{equation*}
\begin{split}
& \mes\left\{\w \in (c,d): \exists x \in (a_1, b_1), |F_1(x,\w)|\le \ve,\ |F_2(x+n_1 \w, \w)| \le \ve\right\}\\
& \lesssim \ve^\vt (d-c)(b_1 - a_1)
\end{split}
\end{equation*}
\end{proof}


\section{Volume of Hyperplane}
Given $0 \ne a = (a_1, \dots, a_N) \in \IR^N$, $C \in \IR$.  Let 
$P(a, C) = \left\{x \in \IR^N, a \cdot x = C\right\}$.  Suppose $a_i \ne 0$.  
For any $\delta > 0$ define
$$
P_{i,\delta}(a, C) = \left\{x \in P(a, C): |x_j| \le \delta\quad \text{for}\ j \ne i\right\}\ .
$$
Then $\rm{Vol}_{N-1} \bigl(P_{i, \delta}(a, C)\bigr) = {\|a\|\over |a_i|} (2\delta)^{N-1}$.

\begin{lemma} Suppose $a_j \ge 0$ for all $j = 1, \dots, N$, $\Sigma a_j = 1$.
For any $C \in \IR$, $\delta, \ve > 0$, let
$$
P_\delta(a, C, \ve) = \left\{\xi \in (\xi_1, \dots, \xi_N): |\xi_j| \le \delta\quad \text{for all}\ j,\ |a\cdot \xi - C| \le \ve\right\}
$$
Then $(2\delta)^{-N} \mes P_\delta(a, C, \ve) \le N{\ve\over \delta}$.
\end{lemma}

\begin{proof} Write $\xi = t{a\over \|a\|} + \eta$ where $\eta\cdot a = 0$.  Then $\xi \cdot a = t\|a\|$.

From the hypothesis, there is $|a_i| \ge {1\over N}$
\begin{align*}
& \xi \in P_\delta(a, C, \ve) \Longrightarrow t \in \Bigl[{C\over \|a\|} - {\ve\over \|a\|}, {C\over \|a\|} + {\ve\over \|a\|}\Bigr] = :J\\
& \mes P_\delta(a, C, \ve) \le \int_J P_{i, \delta}(a, y) dy = {2\ve\over |a_i|}(2 \delta)^{N-1}
\end{align*}
\end{proof}


\begin{thebibliography}{CFKS}

\bibitem[Bj]{bj} K. Bjerkl\"{o}v, \emph{Positive Lyapunov exponent and Minimality for a class of 1-D quasi-periodic Schr\"{o}dinger equations}, preprint.

\bibitem[Bo]{bo} J. Bourgain, \emph{Green's Function Estimates for Lattice Schr\"{o}dinger Operatiors and Applications,} Princeton University Press, 2004.

\bibitem[BG]{bg} J. Bourgain and M. Goldstein, \emph{On nonperturbative localization with quasi-periodic potential}, Ann. of Math. \textbf{152} (2000), 835-879.

\bibitem[CL]{cl} R. Carmona and J. Lacroix, \emph{Spectral Theory of Random Schr\"{o}dinger Operators}, Birkh\"{a}user Boston Inc., 1990.

\bibitem[CFKS]{cfks} H.L. Cycon, R.G. Froese, W. Kirsch, and B. Simon, \emph{Schr\"{o}dinger Operators with Application of Quantum Mechanics and Global Geometry}, Springer-Verlag, 1987.

\bibitem[GS1]{gs1} M. Goldstein and W. Schlag, \emph{H\"{o}lder continuity of the integrated density of states for quasi-periodic Schr\"{o}dinger equations and averages of shifts of subharmonic functions}, Ann. of Math. \textbf{154} (2001), 155-203.

\bibitem[GS2]{gs2} M. Goldstein and W. Schlag, \emph{Equidistribution of zeros of Dirichlet determinants and fine properties of integrated density of states}, preprint.

\bibitem[Kh]{kh} A. Ya. Khinchin, \emph{Continued Fractions}, The Univ. of Chicago Press, 1964.

\bibitem[Kl]{kl} S. Klein, \emph{Anderson localization for the discrete one-dimensional quasi-periodic Schr\"{o}dinger operator with potential defined by a Gevrey-class function}, J. Funct. Anal. \textbf{218} (2005), 255-292.

\bibitem[SoSp]{sosp} E. Sorets and T. Spencer, \emph{Positive Lyapunov Exponents for Schr\"{o}dinger Operators with Quasi-Periodic Potentials}, Comm. Math. Phys. \textbf{142} (1991), 543-566.

\bibitem[StSu]{stsu} G.W. Stewart and J. Sun, \emph{Matrix perturbation theory}, Academic Press, 1990.

\end{thebibliography}
\end{document}